\documentclass[11pt,a4paper]{article}

\usepackage{booktabs} % For professional looking tables
\usepackage{hyperref}
\usepackage{graphicx}
\usepackage{blindtext}
\usepackage[margin=7em]{geometry}
\usepackage{parskip}
\usepackage{multirow}
\usepackage{multicol}
\usepackage{amsmath}
\usepackage{amssymb}
\usepackage{natbib}
\usepackage{enumerate}
\usepackage{tikz}
\usepackage{changepage}
\usetikzlibrary{decorations.markings}
\usepackage[symbol]{footmisc}

\hypersetup{
    colorlinks=false,
    linktoc=all,
}

\newcommand{\eqreflong}[1]{(equation \ref{#1})}

\newcommand{\firstrevision}[1]{{\color{black}#1}}
\newcommand{\comment}[1]{}

\usepackage{amsthm}

\newcommand{\pdv}[2]{\frac{\partial #1}{\partial #2}}

\newcommand{\odv}[2]{\frac{{\rm d} #1}{{\rm d} #2}}

\newcommand{\dx}[1]{{\rm d}#1}

\newcommand{\ndim}[1]{\hat{#1}} % non-dimensionalised scalars
\newcommand{\vnd}[1]{\vec{\ndim{#1}}} % non-dimensionalised vectors
\newcommand{\rom}[1]{\textup{\uppercase\expandafter{\romannumeral#1}}}

\renewcommand{\vec}[1]{\boldsymbol{#1}}  % vector notation
\newcommand{\curl}[1]{\vec{\nabla}\times#1} % curl
\newcommand{\Del}[1]{\vec{\nabla}#1} % gradient
\newcommand{\laplacian}[1]{\nabla^2#1} % laplacian
\newcommand{\bigo}[1]{\mathcal{O}\left(#1\right)} % big o notation
\newcommand{\Rinner}{R_{\rm i}} % Innner core radius (capital)
\newcommand{\Router}{R_{\rm o}} % Outer core radius (capital)
\newcommand{\rinner}{r_{\rm i}} % Innner core radius
\newcommand{\router}{r_{\rm o}} % Outer core radius
\newcommand{\gouter}{g_{\rm o}} % Gravity at outer core radius

\newcommand{\Els}{\Lambda}
\newcommand{\Ek}{{\rm Ek}} % Ekman Number
\newcommand{\Ra}{{\rm Ra}} % Rayleigh Number
\newcommand{\Reynolds}{{\rm Re}}
\newcommand{\Rm}{{\rm Rm}}
\newcommand{\Pm}{{\rm Pm}}
\renewcommand{\Pr}{{\rm Pr}}
\usepackage{siunitx} % use this package module for SI units
\DeclareSIUnit\gauss{G}

\usepackage{caption}
\usepackage{subcaption}
\usepackage[capitalise,noabbrev]{cleveref}
\crefrangelabelformat{subfigure}{#3#1#4~--~#5\crefstripprefix{#1}{#2}#6}

\begin{document}

\title{Magnetoconvection in a spherical shell: Bridging the gap between weak and strong-field dynamo regimes}

\author{L. J. Gostelow \& R. J. Teed}

\date{}

\maketitle

\begin{abstract}
\firstrevision{At moderate levels of convective driving, simulations of dynamos in spherical shells often separate into two broad dipolar regimes} characterised either by their relative magnetic field strength (weak/strong) or by their dominant force balance. These regimes can tend smoothly from one to the other but can also be bistable, a phenomenon which occurs particularly at \firstrevision{low magnetic diffusivity}.
%We show that in either case the transition correlates with a breaking of equatorial symmetry.

Nonlinear simulations of the geodynamo cannot be performed at realistic parameters and hence it is important to ensure that the (strong-field) branch\firstrevision{, appropriate for Earth's core,} is tracked as a distinguished limit is \firstrevision{established towards a suitable} parameterisation from the simulations that we can perform. In order to understand the transition to strong-field dynamos, and better understand the mechanisms that occur in both branches, we report on a series of magnetoconvection simulations\firstrevision{: models in which an external magnetic field modifies the dynamically-generated dynamo.} 

\firstrevision{Our primary input parameters are the magnetic diffusivity, imposed magnetic field strength, and the convective forcing; as these vary, we show how the morphology of the flow can change gradually, as modifications of an existing mode or continuous variations in force composition, and also abruptly, when onset modes become unstable to secondary instabilities and are replaced either by newly-dominant modes, a vacillating regime, or highly nonlinear configurations of the flow. The loss of equatorial symmetry
%, which is a property of some linear onset modes,
is also investigated, as this typically coincides with the weak-to-strong transition, and we discuss why an anti-symmetric instability emerges inside the tangent cylinder along with its implications for the geodynamo.}
\end{abstract}

\section{Introduction} \label{sec:intro}

Dynamos, or self-sustaining magnetic fields, are generated by the convective motions of electrically-conducting fluids in astrophysical bodies and ubiquitous in the Universe; simulations of these systems exhibit a wide range of solutions. Some are essentially kinematic, with the magnetic field generated by—yet only weakly influencing—the flow dynamics. Other states emerge only once the magnetic field becomes sufficiently strong. In simulations modelling Earth's core, three principal branches are now recognised, each associated with distinct magnetic‑field morphologies and strengths \citep{DormyEtAl2018, Dormy2025}. These states are commonly characterised by the dominant balance between physical forces, or more precisely, in incompressible flows, by the balance between the solenoidal components of those forces, thereby excluding pressure, which acts as a Lagrange multiplier \citep{HughesCattaneo2019, TeedDormy2023}. The expected force balance in Earth’s core cannot be reproduced exactly in simulations due to a lack of adequate computing power. Because force balances underpin much of the scaling and structure of dynamo behaviour \citep{Christensen2010}, accurately identifying subtle shifts in these balances—and their consequences for flow and field morphology—is essential for reliable extrapolation from simulations to geophysical conditions.

Nonlinear geodynamo simulations are often complemented by linear convection studies, which are computationally inexpensive and therefore provide a broader view of the families of modes that can arise. In differentially heated spheres and spherical shells, geostrophic modes onset when thermal or compositional forcing overcomes the stabilising effects of rotation and gravity; viscosity also plays a crucial role in mediating the Taylor–Proudman constraint, which demands invariance of flows along the rotation axis \citep{Roberts1968, Busse1970}. The relative importance of viscous and Coriolis forces is quantified by the Ekman number, $\Ek=\nu/(\Omega L^2)$, where $\nu$ is the kinematic viscosity, $\Omega$ the rotation rate, and $L$ a characteristic length scale. \firstrevision{The Rayleigh number, $\Ra$, is the non-dimensional number describing the degree of convective forcing, and the critical Rayleigh number, $\Ra_c$ is the degree of convective forcing required for convection to onset.} Near convective onset, viscosity is the primary mediator of rotational constraints but becomes significant only at small scales; consequently, \firstrevision{in a sufficiently rapidly rotating system,} the large‑scale flow is predominantly axially-independent and takes the form of precessing columnar Rossby waves with heights of the order of the shell and widths scaling as $\Ek^{-1/3}$, consistent with the leading‑order VAC (viscous–Archimedean–Coriolis) balance. \firstrevision{Columns of a single azimuthal mode typically dominate at weakly supercritical Rayleigh number, delaying the onset of other harmonics and potentially leading to multistability \citep{FeudelEtAl2013}.}
The precessing columns bend radially inward around the inner core, enhancing radial heat transport \citep{Busse2002}. Their columnar rotation, together with a thermally driven axial flow, can drive an $\alpha^2$-dynamo provided the shearing rate exceeds the rate of magnetic diffusion, usually quantified by the magnetic Reynolds number \citep[e.g.][]{ProctorGilbert}. In the VAC regime \firstrevision{of a dynamo}, the Lorentz force is absent from the leading-order force balance so variation of the magnetic field does not significantly alter the flow structure. Thus, one path to achieve a dynamo is to reduce magnetic diffusivity $\eta$ (i.e. increase the magnetic Prandtl number, $\Pm=\nu/\eta$), though this can substantially modify the magnetic‑field properties \citep{TeedDormy2025a}.
% (Teed \& Dormy, 2025).

In the presence of a strong magnetic field, magnetic tension resists fluid motion that bends field lines, generally stabilising the fluid against convection; that is, the critical Rayleigh number, which measures the (relative) strength of thermo-compositional forcing required for convective onset, is large. However, in rapidly rotating systems, a magnetic field for which the ratio of the Lorentz and Coriolis forces, the Elsasser number, is of order one, modifies the Taylor-Proudman constraint to allow fully three-dimensional convective (magnetostrophic) modes to onset at significantly smaller critical Rayleigh numbers \citep{Chandrasekhar,Fearn1979b}. Geostrophic and magnetostrophic modes can have similar critical Rayleigh numbers at Elsasser numbers only just below one \citep{HornAurnou2022}; when these modes coexist, the wider array of potential convective modes, with sufficient nonlinearity, can lead to chaotic flows that may promote strong-field dynamos.

The tangent cylinder is the cylindrical surface aligned with the rotation axis that circumscribes the inner core. Dynamics inside the tangent cylinder are often disconnected from dynamics outside the tangent cylinder due to the effect of the Taylor-Proudman constraint. Sakuraba \cite{Sakuraba2002} shows that polar convective modes, that is, modes which peak in amplitude inside the tangent cylinder, can have critical Rayleigh numbers similar to the geostrophic and magnetostrophic modes at order unity Elsasser numbers. Polar convective modes may also onset in nonlinear simulations at critical Rayleigh numbers close to values found in linear theory, when convection is significant only outside the tangent cylinder \citep{Gilman1975,Gilman1977}. Sreenivasan \& Jones \cite{SreenivasanJones2006} use an $f$-plane model to study these modes and show that they onset as axial plumes with oppositely signed vorticity at their upper and lower ends. Unlike equatorial modes, the dominant flow component of a polar convective mode is axial, directly supporting the poloidal magnetic field.

In a spherical shell, polar convective modes can onset with either equatorially symmetric or antisymmetric eigenfunctions with no strong preference for either (Appendix~\ref{app:symmetry}). Equatorial symmetry is the preferred onset configuration for modes outside the tangent cylinder, due to the Taylor-Proudman constraint \citep{Busse1970,Sakuraba2002}, and this symmetry is preserved by the nonlinear terms of the momentum and induction equations. The breaking of equatorial symmetry can thus only be caused by the separate onset of antisymmetric modes, such as polar convective modes, and the change in symmetry will significantly alter the structure of the flow \cite{ChristensenEtAl1999}. In particular, the breaking of equatorial symmetry can allow for the formation of large scale meridional circulation currents that can strengthen the poloidal magnetic field and support its reversals \citep{SarsonJones1999,WichtOlson2004}.

Astrophysical dynamos lie further into the diffusion-free asymptotic realm than can be reached in numerical simulation. Many studies push towards low $\Ek$, $\Pm$, seeking to derive scaling laws independent of $\nu$, $\eta$, and $\kappa$; however, a refinement of this method is to push towards this asymptotic regime in an ordered way, along a distinguished limit \citep{Dormy2016} that preserves a desired property of the simulation, such as the force balance, magnetic field morphology, or reversals. To capture these properties, at moderate Ekman numbers, it is often necessary to use larger than physical values of the magnetic or thermal Prandtl numbers \citep[i.e., weaker than physical thermal or magnetic diffusion, e.g.,][]{Dormy2016, JonesTsang2025}. However, at these larger values of $\Pm$, both weak-field (columnar) and strong-field dipolar solutions can co-exist at the same input parameters. This bistability between dynamo branches means that the resultant state depends on the initial state of the simulation with the strong-field regime emerging due to the triggering of non-geostrophic modes \citep{Dormy2016,TeedDormy2025b}. Separately, bistability has also been shown to occur between the weak-field dipolar and inertia-dominated multipolar branches \citep{SimitevBusse2009}, although it is unclear whether this type of bistability is possible between the strong-field dipolar and multipolar branches \citep{MenuEtAl2020}.

One possible way to circumvent the issue of bistability is to modify the boundary conditions on the magnetic field so that it is either strengthened or weakened \citep{TeedEtAl2015,MasonEtAl2022}. \firstrevision{In such models, a component of the magnetic is fixed so there is less risk of a sudden collapse of the dynamo or a jump to a different branch; simulations can also be run for shorter time, reducing computational demand.} These simulations correspond to magnetoconvection rather than a dynamo, but the properties of the flow are generally similar, provided that the imposed boundary conditions are consistent with those inferred from equivalent dynamo simulations (see \S\ref{sec:3.dynamo}). Another advantage of magnetoconvection is that it allows for the direct determination of the influence of the magnetic field on flow, for example, by promoting convective modes with larger length scales when magnetic stresses replace viscosity as the primary mechanism that breaks the Taylor–Proudman constraint.

In this work, we use magnetoconvection simulations to study \firstrevision{how the varying the magnetic field strength affects the dynamics of the flow, specifically focusing on how this relates to} the weak- to strong-field transition. In \S\ref{sec:formulation}, we outline the governing equations, our choice of parameterisation, and define several important output quantities, including a symmetry parameter, which measures the equatorial symmetry of the flow and magnetic field respectively. We then review suites of hydrodynamic and dynamo simulations from \citep{TeedDormy2025a,TeedDormy2025b} in \S\ref{sec:3}, to understand how these output quantities vary in those simulations and provide context for our magnetoconvection results. In \S\ref{sec:4.columnar}, we present magnetoconvection simulations at three values of $\Pm$, chosen to portray a wide array of possible phase spaces, with respect to the onset of the weak- and strong-field branches.
We still find that bistability can occur for moderately strong imposed fields.
\firstrevision{Finally, we provide a discussion in \S\ref{sec:conclusion}, sketching out the implications of this magnetoconvection study for dynamos in models and in nature, and offer concluding thoughts on the benefits of magnetoconvection studies in \S\ref{sec:conclusion2}.}
\section{Formulation}
\label{sec:formulation}
\subsection{Mathematical and numerical setup}

We consider a spherical shell filled with Boussinesq, electrically conducting fluid of constant density, $\rho_0$, rotating at constant rate $\vec{\Omega}=\Omega\vnd{z}$, in spherical coordinates $(r,\theta,\phi)$ where $r\in[\Rinner,\Router]$ is the radial coordinate, $\theta\in[0,\pi]$ is the colatitude, $\phi\in[0,2\pi)$ is the azimuthal coordinate and $\vnd{z}=\left(\cos{\theta},-\sin{\theta},0\right)$\firstrevision{; the separation of the two shells is defined by the non-dimensional radius ratio, $\chi=\Rinner/\Router=0.35$, which is standard in models of the present-day outer core of Earth. Gravity acts radially and is defined as $\vec{g}=-\gouter\vec{r}$, where $\vec{r}=(1-\chi)r\vnd{r}$, so that $\gouter$ is the gravitational field strength at the outer boundary}, and convection is driven by a constant temperature difference $\Delta T$, imposed at the boundaries. Diffusion is uniform and isotropic, with $\kappa$, $\nu$, and $\eta$, the thermal, viscous, and magnetic diffusivity coefficients, respectively. The equations describing the evolution of the velocity and magnetic fields, $\vec{u}$ and $\vec{B}$, and the temperature, $T$, can be written as
\begin{subequations}
\begin{align}
    \Ek\left(\Pm^{-1}\left(\pdv{}{t}+\vec{u}\cdot\Del\right)-\laplacian\right)\vec{u}&=-\Del p-2\vnd{z}\times\vec{u}+\left(\Del\times\vec{B}\right)\times\vec{B}+\Pm\Ra T\vec{r}\label{eqn:2.01}\\
    \left(\pdv{}{t}+\vec{u}\cdot\Del-\laplacian\right)\vec{B}&=\vec{B}\cdot\Del\vec{u},\label{eqn:2.02}\\
    \left(\pdv{}{t}+\vec{u}\cdot\Del-\Pm\Pr^{-1}\laplacian\right)T&=0,\label{eqn:2.03}
\end{align}
\end{subequations}
where \firstrevision{$p$ is the pressure and} length has been scaled by $D=\Router-\Rinner$, time by $D^2/\eta$, temperature, $\Delta T$, velocity, $\eta/D$, pressure, $\rho_0\eta\Omega$, and magnetic field by $\sqrt{\rho_0\mu_0\eta\Omega}$, where $\mu_0$ is the magnetic permeability. The scaled radial domain is $[\rinner,\router]$ with $\rinner=7/13$ and $\router=20/13$. The system can be described by four independent nondimensional parameters; we use
\begin{equation}
    \Ra=\frac{\gouter\alpha\Delta T D}{\Omega\nu},\quad \Ek=\frac{\nu}{\Omega D^2},\quad \Pr=\frac{\nu}{\kappa},\quad \Pm=\frac{\nu}{\eta},
\end{equation}
which are the Rayleigh, Ekman, Prandtl, and magnetic Prandtl numbers, respectively. The Prandtl number is fixed at $\Pr=1$. We also fix the Ekman number, $\Ek=10^{-4}$. Although this is far from the geophysical estimate of $\Ek=\bigo{10^{-15}}$, it is sufficiently small to obtain rotationally dominated and interesting dynamics, but large enough to allow for a wide survey of the parameter space. 

With $\Pr$, $\Ek$, and $\chi$ fixed, we will present results from three types of simulation: The first are hydrodynamic simulations, the equation for which can be derived from \eqref{eqn:2.01}-\eqref{eqn:2.03} with $\vec{B}\equiv \vec{0}$ and $\Pm=1$, where the latter criterion changes the effective nondimensionalisation, and the second are dynamo simulations for which electrically-insulating boundary conditions are used; these two types of simulation correspond to the same dataset presented by \cite{TeedDormy2025b,TeedDormy2025a}. These are presented in \S\ref{sec:3} as motivation for and comparison with the main set of simulations: magnetoconvection simulations, in which the radial component of the magnetic field is fixed at the outer boundary (CMB), and the inner boundary (ICB) is kept with insulating conditions. \firstrevision{In such simulations the strength of the radial, dipolar magnetic field at the outer boundary is imposed through an additional input parameter
\begin{equation}
  p_{10}=P_1^0(\router), 
\end{equation}
the coefficient of the spherical harmonic of degree $l$ and order $m$ of the poloidal magnetic field at the outer boundary; further details are provided in Appendix \ref{app:MagBCs}.
Appropriate values of $p_{10}$} are chosen by comparison with dynamo simulations at the end of \S\ref{sec:3}.
In our dynamo and magnetoconvection simulations, magnetic Prandtl numbers in the range $[1,12]$ are used, which, given the other input parameters, covers a region of parameter space that allows for different bifurcations and solution branches in dynamo solutions \cite{TeedDormy2025b}. In particular, large $\Pm$ values in this range admit separate weak- and strong-field dipolar branches with bistability between them, and such solutions correspond to those envisaged for a distinguished limit \cite{Dormy2016}.

%No-slip velocity conditions are believed to be correct both the ICB and CMB, however, it has been suggested that stress-free boundary layers could be preferred when the Ekman number is larger in simulations to reduce the effect of Ekman boundary layers \citep[e.g.][]{SimitevBusse2009} \robnote{Do we need to say this?}. A recent survey has investigated the effect of no-slip against stress-free boundaries on the onset of convection and they find that stress-free boundaries decrease the onset of convection by a few percent \citep{SeeleyEtAl2025} \robnote{This actually depends on input parameters but your statement is true for the values of Pr and $\chi$ we are using}. We opt for no-slip boundaries for comparison with the majority of literature.

With $\Pr$, $\Ek$, and $\chi$ fixed, and using no-slip boundary conditions, the onset of \emph{hydrodynamic} convection occurs at a fixed critical Rayleigh number of
\begin{equation}
    \Ra_{c}^{h}\simeq69.4,
\end{equation}
which can be obtained from a linear stability analysis \cite[e.g.~][]{SeeleyEtAl2025}. For simplicity we present our results using the supercriticality index,
\begin{equation}
    \Ra'=\Ra/\Ra_{c}^{h},\label{eqn:2.RayleighCritDef}
\end{equation}
which is defined relative to the critical Rayleigh number of the hydrodynamic ($m=7$) onset mode. It is important to note, of course, that the true onset of (magneto)convection varies with the magnetic field strength and the magnetic Prandtl number \citep{Chandrasekhar}. In particular, the onset of magnetoconvection is reduced when \firstrevision{the Elsasser number,} $\Els\gtrsim\Ek^{1/3}$, reaching a minimum when $\Els\simeq\bigo{1}$; that is, when a balance is reached between the reduction of the geostrophic (Taylor-Proudman) constraint, which is destabilising, and the stabilising effect of magnetic tension \citep{Fearn1979b,Sakuraba2002,JonesEtAl2003}.

\firstrevision{The governing equations, \eqref{eqn:2.01}-\eqref{eqn:2.03}, are solved numerically using the Leeds spherical dynamo code \cite{WillisEtAl2007}. Convergence of simulations is determined by tracking the kinetic and magnetic energy variation to a statistically steady state. 
%For weakly supercritical simulations, this is straightforward to determine; however, for more chaotic simulations, a straightforward criterion can be more difficult to enforce without introducing unnecessary compute time.
Threshold simulations, i.e.~those at regime boundaries, are identified from a broader parameter sweep, and are run for longer to ensure that they are indeed stable. One advantage of using magnetoconvection simulations, rather than dynamo, is that they can generally be run for less time since there is an ever-present seed field that prevents absolute magnetic field collapse.
Full details of the simulations run, along with figure data, are available in the dataset connected to this paper \cite{GostelowTeed2026data}.
%, which can be found at \hyperlink{https://researchdata.gla.ac.uk/2221/}{https://researchdata.gla.ac.uk/2221/}.

While we find that there are several different phase transitions in the parameter space (which are investigated later), a central motivation of this work is to study the weak-to-strong field transition and the possible region of bistability around it. To properly capture this transition in magnetoconvection we initialise simulations from nearby states in parameter space, which are classified as either weak- or strong-field through their force hierarchy, for example. A parameter, typically $\Ra$ or $p_{10}$, is then varied, tracking out a branch of solutions, until a transition to a new branch clearly occurs. We then trace the new branch back in the opposite direction, varying the same parameter, and in this way can capture regions of bistability; found in this case at $\Pm=5$ and $\Pm=12$, but not $\Pm=1$ (see \S\ref{sec:4.columnar}). This method has been employed in previous studies, e.g.~\cite{TeedDormy2025b}, and also typifies the procedure in the remainder of parameter space for other transitions.}

\subsection{Output quantities}
\label{sec:outputs}

The Elsasser number measures the relative sizes of the Lorentz and Coriolis forces. In rapidly rotating systems, for which the Coriolis force is an essential part of the dominant balance, the Elsasser number is expected to be of order unity if the Lorentz force is also part of the dominant balance. The Elsasser number is traditionally approximated as the square of the non-dimensional magnetic field strength,
\begin{equation}
    \frac{|\left(\curl{\vec{B}}\right)\times\vec{B}|}{2\rho_0\mu_0\left|\vec{\Omega}\times\vec{u}\right|}\sim\Els=\frac{\left<B\right>^2}{2\rho_0\mu_0\eta\Omega},\label{eqn:2.ElsasserDef}
\end{equation}
where $\left<\cdot\right>$ denotes a suitable measure of size, chosen here as $\left<\cdot\right>=\left(\int_{V}\left|\cdot\right|^2~\dx{V}\right)^{1/2}$, which is the $L_2$ norm. \firstrevision{This quantity is order unity (or larger) for both weak- and strong-field solutions, and several authors \cite{SoderlundEtAl2012,Dormy2016} suggest that a more refined estimate is,}
\begin{equation}
    \Els'=\frac{\left<B\right>^2}{2\Omega\mu_0\rho_0\left<U\right>\ell_{\eta}}=\Els\frac{L}{\Rm \ell_{\eta}},\label{eqn:elsasserprime}
\end{equation}
where $\Rm=\left<U\right>D/\eta$ is the traditional magnetic Reynolds number and
\begin{equation}
    \ell_{\eta}^{2}=\frac{\left<\vec{B}\right>^2}{\left<\vec{\curl{\vec{B}}}\right>^2}\label{eqn:scale_lB}
\end{equation}
is the magnetic diffusion scale, which is the scale at which Ohmic dissipation occurs and thus determines the strength of the magnetic field through power balance when viscous energy losses are negligible \citep[e.g.][]{OrubaDormy2014}. Using $\Els'$ gives a more refined estimate of the dominant balance and generally captures the difference between weak- and strong-field solutions \citep{TeedDormy2023}; an even more accurate estimate of the balance can be found by directly calculating the forces or their curls \citep{TeedDormy2025a}.

Since forces can vary across length scales, such as viscosity, which predominantly enters at small scales, so can the force balances. We make use of force spectra plots for which each force is plotted as a function of the spherical degree $l$ and averaged over a long time-scale. Following the discussion above, we use the curl of forces, which is numerically more straightforward to calculate than the solenoidal component, but still removes gradient terms. Forces are then divided by $\sqrt{l\left(l+1\right)}$ to remove the imbalance caused by the angular derivative - see \cite{TeedDormy2023}.

A key change in the flow at weak nonlinearity as thermal forcing is increased and convection becomes more vigorous is the breaking of the Taylor-Proudman constraint, that is, the trend away from a largely $z$-invariant flow. Soderlund et al.~\cite{SoderlundEtAl2012} suggest that the anisotropy of such a flow can be measured by its columnarity, the relative size and $z$-independence of the $z$-component of the vorticity. We calculate this as
\begin{equation}
    C_{\omega z}=\frac{\left<h^{-1/2}\left|\int_{z}\omega_{z}^{\prime}\ \dx{z}\right|\right>}{\left<\vec{\omega}^{\prime}\right>},
    \label{eqn:2.columnarity}
\end{equation}
where $h=2\sqrt{\router^2-s^2}$ and $\vec{\omega}^{\prime}$ corresponds to the vorticity after the zonal flow (the axisymmetric component of \firstrevision{the vorticity, $\vec{\omega}=\curl\vec{u}$}) is removed. The norm in the numerator only integrates over the remaining two dimensions, $s$ and $\phi$, for $s\in[\rinner,\router]$, outside the tangent cylinder. This measure is also used by \cite{MasonEtAl2022} who define it in a qualitatively similar way. If $\vec{u}$ is non-zero, then $C_{\omega z}$ is therefore a scalar which will vary between a third, for isotropic flow, and almost unity, for flow which is $z$-independent except in boundary layers where some $z$-dependence is essential.

We also define a parameter, $s_u$, that reports the mirror symmetry of $\vec{u}$ about the equator,
\begin{equation}
    s_{u}=\frac{\left<\vec{u}_{s}\right>-\left<\vec{u}_{as}\right>}{\left<\vec{u}_{s}\right>+\left<\vec{u}_{as}\right>}.\label{eqn:2.symmetry}
\end{equation}
where
\begin{equation}
    \vec{u}_{s}=\frac{\vec{u}\left(\theta\right)+\vec{u}^{r}\left(\pi-\theta\right)}{2}\qquad\vec{u}_{as}=\frac{\vec{u}\left(\theta\right)-\vec{u}^{r}\left(\pi-\theta\right)}{2},\tag{\theequation a,b}
\end{equation}
and $\vec{u}^{r}$ denotes the reflection of $\vec{u}$ across the equator. The equivalent for the magnetic field, $s_{B}$, can also be defined similarly. $s_u$ and $s_B$ lie in the range $[-1,+1]$ where $+1$ corresponds to a perfectly equatorially symmetric vector field and $-1$ to a perfectly anti-symmetric one. An equivalent definition is to calculate $\left<\vec{u}_{s}\right>$ from the energy of the poloidal modes with $l-m$ even and the toroidal modes with $l-m$ odd; $\left<\vec{u}_{as}\right>$ is then the remainder of the total energy.

\firstrevision{The kinetic or magnetic energy can be separated 
%into various constituent parts. 
into their components perpendicular and parallel to the rotation axis, as well as the zonal component. These are defined, for the kinetic energy, by}
\begin{subequations}
\begin{align}
    \text{KE}_{\text{zonal}}&=\frac{1}{2}\left<\left(u_\phi\right)_{m=0}\right>^2\,,\\
    \text{KE}_{\parallel}&=\frac{1}{2}\left<u_{z}\right>^2\,,\\
    \text{KE}_{\perp}&=\frac{1}{2}\left<\left|\vec{u}\right|\right>^2-\text{KE}_{\text{zonal}}-\text{KE}_{\parallel}\,,
\end{align}\label{eqn:2.kineticenergy}
\end{subequations}
\firstrevision{and equivalently for the magnetic energy by using the magnetic field in definitions.}

\subsection{Minimal field configuration}
\label{sec:2.fieldconfig}

In the magnetoconvection setup, the ICB is assumed to be electrically insulating, whilst the field at the outer boundary is specified so that
\begin{equation}
    \vnd{r}\cdot\vec{B}=\frac{2\firstrevision{p_{10}}}{\router}\cos{\theta}\vnd{r};
\end{equation}
see Appendix \ref{app:MagBCs}. Two possible divergence- and curl-free solutions are a uniform axial field
\begin{equation}
    \vec{B}_{\text{axial}}=\frac{2\firstrevision{p_{10}}}{\router}\vnd{z},\label{eqn:2.axialB}
\end{equation}
and a dipolar field,
\begin{equation}
    \vec{B}_{\text{dip}}=\frac{\router^{2}\firstrevision{p_{10}}}{r^3}\left(2\cos\theta\vnd{r}+\sin\theta\vnd{\theta}\right).\label{eqn:2.dipolarB}
\end{equation}
The total (dimensional) magnetic energy is given by
\begin{equation}
    \text{ME}^{*}=\rho_0\eta^2D\frac{\Pm}{\Ek}\int_{V}\frac{B^2}{2}~\dx{V},
\end{equation}
where $V$ corresponds to the simulated domain, the spherical shell. Hence, the non-dimensional magnetic energy is
\begin{equation}
    \text{ME}=\frac{\Pm}{\Ek}\int_{V}\frac{B^2}{2}~\dx{V}.
\end{equation}
For the two magnetic field solutions \eqref{eqn:2.axialB} and \eqref{eqn:2.dipolarB}, the expression for the magnetic energy is:
\begin{align}
    \text{ME}_{\text{axial}}&=\firstrevision{p_{10}}^2\frac{\Pm}{\Ek}\frac{8\pi}{3\router^2}\left(\router^{3}-\rinner^{3}\right),\label{eqn:2.MEaxial}\\
    \text{ME}_{\text{dip}}&=\firstrevision{p_{10}}^2\frac{\Pm}{\Ek}\frac{4\pi \router^4}{3}\left(\rinner^{-3}-\router^{-3}\right).\label{eqn:2.MEdipolar}
\end{align}
The energy of the axial magnetic field is smaller for $\chi<0.794$ and hence for $\chi=0.35$ the axial field configuration is preferred. For example, in the case $\Pm=1$, $\Ek=10^{-4}$, $\firstrevision{p_{10}}=1.0$ (cf. Figure \ref{fig:4.TorPol}), we have
\begin{align}
    \text{ME}_{\text{axial}}&=1.23\times10^{5},\\
    \text{ME}_{\text{dip}}&=1.44\times10^{6}.
\end{align}
The poloidal magnetic energy in these simulations is $\text{ME}\approx1.3\times10^{5}$, which corresponds closely with an axial magnetic field.
\section{Hydrodynamic and dynamo simulations}
\label{sec:3}

In this section, we review previous work on nonlinear hydrodynamic and dynamo simulations and calculate some of the diagnostic quantities that we have identified in \S\ref{sec:outputs}.

\subsection{Weakly-nonlinear hydrodynamic convection}
\label{sec:3.hydro}

\begin{figure}
    \centering
    \begin{subfigure}{0.49\linewidth}
        \includegraphics[trim=0cm 0.35cm 0cm 0cm,clip,width=\linewidth]{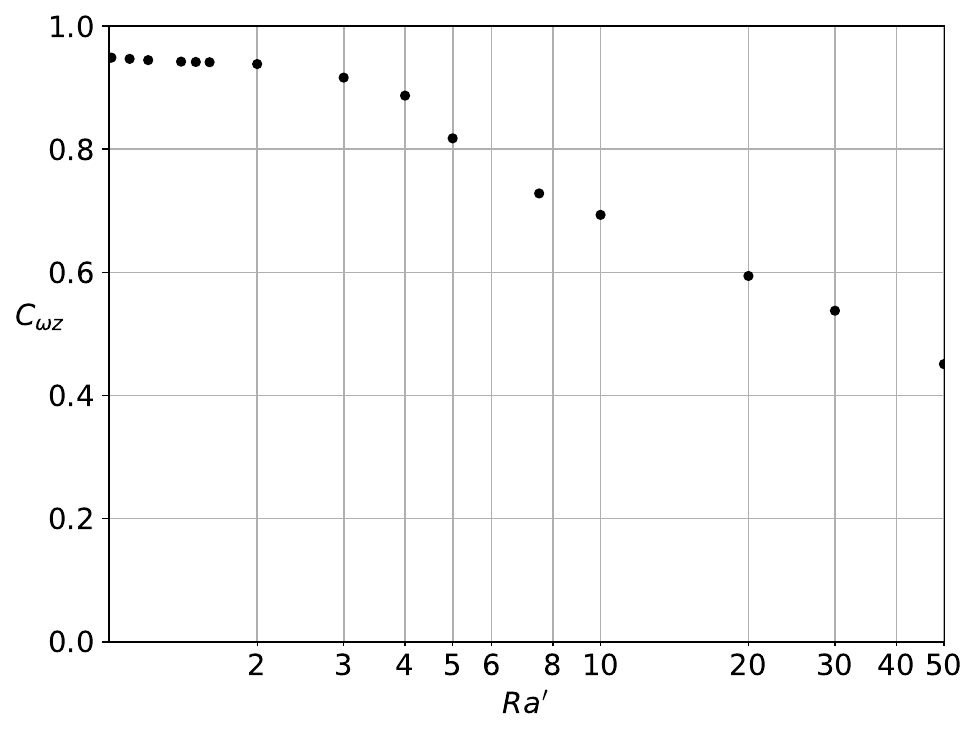}
        \caption{Columnarity, $C_{\omega z}$}
        \label{fig:Columnarity_hydro}
    \end{subfigure}
    \begin{subfigure}{0.49\linewidth}
        \includegraphics[trim=0cm 0.35cm 0cm 0cm,clip,width=\linewidth]{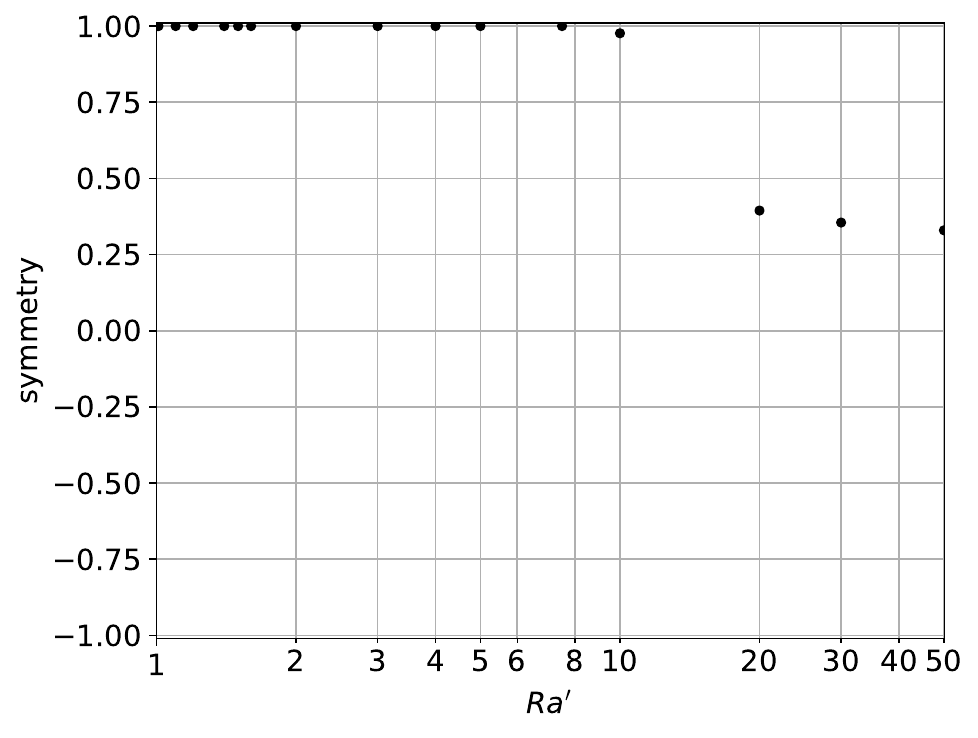}
        \caption{Symmetry, $s_u$}
        \label{fig:Symmetry_hydro}
    \end{subfigure}
    \begin{subfigure}{0.49\linewidth}
        \includegraphics[trim=0cm 0.35cm 0cm 0cm,clip,width=\linewidth]{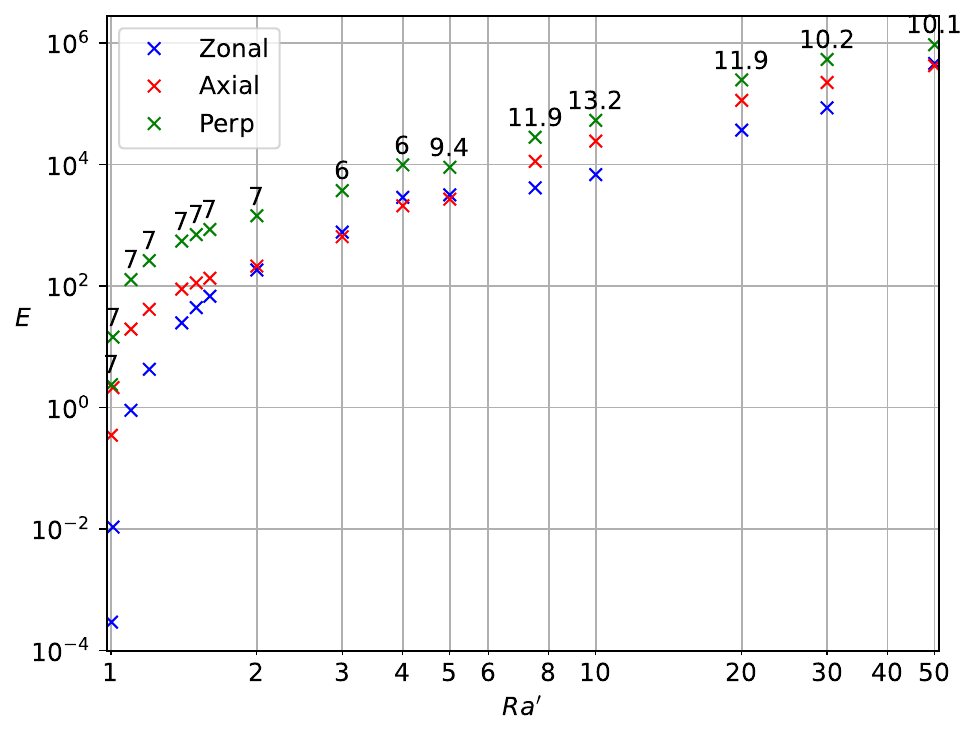}
        \caption{Components of the kinetic energy}
        \label{fig:Ecomps_hydro}
    \end{subfigure}
    \caption{Global diagnostics of hydrodynamic simulations for $\Ra'$ in the range $(1,50]$: (a) Columnarity \eqreflong{eqn:2.columnarity}; (b) Flow symmetry \eqreflong{eqn:2.symmetry}; (c) Kinetic energy components \eqreflong{eqn:2.kineticenergy}. \firstrevision{The number above each trio of points in (c) denotes the dominant azimuthal mode of the kinetic energy spectrum which is defined as a single mode if there is at least five times more energy in that mode than the secondary mode and then displayed as an integer, or else as the mean harmonic of the kinetic energy spectrum (as in e.g. \cite{ChristensenAubert2006}), which excludes the zero mode, and is then displayed to one decimal place to distinguish the calculation.}}
    \label{fig:Hydro_params}
\end{figure}

%\lukenote{ONSET AND COLUMNAR REGIME}
In the hydrodynamic problem (at $\Ek=10^{-4}$, $\Pr=1$), the first mode to become unstable is an $m=7$ mode so the flow generally takes the form of $m=7$ convective columns for $\Ra'\gtrsim1$. Additional wavenumbers become unstable with Hopf bifurcations as the Rayleigh number increases, several of which can be simultaneously stable, which can lead to a dependence of the long-term state on the initial conditions \citep[e.g.,][]{FeudelEtAl2013,SanchezEtAl2013,FeudelEtAl2017}. \firstrevision{Figure \ref{fig:Hydro_params} shows global diagnostics for several hydrodynamic simulations as the driving strength is increased above the onset of convection.} For $\Ra'\in[1,2]$, the flow is weakly nonlinear and hence inertia is small; thus, the remaining solenoidal terms define the dynamics through a VAC (viscous-Archimedean-Coriolis) balance. From this, a scaling for the Reynolds number,
\begin{equation}
    \Reynolds\sim\left[\Ra'\left(\Ra'-1\right)\right]^{1/2}\label{eqn:Re_scaling}
\end{equation}
can be derived \citep{GilletJones2006,GastineEtAl2016}. The weaker zonal component of the flow is driven by Reynolds stresses and so is nonlinearly dependent on the strength of the columnar flow \citep{Busse2002,Aubert2005}; thus, it is initially small, but grows more rapidly than the other flow components until the associated kinetic energy becomes larger than the axial component at around $\Ra'=2$ (Figure \ref{fig:Ecomps_hydro}).

%\lukenote{COLUMNARITY IN COLUMNAR REGIME}
For $\Ra'\lesssim2.0$ the columnarity is approximately constant, $C_{\omega z}\approx0.95$ (Figure \ref{fig:Columnarity_hydro}). This raw value also be approximated by
\begin{equation}
    C_{\omega z}\sim\frac{2U_{\perp}/\ell_\perp}{2U_\perp/\ell_\perp+U_\parallel/\ell_\parallel}=\frac{2}{2+\frac{U_\parallel}{U_\perp}\frac{\ell_\perp}{\ell_\parallel}}\,,\label{eqn:3.cwz_wnl}
\end{equation}
where $U_{\parallel}/U_\perp\approx\sqrt{0.2}$ can be determined from the energy balance (Figure \ref{fig:Ecomps_hydro}), and we can take $\ell_\parallel=2\sqrt{\router^{2}-s_{c}^2}$, the height of the sphere at the critical radius, $s_c\approx0.42\router$ \firstrevision{\citep{SeeleyEtAl2025}}, and $\ell_\perp=2s_c\pi/m_c$, where $m_c=7$ is the mode of onset. Using these, we obtain an estimate of $C_{\omega z}\approx0.955$, which is slightly larger than the value calculated numerically. As the Rayleigh number increases, the flow becomes unstable further away from the critical radius and the convective columns increase their radial extent, slightly increasing $\ell_\perp$, and therefore slightly decreasing $C_{\omega z}$. The preferred onset modes are equatorially symmetric \citep{Busse1970}, and hence the flow remains equatorially symmetric since the nonlinear terms preserve this symmetry (Figure \ref{fig:Symmetry_hydro}); antisymmetric modes must onset separately.

%\lukenote{INSTABILITY OF COLUMNAR REGIME}
Increasing the Rayleigh number above $\Ra'=2$ eventually leads to an instability of the axial columns which results in a steady decline in $C_{\omega z}$; this also allows other modes to grow and eventually dominate over the primary onset mode \firstrevision{as seen by the change in the dominant azimuthal mode in Figure \ref{fig:Ecomps_hydro}}. The columnarity, Figure \ref{fig:Columnarity_hydro}, can also be compared with Figure 3a of \cite{SoderlundEtAl2012}, who use $\chi=0.4$, i.e. a slightly larger inner core radius, and $\Ek\approx1.7\times10^{-4}$\footnote[2]{\cite{SoderlundEtAl2012} use $\Ek=10^{-4}$, but define their Ekman number as $\Ek=\nu/2\Omega D^2$, where $D$ is the distance between the outer and inner boundaries and also use $\chi=0.4$; one way to convert is to fix $\router$, the outer radius, which gives the scale factor $2\times(1-0.4)^2/(1-0.35)^2\approx1.7$.}. This leads to a more rapid decrease in $C_{\omega z}$ with $\Ra'$ since the preferred azimuthal wavenumber of convection, $m_c$, grows due to the larger inner core \citep[e.g.~][]{SeeleyEtAl2025}, and hence the convective columns become unstable to centrifugal instabilities at lower $\Ra'$ \firstrevision{as the aspect ratio between the horizontal radius and columnar height decreases \citep{SeshasayananGallet2020,vanKanAlexaxis2022}}.

\begin{figure}
    \centering
    \begin{subfigure}{0.32\linewidth}
        \centering
        \includegraphics[width=\linewidth]{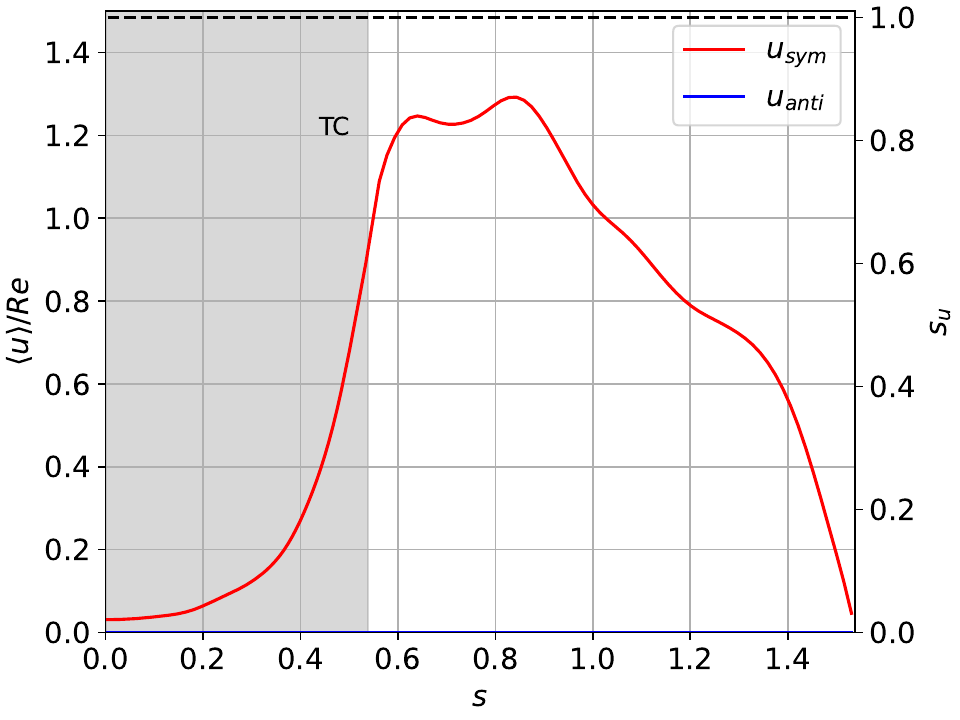}
        \caption{$\Ra'=7.5$}
        \label{fig:3.symmetry_over_s_a}
    \end{subfigure}
    \begin{subfigure}{0.32\linewidth}
        \centering
        \includegraphics[width=\linewidth]{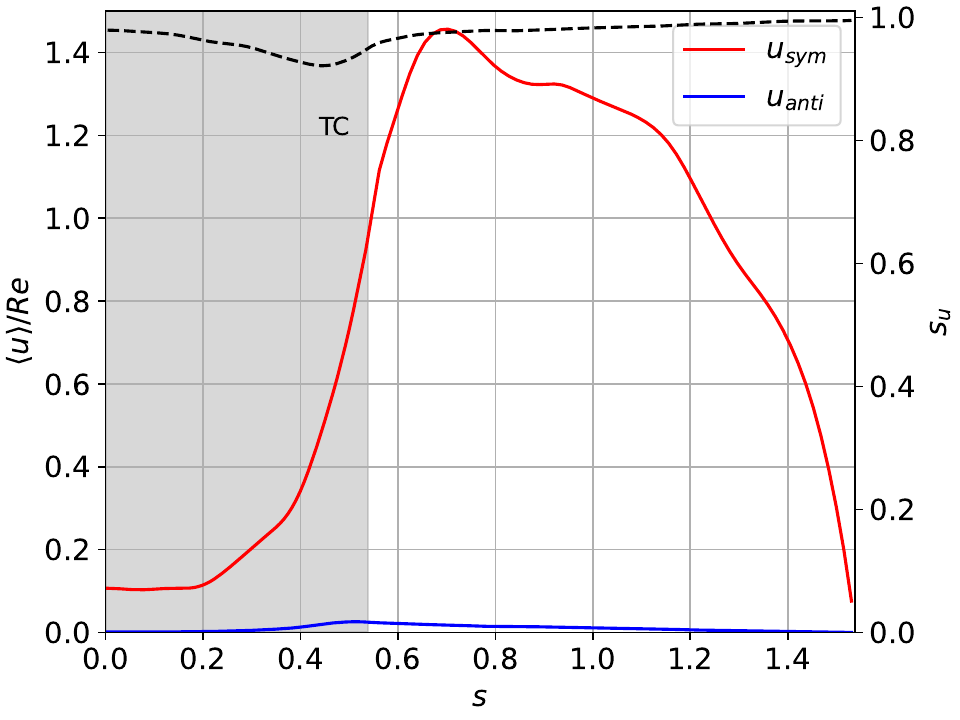}
        \caption{$\Ra'=10$}
        \label{fig:3.symmetry_over_s_b}
    \end{subfigure}
    \begin{subfigure}{0.32\linewidth}
        \centering
        \includegraphics[width=\linewidth]{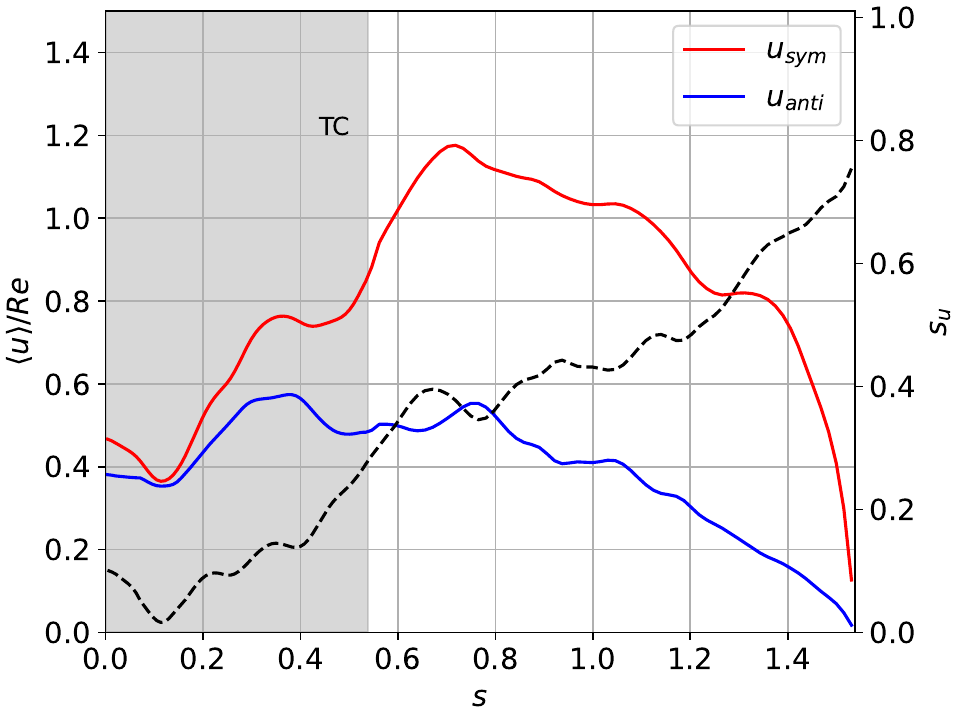}
        \caption{$\Ra'=20$}
        \label{fig:3.symmetry_over_s_c}
    \end{subfigure}
    \caption{\firstrevision{Equatorial (anti-)symmetry of the flow in hydrodynamic simulations at three different Rayleigh numbers and $\Ek=10^{-4}$, $\Pr=1$. Plotted are the equatorially symmetric and anti-symmetric components of the RMS velocity (blue/red solid, left-axis; scaled with the Reynolds number to give a value around unity) and their ratio, which is the degree of equatorial symmetry (black dashed, right-axis), calculated over concentric cylinders to show their radial dependence. The shaded region denotes the region inside the tangent cylinder.}}
    \label{fig:3.symmetry_over_s}
\end{figure}

%\lukenote{ANTISYMMETRIC ONSET}
\firstrevision{At $\Ra'\approx10$ the symmetry, $s_{u}$, suddenly decreases due to the convective onset of equatorially antisymmetric modes. Gilman \cite{Gilman1977} describes a similar situation for the solar convection envelope; since the eigenfunction of the equatorially symmetric onset mode, and hence the resultant convection columns, are localised to the region outside the tangent cylinder, the polar modes, whose eigenfunctions peak inside the tangent cylinder, onset at approximately the same Rayleigh number as expected from linear stability theory around a static basic state; meanwhile, the non-dominant equatorial modes are strongly suppressed by the $m_c=7$ mode and onset well beyond the linear value. Gastine \& Aurnou \cite{GastineAurnou2023} have also raised the importance of the polar mode, discussing it in the context of significantly increasing the heat flux; this occurs since convection cells can develop in the tangent cylinder whose radial flow is less affected by the Coriolis force. We shall see later that this increased radial flow also affects the generation of the poloidal field. There is no associated jump in $C_{\omega z}$ at $\Ra'\approx10$ as it is calculated only over the region outside the tangent cylinder. If equatorial symmetry is instead calculated over a series of concentric cylinders, then its radial variation can be determined and this is plotted as the black dashed curve in Figure~\ref{fig:3.symmetry_over_s}, which is the ratio between the blue and red curves, the mean symmetric and anti-antisymmetric energy in each of these concentric cylinders, respectively. As the Rayleigh number is increased over the transition, from $\Ra'=7.5$ to $\Ra'=20$ (Figure~\ref{fig:3.symmetry_over_s_a}-\ref{fig:3.symmetry_over_s_c}), antisymmetric modes first onset (for $\Ra'\approx10$) just within the tangent cylinder
%($s\lesssim7/13\approx0.538$)
and become a significant component of the kinetic energy inside the tangent cylinder at larger $\Ra'$, whilst symmetric modes remain dominant outside.}

\subsection{Dynamo models}
\label{sec:3.dynamo}

\begin{figure}
    \centering
    \includegraphics[width=0.8\linewidth]{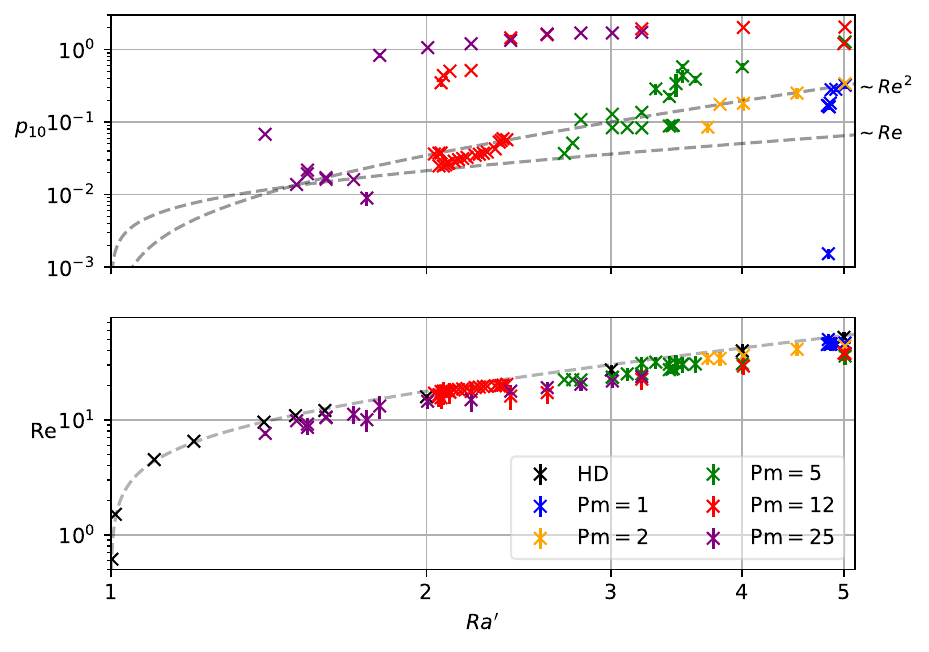}
    \caption{\firstrevision{Top: The $(l,m)=(1,0)$ component of the poloidal scalar, $p_{10}$, at the outer boundary, for dynamo simulations with varied $\Pm$ and $\Ra'$ close to the onset of (hydrodynamic) convection for $\Ek=10^{-4}$ and $\Pr=1$. Bottom: The Reynolds number for the same set of simulations. On both plots dashed curves correspond to the theoretical scaling of $\Reynolds$ \cite{GastineEtAl2016} given by \eqref{eqn:Re_scaling} with an additional curve for $\Reynolds^2$ on the top plot.} \citep[This data is derived from simulations conducted by][]{TeedDormy2025b,TeedDormy2025a}.}
    \label{fig:Ra_vs_p10_dynamos}
\end{figure}

\begin{table}
    \centering
    \begin{tabular}{c|c|c|c|c|c}
        \Pm & 1 & 2 & 5 & 12 & 25\\\hline
        \Ra$^\prime$ & 4.82 & 3.7 & 2.7 & 2.03 & 1.4\\
        \Rm & 45.5 & 67.5 & 112 & 205 & 190\\
    \end{tabular}
    \caption{Time-averaged value of the magnetic Reynolds number, $\Rm$, for Rayleigh numbers, $\Ra'$, just above the onset of dynamo action in dynamo (electrically-insulating outer boundary) simulations at a range of magnetic Prandtl numbers, $\Pm$ ($\Pr=1$, $\Ek=10^{-4}$). At larger magnetic diffusion ($\Pm=1,2,5,12$) the flow and magnetic fields correspond to a weak-field dipolar solution, whilst the $\Pm=25$ case corresponds to a hemispherical dynamo - see \cite{TeedDormy2025a}.}
    \label{tab:DynMagReynolds}
\end{table}

% \begin{figure}
%     \centering
%     \begin{subfigure}{0.32\linewidth}
%         \includegraphics[width=\linewidth]{Figures/geo_11d_E1e-4_1.6Rac_Pr1_Pm25_17_0001_eqsect_u_r.png}
%         \caption{$\Pm=25$, $\Ra'=1.6$}
%         \label{fig:zsect_weakly_nonlinear_a}
%     \end{subfigure}
%     \begin{subfigure}{0.32\linewidth}
%         \includegraphics[width=\linewidth]{Figures/geo_11d_E1e-4_2.06Rac_Pr1_Pm12_01_0009_eqsect_u_r.png}
%         \caption{$\Pm=12$, $\Ra'=2.06$}
%         \label{fig:zsect_weakly_nonlinear_b}
%     \end{subfigure}
%     \begin{subfigure}{0.32\linewidth}
%         \includegraphics[width=\linewidth]{Figures/geo_11d_E1e-4_3.43Rac_Pr1_Pm5_WD_04FB_0002_eqsect_u_r.png}
%         \caption{$\Pm=5$, $\Ra'=3.43$}
%     \end{subfigure}
%     \begin{subfigure}{0.32\linewidth}
%         \includegraphics[width=\linewidth]{Figures/geo_11d_E1e-4_4Rac_Pr1_Pm2_05FB_0002_eqsect_u_r.png}
%         \caption{$\Pm=2$, $\Ra'=4.0$}
%     \end{subfigure}
%     \begin{subfigure}{0.32\linewidth}
%         \includegraphics[width=\linewidth]{Figures/geo_11d_E1e-4_4.85Rac_Pr1_Pm1_06FB_0003_eqsect_u_r.png}
%         \caption{$\Pm=1$, $\Ra'=4.85$}
%     \end{subfigure}
%     \caption{$u_r/\Pm$ for dynamo simulations in the ``weak-field" regime, just above the onset of dynamo action.}
%     \label{fig:zsect_weakly_nonlinear}
% \end{figure}

%\lukenote{DYNAMO ONSET}
\firstrevision{In equivalent dynamo problems (at $\Ek=10^{-4}$, $\Pr=1$, for a range of values of $\Pm$), provided $\Pm$ is large enough, two clear branches (weak- and strong-field) can be distinguished \citep[as previously demonstrated by ][]{TeedDormy2025a,TeedDormy2025b}. This can be observed in Figure~\ref{fig:Ra_vs_p10_dynamos}, where $p_{10}$ is used as a proxy for the magnetic field strength (and for comparison with the later magnetoconvection simulations). As the convective driving is increased, the Reynolds number grows approximately as described by \eqref{eqn:Re_scaling}, which we verified for the hydrodynamic and, perhaps surprisingly, for both dynamo branches in Figure~\ref{fig:Ra_vs_p10_dynamos}. When a magnetic seed field is introduced, it can be amplified provided that the magnetic Reynolds number, $\Rm$, is sufficiently large \cite{DormySoward}. In the absence of other effects, we would expect (and do find) that increasing the magnetic Prandtl number, $\Pm$, increases $\Rm$ and therefore reduces the Rayleigh number for the onset of dynamo action. Table~\ref{tab:DynMagReynolds} records the values of $\Ra$ at which dynamo action onsets and the corresponding value of $\Rm$ in each dynamo simulation. Although increasing $\Pm$ is effective in reducing the Rayleigh number of dynamo onset, it is clear that the magnetic Reynolds number cannot predict the onset of dynamo action - the morphology of the flow is also an important factor.

%\lukenote{WEAKLY NONLINEAR FIELD SATURATION}
Viscosity, buoyancy, and the Coriolis force have important roles in organising the flow, however the magnetic field has an increasing role in organising the flow as it increases in strength, and plays an important role in breaking the Taylor-Proudman constraint, opening the door for non-axially-independent structures, which can often be more efficient in convection or in driving dynamos. Fearn \cite{Fearn1979b} finds that for asymptotically small $\Ek$
\begin{equation}
    \left|B\right|\gtrsim\bigo{\Ek^{1/6}}
\end{equation}
is sufficiently large as to alter the linear onset of convection (though Mason et al. \cite{MasonEtAl2022} find that $\Ek=10^{-4}$ is not sufficiently small to be asymptotic and that a larger field, $\left|B\right|=\bigo{1}$ is required to alter the critical Rayleigh number). On each dynamo branch $p_{10}$ grows initially after dynamo onset to an ultimate scaling with $\Ra'$, which is independent of $\Pm$. On the strong-field branch, the saturation of the magnetic field strength is determined by the dominant MAC balance and we find that $p_{10}$ is approximately constant ($1$-$2$) over the limited range displayed here. This is in contrast to the (modified) Elsasser number, measuring the nondimensional magnetic energy, which can vary significantly with $\Pm$ and $\Ra'$ \cite{Dormy2016,TeedDormy2025b}. On the weak-field branch, the magnetic field strength is generally weaker than Fearn or Mason's criterion for it to have an impact on the linear onset of convection; the nonlinear saturation of the magnetic field, in this case, is determined at second-order, that is, the magnetic field grows until there is a significant disruption to the primary flow. We find empirically that $p_{10}$ scales quadratically with the Reynolds number on the weak-field branch, as shown by the dashed curve in Figure~\ref{fig:Ra_vs_p10_dynamos}. Such a scaling is unlikely to persist at larger Rayleigh numbers, however, as the flow becomes more vigorous and inertia replaces viscosity in the primary force balance (which changes from VAC to CIA \cite{TeedDormy2023}).}

%\lukenote{BATMAN: THE RISE OF INERTIA (AND LENGTH SCALES)}
As the Rayleigh number increases ($\Ra'\gtrsim5$), inertia grows and convective rolls become unstable, \firstrevision{and a new branch of inertially-influenced solutions can be reached}; these can also sustain a dipolar magnetic field \cite{TeedDormy2025b}.
At larger $\Ra'$, a distinct multipolar branch materialises due to the smaller inertial(/Coriolis) scales (compared with the (VAC) onset scales). However, for sufficiently large $\Pm$, the magnetic field of the strong-field branch can overcome resistance to large scales perpendicular to the rotation axis, generally increasing length scales on this MAC branch \cite{DormyEtAl2018}. The larger length scales on this branch reduce the energy lost to diffusion and hence increase the strength of the dipolar field; this feedback mechanism is another potential avenue for bistability.

The energy in zonal flow, which is on par with the other two components in hydrodynamic convection, is found to drop significantly in dynamo simulations, since it is resisted by the presence of a large scale poloidal field \citep{Aubert2005}. We investigate this aspect of the problem in \S\ref{sec:4.columnar}, in our range of magnetoconvection simulations. In the absence of a significant $\Omega$-effect from a zonal flow, the dynamo mechanism on all three branches must be of an $\alpha^2$-type. A straightforward balance shows that the toroidal component of the field should therefore be of comparable strength to the poloidal field \citep{RobertsSoward1992}. Teed \& Dormy \cite{TeedDormy2025a} verify this in the strong-field regime, but find that the toroidal field can be several times larger in the weak-field regime; a feature that can perhaps be explained by the disparity between the strengths of the axial and columnar flows in that case.
%\robnote{I'm somewhat sceptical about invoking mean-field theory ideas here but it may serve a purpose.}\lukenote{Discuss - This is (probably) the only real comment like this I have. I think it's an interesting point to keep in mind - the difference between the toroidal and poloidal parts of the velocity and magnetic fields - e.g. in the time evolution curves of the transition. I suppose this sentence was originally a pre-empt for a longer discussion on the two components of the Reynolds number. Since we haven't ended up using that, maybe this isn't necessary, but I don't think it detracts.}

\begin{figure}
    \centering
    % \begin{subfigure}{0.49\linewidth}
    %     \includegraphics[trim=-0.3cm 0cm 0cm 0cm,clip,width=\linewidth]{Figures/Columnarity_Pm5_dynamos.pdf}
    %     \caption{Columnarity, $C_{\omega z}$, $\Pm=5$}
    %     \label{fig:Columnarity_dynamos_pm5}
    % \end{subfigure}
    % \begin{subfigure}{0.49\linewidth}
    %     \includegraphics[trim=-0.3cm 0cm 0cm 0cm,clip,width=\linewidth]{Figures/Columnarity_Pm12_dynamos.pdf}
    %     \caption{Columnarity, $C_{\omega z}$, $\Pm=12$}
    %     \label{fig:Columnarity_dynamos_pm12}
    % \end{subfigure}
    \begin{subfigure}{\linewidth}
        \includegraphics[trim=-0.3cm 0cm 0cm 0cm,clip,width=\linewidth]{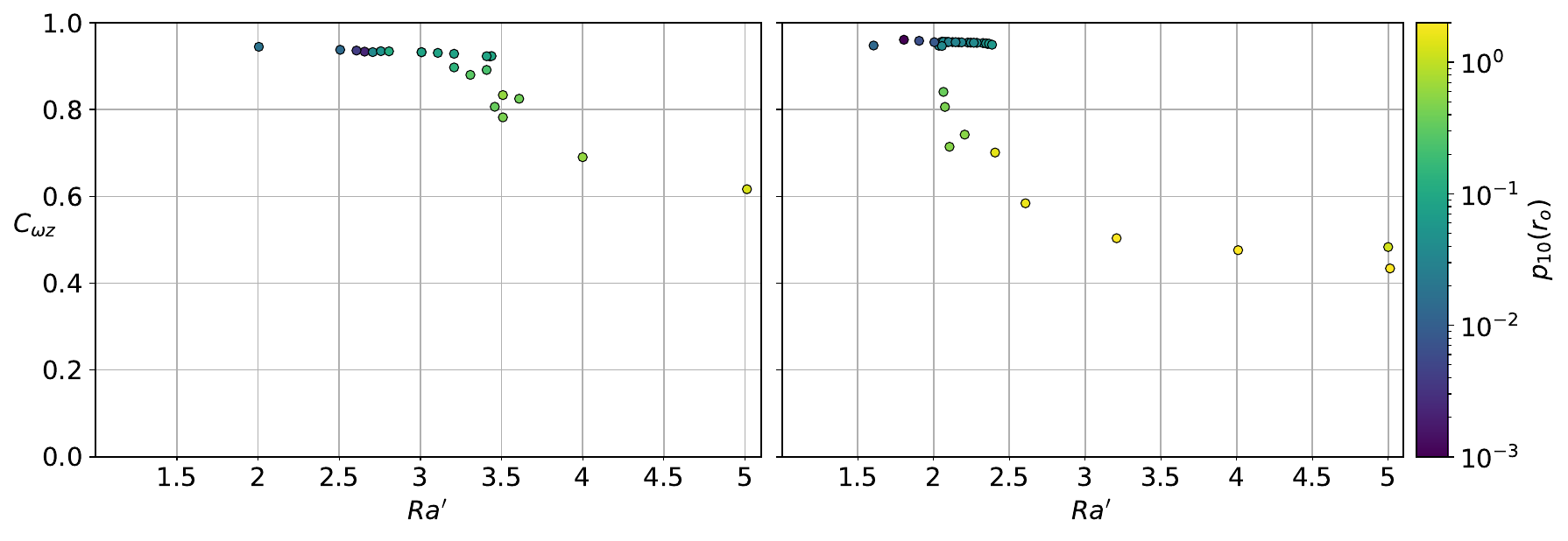}
        \caption{Columnarity, $C_{\omega z}$ \eqreflong{eqn:2.columnarity}; $\Pm=5$ (left) and $\Pm=12$ (right)}
        \label{fig:Columnarity_dynamos}
    \end{subfigure}
    % \begin{subfigure}{0.49\linewidth}
    %     \includegraphics[trim=0.3cm 0cm 0cm 0cm,clip,width=\linewidth]{Figures/Symmetry_u_Pm5_dynamos.pdf}
    %     \caption{Symmetry, $s_u$, $\Pm=5$}
    %     \label{fig:Symmetry_u_dynamos_pm5}
    % \end{subfigure}
    % \begin{subfigure}{0.49\linewidth}
    %     \includegraphics[trim=0.3cm 0cm 0cm 0cm,clip,width=\linewidth]{Figures/Symmetry_u_Pm12_dynamos.pdf}
    %     \caption{Symmetry, $s_u$, $\Pm=12$}
    %     \label{fig:Symmetry_b_dynamos_pm5}
    % \end{subfigure}
    \begin{subfigure}{\linewidth}
        \includegraphics[trim=0.3cm 0cm 0cm 0cm,clip,width=\linewidth]{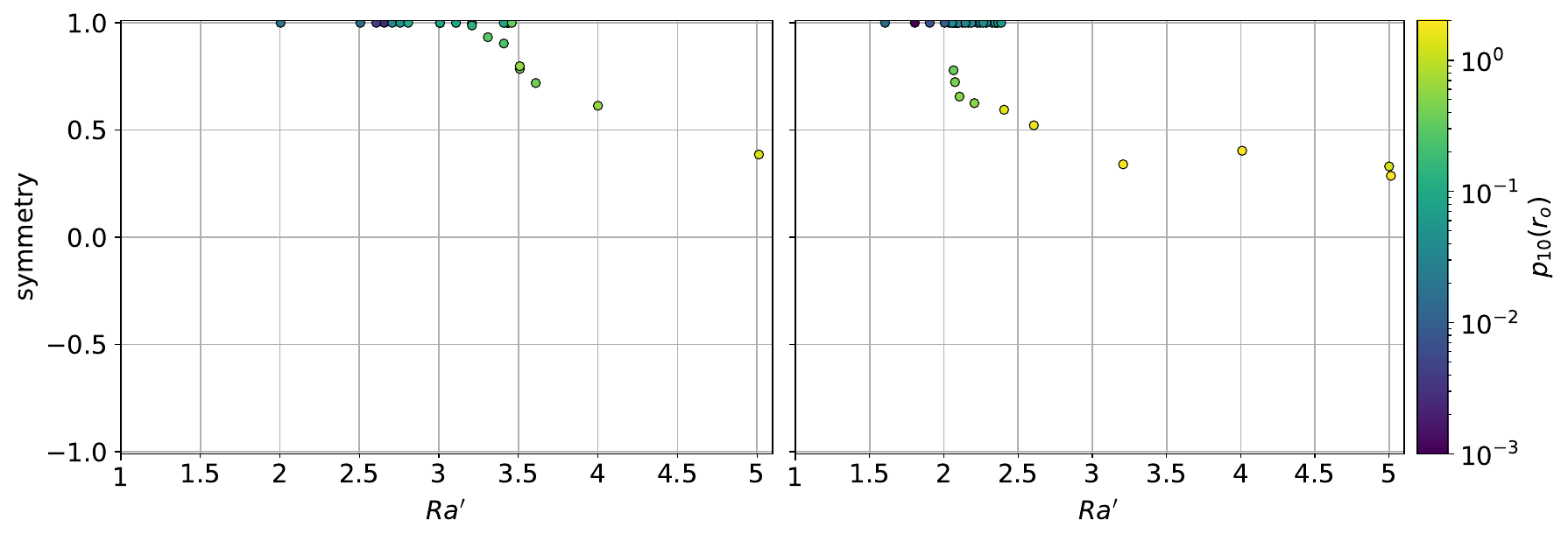}
        \caption{Symmetry, $s_u$ \eqreflong{eqn:2.symmetry}; $\Pm=5$ (left) and $\Pm=12$ (right)}
        \label{fig:Symmetry_u_dynamos}
    \end{subfigure}
    % \begin{subfigure}{0.49\linewidth}
    %     \includegraphics[trim=0.3cm 0cm 0cm 0cm,clip,width=\linewidth]{Figures/Symmetry_B_Pm5_dynamos.pdf}
    %     \caption{Symmetry, $s_B$, $\Pm=5$}
    %     \label{fig:Symmetry_u_dynamos_pm12}
    % \end{subfigure}
    % \begin{subfigure}{0.49\linewidth}
    %     \includegraphics[trim=0.3cm 0cm 0cm 0cm,clip,width=\linewidth]{Figures/Symmetry_B_Pm12_dynamos.pdf}
    %     \caption{Symmetry, $s_B$, $\Pm=12$}
    %     \label{fig:Symmetry_b_dynamos_pm12}
    % \end{subfigure}
    \begin{subfigure}{\linewidth}
        \includegraphics[trim=0.3cm 0cm 0cm 0cm,clip,width=\linewidth]{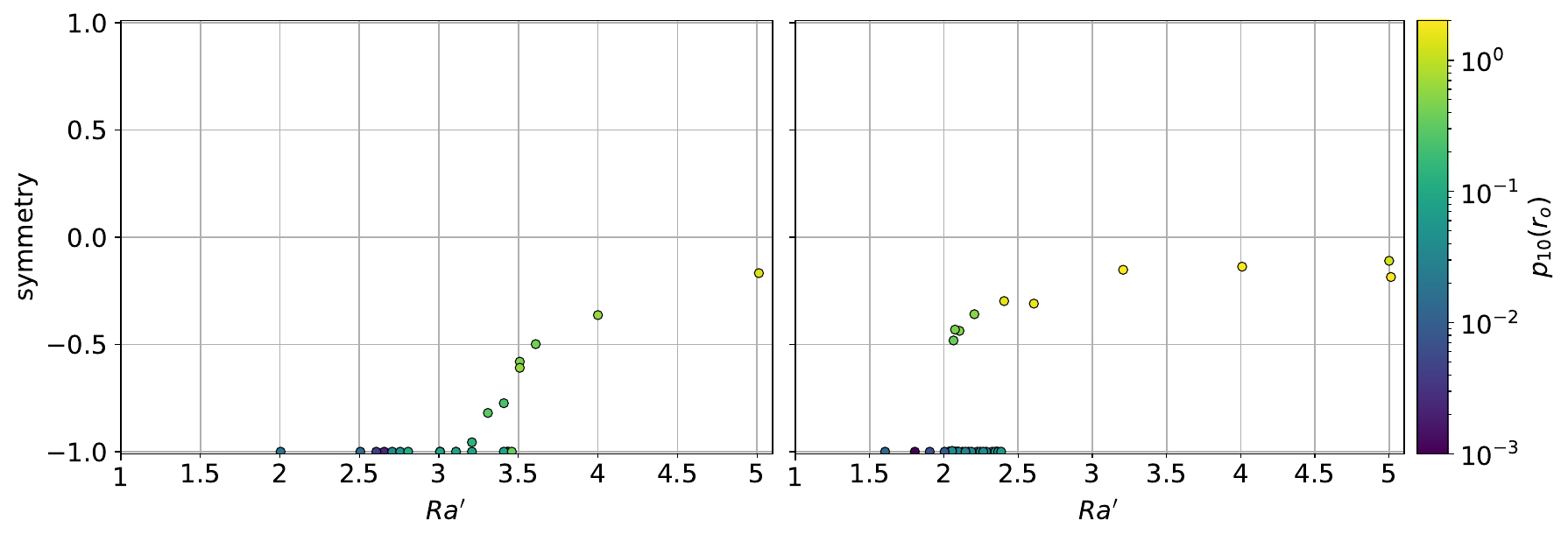}
        \caption{Symmetry, $s_B$ \eqreflong{eqn:2.symmetry}; $\Pm=5$ (left) and $\Pm=12$ (right)}
        \label{fig:Symmetry_b_dynamos}
    \end{subfigure}
    \caption{Global diagnostics of dynamo simulations for $\Pm=5,12$ and $\Ra'\in(1,5]$ where points are coloured by $p_{10}$.}
    \label{fig:Dynamo_params}
\end{figure}

%\lukenote{INTRODUCING THE SIGNIFICANCE OF EQUATORIAL SYMMETRY}
Figure \ref{fig:Dynamo_params} shows the columnarity and symmetry in dynamo simulations for $\Ra'$ between the onset of dynamo action and $\Ra'=5$ for the two values of $\Pm$ that exhibit the clearest bistability between weak- and strong-field (dipolar) solutions. The sharp drops in equatorial symmetry at $\Ra'=3.3$, for $\Pm=5$, and $\Ra'=2.1$, for $\Pm=12$, indicate that an important element in the onset of strong-field convection is the breaking of equatorial symmetry which, as in hydrodynamic convection, seems to occur due to the destabilisation of antisymmetric modes of convection. Polar convective modes are significantly destabilised by the presence of an axial magnetic field since the wavelength of the onset mode perpendicular to the rotation axis is short, minimising the effect of rotation, which is felt more strongly inside the tangent cylinder; this also corresponds to the form required to minimise the bending of axial field lines \citep{Sakuraba2002}. The columnarity drops at the onset of the polar modes due to the growth of the magnetic field which destabilises convective rolls in the flow outside the tangent cylinder.

%\lukenote{FIELD STRENGTH FOR MCV SIMS}
With our magnetoconvection simulations we aim to replicate the strength of the radial magnetic field on the outer boundary separately for weak- and strong-field solutions. Thus, to determine the appropriate magnetic field strength to apply in magnetoconvection simulations, we refer to Figure \ref{fig:Ra_vs_p10_dynamos}, which shows the value of $p_{10}$ (the $(l,m)=(1,0)$ component of the poloidal scalar) at the outer boundary in a set of dynamo simulations. To represent the weak-field dipolar solution branch we choose $p_{10}=0.1$, although, as we will see, we can still reach the strong-field dipolar branch with this boundary condition with sufficiently large $\Pm$, $\Ra'$. For the strong-field dipolar branch, we see that $p_{10}$ generally lies in the range $(3\times10^{-1},2\times10^{0})$, and increases with $\Ra'$ and $\Pm$. We choose to use $p_{10}=0.5$ and $p_{10}=1$ to reflect the strong-field branch.

\section{Magnetoconvection simulations}
\label{sec:4.columnar}

\begin{figure}
    \centering
    % \begin{subfigure}{0.32\linewidth}
    %     \centering
    %     \includegraphics[trim=0cm 0.3cm 0cm 0.1cm,clip,width=\linewidth]{Figures/Columnarity_Pm1.pdf}
    %     \caption{Columnarity, $C_{\omega z}$, $\Pm=1$}
    %     \label{fig:4.Columnarity1}
    % \end{subfigure}
    % \begin{subfigure}{0.32\linewidth}
    %     \centering
    %     \includegraphics[trim=0cm 0.3cm 0cm 0.1cm,clip,width=\linewidth]{Figures/Columnarity_Pm5.pdf}
    %     \caption{Columnarity, $C_{\omega z}$, $\Pm=5$}
    %     \label{fig:4.Columnarity5}
    % \end{subfigure}
    % \begin{subfigure}{0.32\linewidth}
    %     \centering
    %     \includegraphics[trim=0cm 0.3cm 0cm 0.1cm,clip,width=\linewidth]{Figures/Columnarity_Pm12.pdf}
    %     \caption{Columnarity, $C_{\omega z}$, $\Pm=12$}
    %     \label{fig:4.Columnarity12}
    % \end{subfigure}
    \begin{subfigure}{\linewidth}
        \centering
        \includegraphics[trim=-0.55cm 0.3cm 0cm 0.1cm,clip,width=\linewidth]{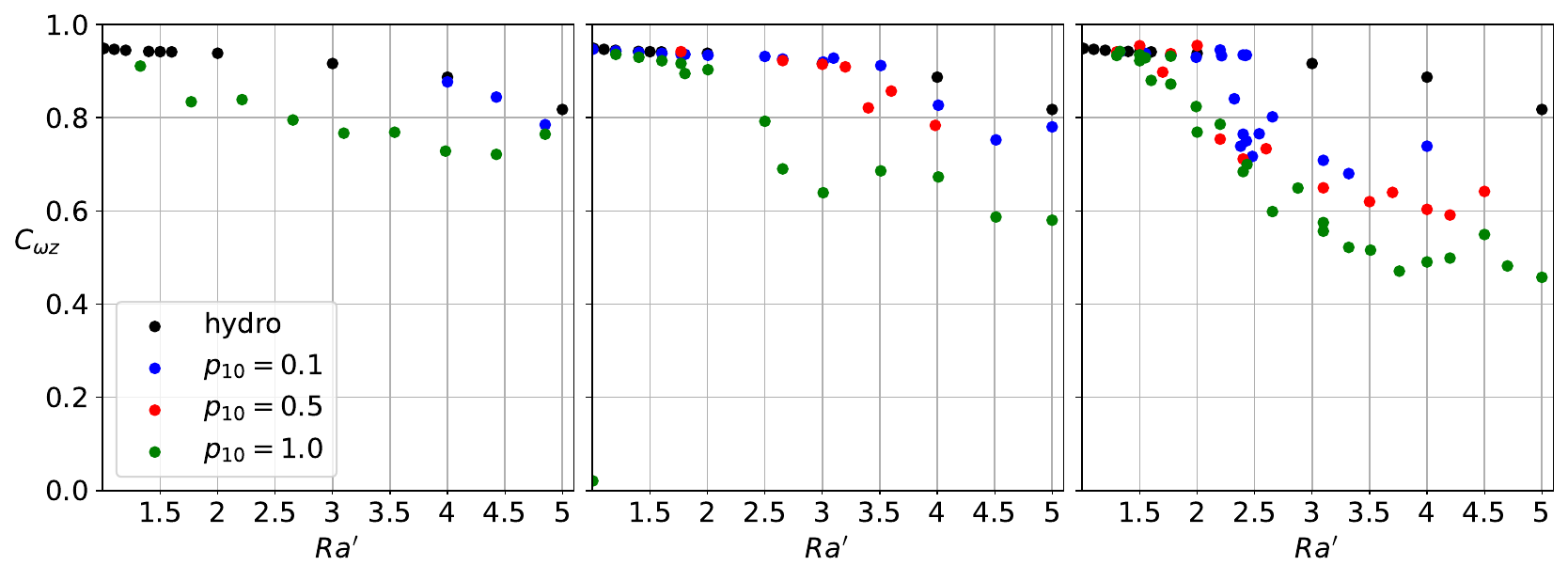}
        \caption{Columnarity, $C_{\omega z}$ \eqreflong{eqn:2.columnarity}; $\Pm=1,5,12$ (left-to-right)}
        \label{fig:4.Columnarity}
    \end{subfigure}
    % \begin{subfigure}{0.32\linewidth}
    %     \centering
    %     \includegraphics[trim=0cm 0.3cm 0cm 0.1cm,clip,width=\linewidth]{Figures/Symmetry_u_Pm1.pdf}
    %     \caption{Symmetry of $u$, $s_u$, $\Pm=1$}
    %     \label{fig:4.Symmetryu1}
    % \end{subfigure}
    % \begin{subfigure}{0.32\linewidth}
    %     \centering
    %     \includegraphics[trim=0cm 0.3cm 0cm 0.1cm,clip,width=\linewidth]{Figures/Symmetry_u_Pm5.pdf}
    %     \caption{Symmetry of $u$, $s_u$, $\Pm=5$}
    %     \label{fig:4.Symmetryu5}
    % \end{subfigure}
    % \begin{subfigure}{0.32\linewidth}
    %     \centering
    %     \includegraphics[trim=0cm 0.3cm 0cm 0.1cm,clip,width=\linewidth]{Figures/Symmetry_u_Pm12.pdf}
    %     \caption{Symmetry of $u$, $s_u$, $\Pm=12$}
    %     \label{fig:4.Symmetryu12}
    % \end{subfigure}
    \begin{subfigure}{\linewidth}
        \centering
        \includegraphics[trim=0cm 0.3cm 0cm 0.1cm,clip,width=\linewidth]{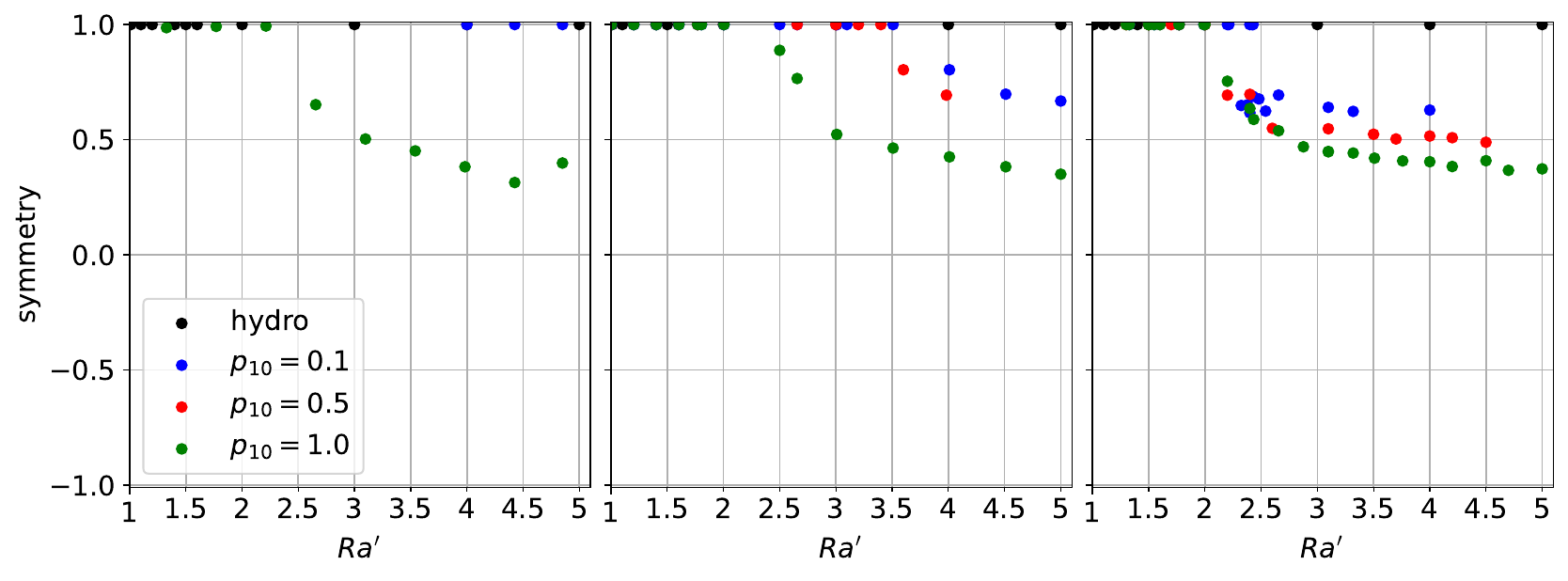}
        \caption{Symmetry of $u$, $s_u$ \eqreflong{eqn:2.symmetry}; $\Pm=1,5,12$ (left-to-right)}
        \label{fig:4.Symmetryu}
    \end{subfigure}
    % \begin{subfigure}{0.32\linewidth}
    %     \centering
    %     \includegraphics[trim=0cm 0.3cm 0cm 0.1cm,clip,width=\linewidth]{Figures/Symmetry_B_Pm1.pdf}
    %     \caption{Symmetry of $B$, $s_B$, $\Pm=1$}
    %     \label{fig:4.SymmetryB1}
    % \end{subfigure}
    % \begin{subfigure}{0.32\linewidth}
    %     \centering
    %     \includegraphics[trim=0cm 0.3cm 0cm 0.1cm,clip,width=\linewidth]{Figures/Symmetry_B_Pm5.pdf}
    %     \caption{Symmetry of $B$, $s_B$, $\Pm=5$}
    %     \label{fig:4.SymmetryB5}
    % \end{subfigure}
    % \begin{subfigure}{0.32\linewidth}
    %     \centering
    %     \includegraphics[trim=0cm 0.3cm 0cm 0.1cm,clip,width=\linewidth]{Figures/Symmetry_B_Pm12.pdf}
    %     \caption{Symmetry of $B$, $s_B$, $\Pm=12$}
    %     \label{fig:4.SymmetryB12}
    % \end{subfigure}
    \begin{subfigure}{\linewidth}
        \centering
        \includegraphics[trim=0cm 0.3cm 0cm 0.1cm,clip,width=\linewidth]{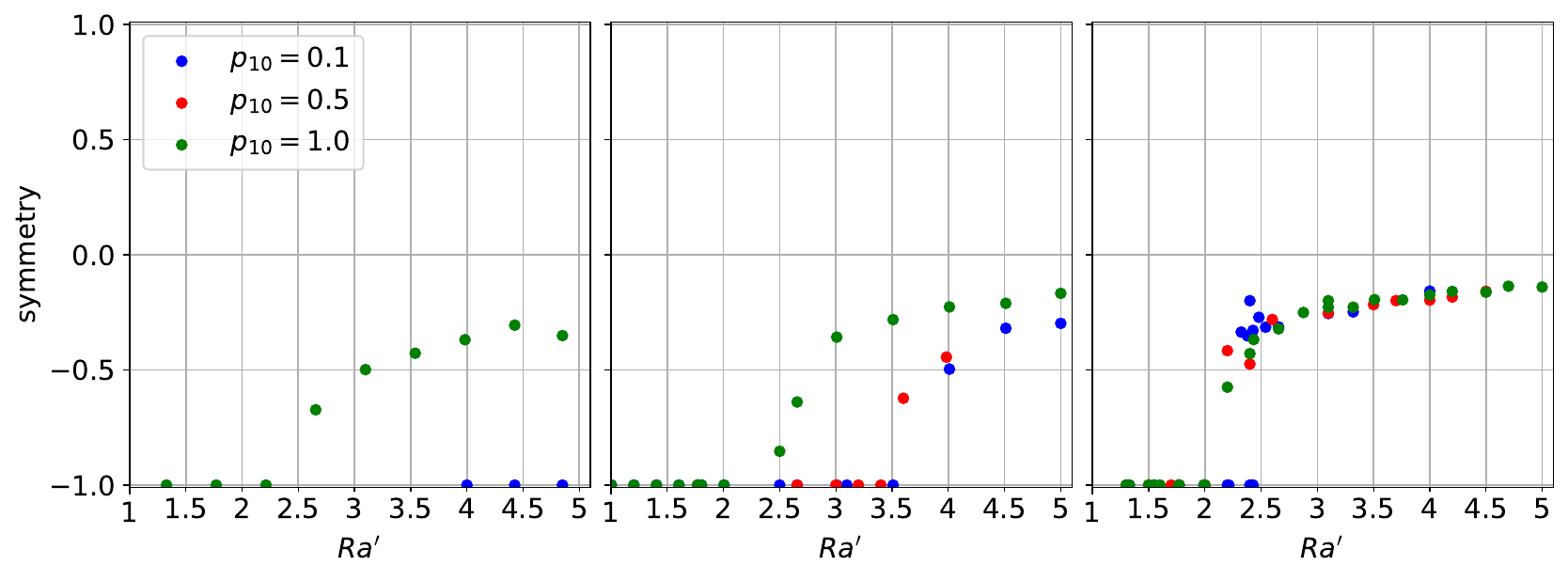}
        \caption{Symmetry of $B$, $s_B$ \eqreflong{eqn:2.symmetry}; $\Pm=1,5,12$ (left-to-right)}
        \label{fig:4.SymmetryB}
    \end{subfigure}
    \caption[]{Global diagnostics of magnetoconvection simulations for varied $\Pm$ and $p_{10}$ over $\Ra'$ in the range $(1,5]$ \emph{(Continued on next page)}.}
    \label{fig:4.Diagnostics}
\end{figure}

\begin{figure}
    \ContinuedFloat
    \centering
    % \begin{subfigure}{0.32\linewidth}
    %     \centering
    %     \includegraphics[trim=0cm 0.3cm 0cm 0.1cm,clip,width=\linewidth]{Figures/TorPol_u_Pm1.pdf}
    %     \caption{Kinetic energy, $\Pm=1$}
    %     \label{fig:4.TorPolu1}
    % \end{subfigure}
    % \begin{subfigure}{0.32\linewidth}
    %     \centering
    %     \includegraphics[trim=0cm 0.3cm 0cm 0.1cm,clip,width=\linewidth]{Figures/TorPol_u_Pm5.pdf}
    %     \caption{Kinetic energy, $\Pm=5$}
    %     \label{fig:4.TorPolu5}
    % \end{subfigure}
    % \begin{subfigure}{0.32\linewidth}
    %     \centering
    %     \includegraphics[trim=0cm 0.3cm 0cm 0.1cm,clip,width=\linewidth]{Figures/TorPol_u_Pm12.pdf}
    %     \caption{Kinetic energy, $\Pm=12$}
    %     \label{fig:4.TorPolu12}
    % \end{subfigure}
    \begin{subfigure}{\linewidth}
        \centering
        \includegraphics[trim=0cm 0.3cm 0cm 0.1cm,clip,width=\linewidth]{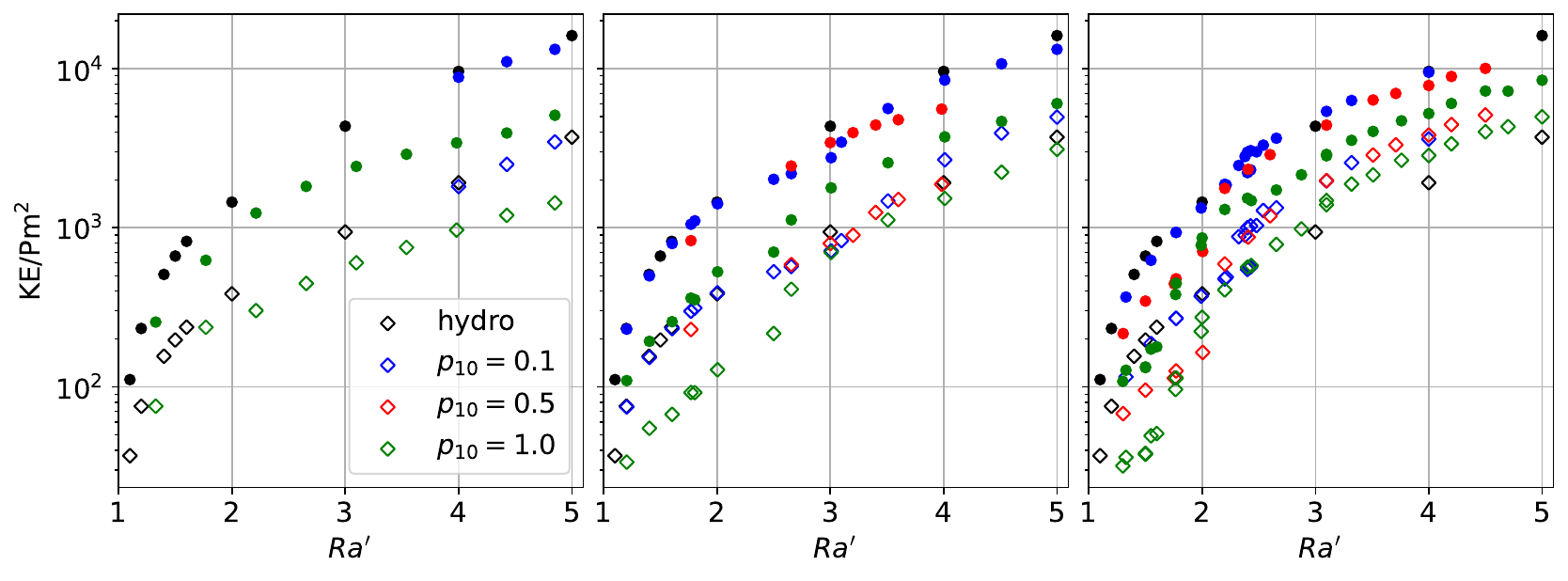}
        \caption{Kinetic energy, $\Pm=1,5,12$ (left-to-right), split into poloidal (unfilled diamonds) and toroidal (filled circles) components.}
        \label{fig:4.TorPolu}
    \end{subfigure}
    % \begin{subfigure}{0.32\linewidth}
    %     \centering
    %     \includegraphics[trim=0cm 0.3cm 0cm 0.1cm,clip,width=\linewidth]{Figures/TorPol_B_Pm1.pdf}
    %     \caption{Magnetic energy, $\Pm=1$}
    %     \label{fig:4.TorPol1}
    % \end{subfigure}
    % \begin{subfigure}{0.32\linewidth}
    %     \centering
    %     \includegraphics[trim=0cm 0.3cm 0cm 0.1cm,clip,width=\linewidth]{Figures/TorPol_B_Pm5.pdf}
    %     \caption{Magnetic energy, $\Pm=5$}
    %     \label{fig:4.TorPol5}
    % \end{subfigure}
    % \begin{subfigure}{0.32\linewidth}
    %     \centering
    %     \includegraphics[trim=0cm 0.3cm 0cm 0.1cm,clip,width=\linewidth]{Figures/TorPol_B_Pm12.pdf}
    %     \caption{Magnetic energy, $\Pm=12$}
    %     \label{fig:4.TorPol12}
    % \end{subfigure}
    \begin{subfigure}{\linewidth}
        \centering
        \includegraphics[trim=0cm 0.3cm 0cm 0.1cm,clip,width=\linewidth]{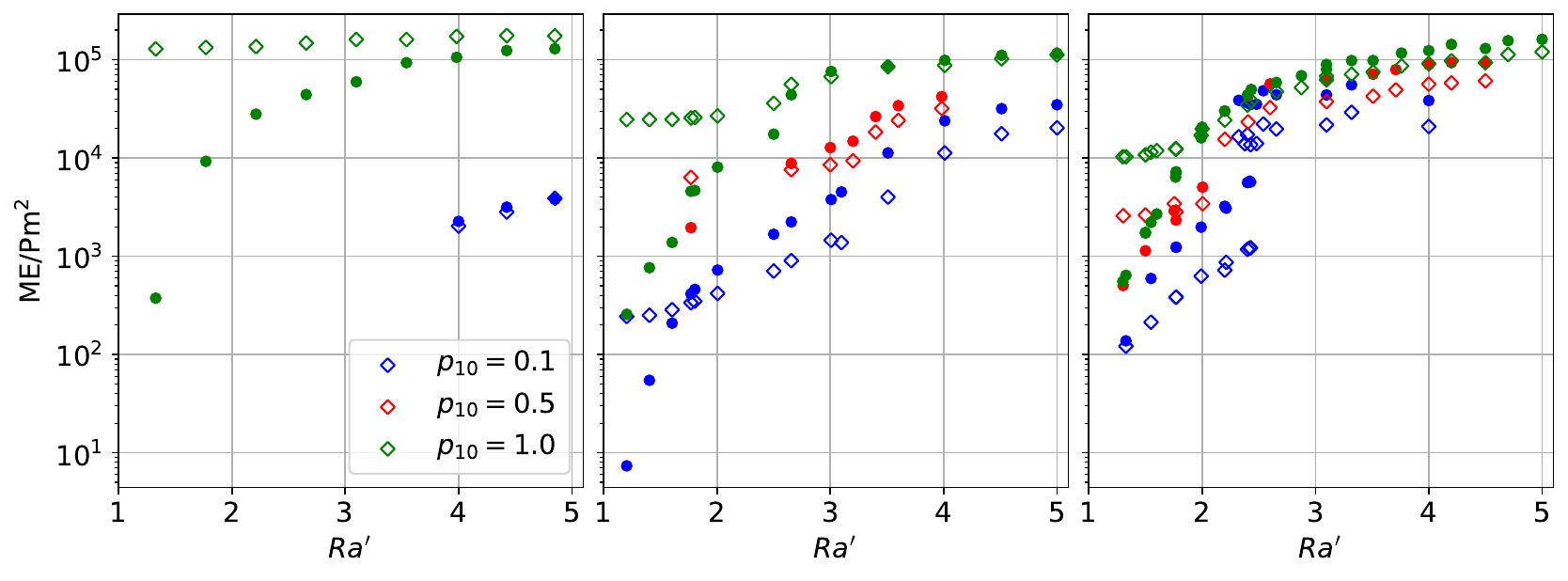}
        \caption{Magnetic energy, $\Pm=1,5,12$ (left-to-right), split into poloidal (unfilled diamonds) and toroidal (filled circles) components.}
        \label{fig:4.TorPol}
    \end{subfigure}
    \caption{Global diagnostics of magnetoconvection simulations for varied $\Pm$ and $p_{10}$ over $\Ra'$ in the range $(1,5]$ \emph{(continued)}. Black points of the kinetic energy refer to hydrodynamic simulations which are independent of $\Pm$. The energies of magnetoconvection simulations are rescaled by a factor of $\Pm^2$ to align them between axes, this is equivalent to using a weak-field non-dimensionalisation.}
\end{figure}

In this section, we present results from magnetoconvection simulations at $\Pm=1,5,12$ with fixed $\Ek=10^{-4}$, $\Pr=1$, Rayleigh numbers $\Ra'\in(1,5]$, and imposed axial field strengths $p_{10}\in[0.0,2.5]$. \firstrevision{The smallest values of $\Ra'$ would be insufficient to permit dynamo action \citep{ChristensenAubert2006} but are vigorous enough to amplify the weak field applied at the outer boundary. Another concern might be that a moderate-strength magnetic field could dampen convection; however, linear magnetoconvection studies show that this effect is small at most, and that typically the addition of a magnetic field ($\Els\gtrsim1$) can reduce the critical Rayleigh number for the onset of convection \cite{MasonEtAl2022}.} 

Figure~\ref{fig:4.Diagnostics} summarises the global diagnostics across this parameter space, including columnarity, equatorial symmetry of the flow and magnetic field, and the total kinetic and magnetic energies. \firstrevision{One immediate observation is that there are regions of bistability where, for the same parameter inputs, distinct values of some output quantities are found. As in dynamo simulations, these regions are found only at $\Pm=5,12$, not at $\Pm=1$.}

Figure~\ref{fig:4.Columnarity} shows the columnarity $C_{\omega z}$ of the bulk flow outside the tangent cylinder. Values close to unity correspond to nearly $z$-independent, columnar convection, while $C_{\omega z}\firstrevision{\ll1}$ would indicate fully three-dimensional isotropic flow. \firstrevision{Soderlund et al. \cite{SoderlundEtAl2012} introduce columnarity to detect columnar convection, hence why the value is defined using integrals over the bulk, since convection typically takes on a different form in the tangent cylinder, where the radial direction aligns more closely with the rotation axis. In their dynamo simulations for $\Pm\in(0,2]$ \cite{SoderlundEtAl2012} introduce a threshold of $C_{\omega z}\approx0.5$ for the transition away from columnar convection, and find that the magnetic field has little effect on its value compared with purely hydrodynamic solutions. Across our suite we find that several magnetoconvection branches (coloured points) correspond closely with columnarity of dynamo and hydrodynamic simulations (cf.~Fig.~\ref{fig:Columnarity_hydro}); this alignment is strongest for small $\Ra'$ when $p_{10}$ and $\Pm$ are both small, e.g.~$p_{10}=0.1$ and $\Pm=0.5$, indicating that the induced small-scale field is more important than the imposed large-scale field for decreasing columnarity. When the leading order balance is between the Lorentz and Coriolis forces, the double-curl of eq.~\eqref{eqn:2.01} yields
\begin{equation}
    2\pdv{\vec{\omega}}{z}\approx\curl\curl{\left((\curl{\vec{B}})\times\vec{B}\right)},
\end{equation}
indicating that the effect on columnarity is proportional to the cube of the scale of the induced magnetic field.}

Figures~\ref{fig:4.Symmetryu}–\ref{fig:4.SymmetryB} show the equatorial symmetry of the velocity and magnetic fields. The flow remains predominantly equatorially symmetric ($s_{u}>0$) for all cases considered, consistent with the persistence of rotational control, but increasing $\Pm$ or \firstrevision{adding even a small seed magnetic field} substantially lowers the Rayleigh number at which equatorially asymmetric states become accessible, compared with the hydrodynamic case where symmetry breaking occurs only for $\Ra'\approx10$ (Figure \ref{fig:Symmetry_hydro}). As $\Ra'$ increases, $s_{u}$ approaches an apparent asymptotic value near $s_{u}\simeq1/3$, although confirmation of this behaviour would require access to higher Rayleigh numbers. \firstrevision{We expect that this asymptotic limit, if it exists, would depend on several features that are held constant in this study, such as the velocity boundary conditions, the morphology of the imposed magnetic field, and the size of the inner core.}

\firstrevision{At weak supercriticality, the symmetry of the magnetic field ($s_{B}=-1$) is precisely opposite that of the velocity field. As $\Ra'$ increases, the antisymmetry of the magnetic field is broken with the symmetry of the velocity field though, as in dynamo simulations, the magnetic field is typically closer to asymmetry than the velocity field. Given that the imposed field is perfectly antisymmetric, it might be expected that a stronger imposed field would generally decrease $s_{B}$, however, this is not found to be the case; at $\Pm=12$ the symmetry of the field is almost independent of $p_{10}$, whilst at $\Pm=5$ a stronger imposed field leads to a value of $s_{B}$ which is closer zero, perhaps indicating the role of the stronger magnetic field in disrupting the flow and, consequently, leading to asymmetry in the field.}

The kinetic and magnetic energies exhibit a more complex dependence on $\Ra'$, $\Pm$, and $p_{10}$. At small $\Ra'$ the kinetic energy is small and the magnetic energy is dominated by the imposed poloidal field (\S\ref{sec:2.fieldconfig}); \firstrevision{however attempting to separate out the induced field is not a straightforward task as \cite{MasonEtAl2022}, e.g., have shown that the induced field can act either to support or diminish the imposed field.} As $\Ra'$ increases both energies increase monotonically \firstrevision{and several sharp changes also coincide with the loss of equatorial symmetry, signalling transitions between distinct flow regimes.} These transitions are examined in detail for each value of $\Pm$ in the following subsections.

\subsection{Magnetoconvection with \texorpdfstring{$\Pm=1$}{Pm=1}}
\label{sec:4.pm1}

\begin{figure}
    \centering
    % \begin{subfigure}{0.49\linewidth}
    %     \includegraphics[width=\linewidth]{Figures/mcv_OD1.0_E1e-04_1.324Rac_Pm1_lforces_compcurl_wobl_odd.pdf}
    %     \caption{Odd $l$-modes, $\Ra'=1.3$, $p_{10}=1.0$}
    % \end{subfigure}
    \begin{subfigure}{0.49\linewidth}
        \includegraphics[width=\linewidth]{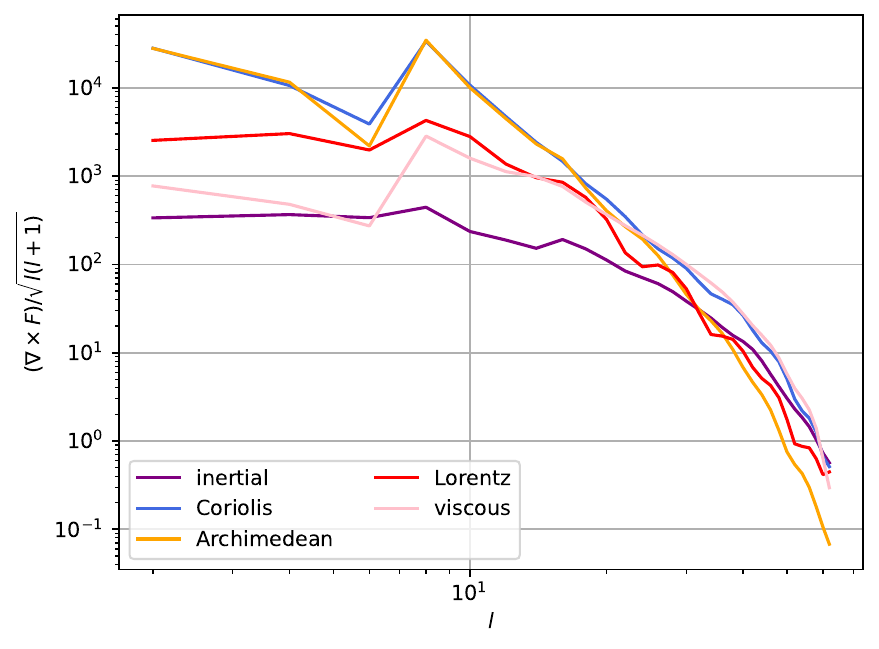}
        \caption{Even $l$-modes, $\Ra'=1.3$, $p_{10}=1.0$}
        \label{fig:4.ForceBal_Pm1_a}
    \end{subfigure}
    % \begin{subfigure}{0.49\linewidth}
    %     \includegraphics[width=\linewidth]{Figures/mcv_OD1.0_E1e-04_1.766Rac_Pm1_lforces_compcurl_wobl_odd.pdf}
    %     \caption{Odd $l$-modes, $\Ra'=1.75$, $p_{10}=1.0$}
    % \end{subfigure}
    \begin{subfigure}{0.49\linewidth}
        \includegraphics[width=\linewidth]{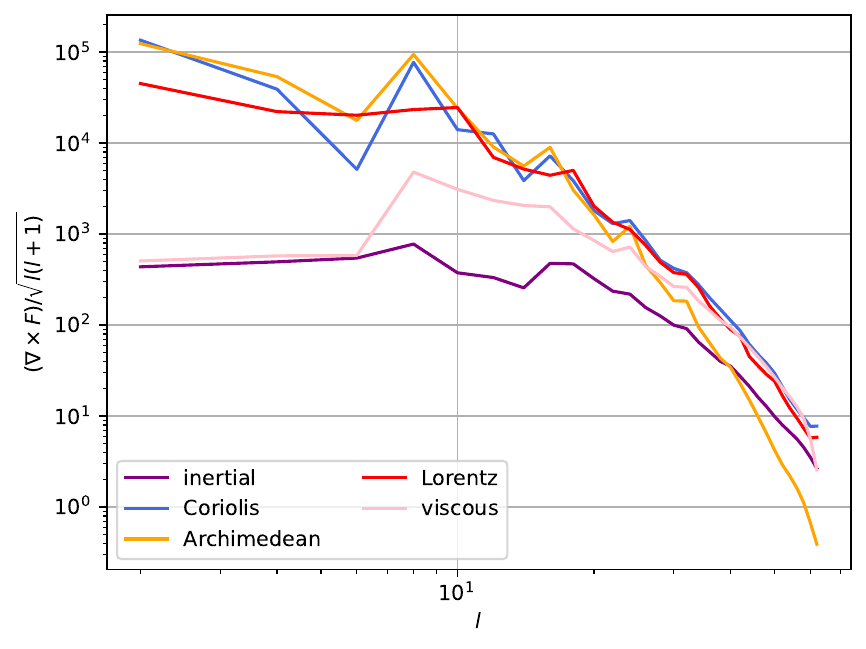}
        \caption{Even $l$-modes, $\Ra'=1.75$, $p_{10}=1.0$}
        \label{fig:4.ForceBal_Pm1_d}
    \end{subfigure}
    \begin{subfigure}{0.49\linewidth}
        \includegraphics[width=\linewidth]{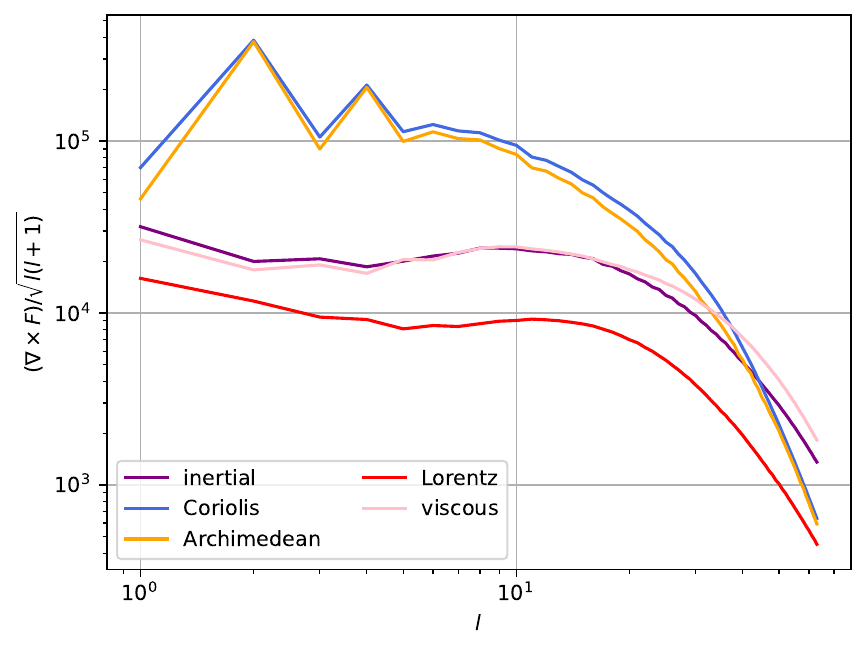}
        \caption{$l$-modes, $\Ra'=4.85$, $p_{10}=0.1$}
        \label{fig:4.ForceBal_Pm1_e}
    \end{subfigure}
    \begin{subfigure}{0.49\linewidth}
        \includegraphics[width=\linewidth]{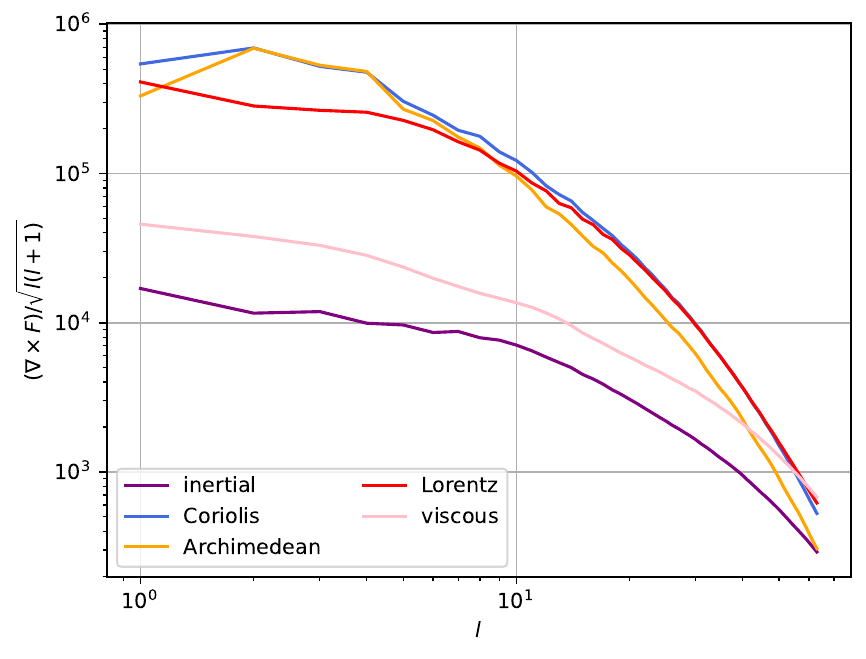}
        \caption{$l$-modes, $\Ra'=4.85$, $p_{10}=1.0$}
        \label{fig:4.ForceBal_Pm1_f}
    \end{subfigure}
    \caption{Time-averaged compensated-curl of forces \citep[see][]{TeedDormy2023} in simulations at $\Pm=1$, $p_{10}=0.1,1.0$, at three values of $\Ra'$. Only even $l$-modes are presented in the first two cases due to the strong equatorial symmetry of the buoyancy force and the prominence of the $m=8$ convective mode and $m=0$ zonal mode leading to a dearth of odd-$l$ buoyancy modes.}
    \label{fig:4.ForcesPm1}
\end{figure}

\begin{figure}
    \centering
    \begin{subfigure}{0.39\linewidth}
        \includegraphics[width=\linewidth]{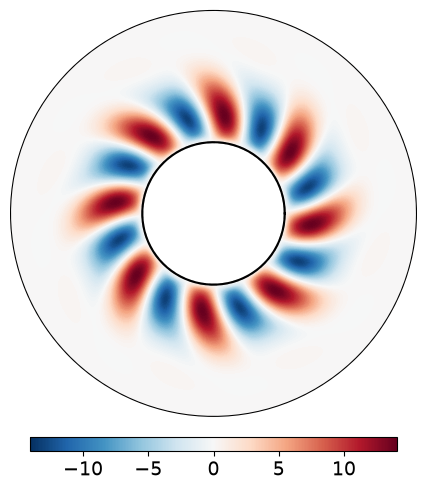}
        \caption{$\vnd{r}\cdot\vec{u}$, $\Ra'=1.3$}
    \end{subfigure}
    \begin{subfigure}{0.39\linewidth}
        \includegraphics[width=\linewidth]{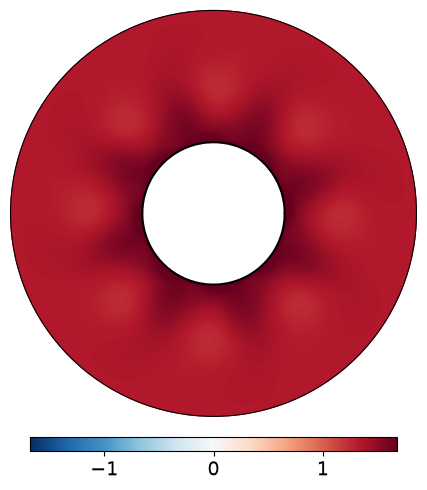}
        \caption{$B_z$, $\Ra'=1.3$}
    \end{subfigure}
    \begin{subfigure}{0.39\linewidth}
        \includegraphics[width=\linewidth]{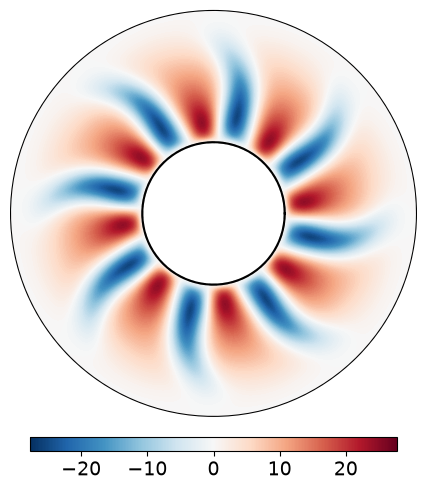}
        \caption{$\vnd{r}\cdot\vec{u}$, $\Ra'=1.75$}
    \end{subfigure}
    \begin{subfigure}{0.39\linewidth}
        \includegraphics[width=\linewidth]{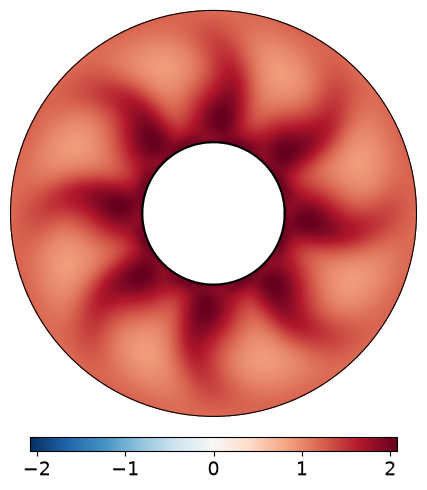}
        \caption{$B_z$, $\Ra'=1.75$}
    \end{subfigure}
    \begin{subfigure}{0.39\linewidth}
        \includegraphics[width=\linewidth]{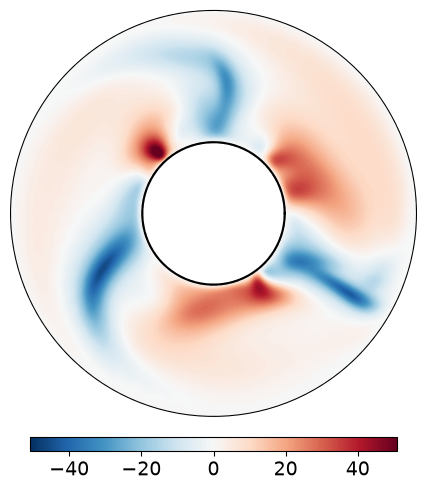}
        \caption{$\vnd{r}\cdot\vec{u}$, $\Ra'=4.85$}
    \end{subfigure}
    \begin{subfigure}{0.39\linewidth}
        \includegraphics[width=\linewidth]{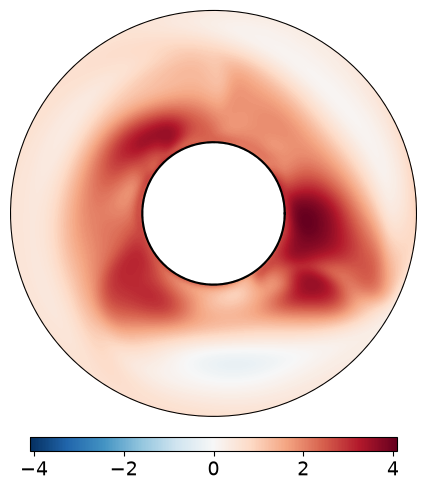}
        \caption{$B_z$, $\Ra'=4.85$}
    \end{subfigure}
    \caption{Equatorial sections for runs with $\Pm=1$ and $p_{10}=1$.}
    \label{fig:4.ColumnarPm1}
\end{figure}

At $\Pm=1$, the system lies only slightly above the threshold for self-sustained dynamo action, with dynamo solutions appearing only for $\Ra'\gtrsim4.4$ (Figure~\ref{fig:Ra_vs_p10_dynamos}). Over the Rayleigh number range considered here, the magnetic field is therefore primarily controlled by the imposed field, particularly for $p_{10}=1.0$. \firstrevision{Global diagnostics for these simulations are plotted in panels of the first column of Figure~\ref{fig:4.Diagnostics} and give no hint of weak/strong-field bistability in this series. This conforms with expectations since the induced magnetic field is small relative to the imposed one at low $\Pm$. Although there are no abrupt transitions at this $\Pm$, some interesting regimes can be identified which are discussed below in turn, referring to the force spectra and flow snapshots which are collected in Figures~\ref{fig:4.ForcesPm1} and \ref{fig:4.ColumnarPm1}, respectively.}

For weak imposed fields ($p_{10}=0.1$), the Lorentz force remains subdominant relative to the Coriolis, viscous, and buoyancy forces, and the flow closely resembles the hydrodynamic columnar state. The preferred azimuthal mode is $m=7$, columnarity is only slightly reduced relative to the hydrodynamic case (Figure~\ref{fig:4.Columnarity}), and the flow remains equatorially symmetric throughout the explored range of $\Ra'$ (Figure~\ref{fig:4.Symmetryu}). As $\Ra'$ increases, inertial effects strengthen \firstrevision{faster than} viscous contributions, with approximate parity between inertia and viscosity reached near $\Ra'=4.85$, where inertia slightly dominates at low degrees and viscosity at high degrees due to its additional derivative (Figure \ref{fig:4.ForceBal_Pm1_e}).

For $p_{10}=1.0$ the field is dominated by the imposed poloidal field and in the range of $\Ra'$ shown, the energy of the toroidal magnetic field does not grow as large as the poloidal energy, although it increases significantly as $\Ra'$ increases. At $\Ra'=1.3$, the flow takes a form similar to the hydrodynamic columns and remains largely $z$-independent (as indicated by $C_{\omega z}$ in Figure~\ref{fig:4.Columnarity}). The imposed magnetic field is weakly swept by the columnar flow and is strongest in the cold, inward, convective plumes (Figures~\ref{fig:4.ColumnarPm1}a,b). The force balance for this run is shown in Figure~\ref{fig:4.ForceBal_Pm1_a}. Odd and even $l$-modes are shown separately because the strong equatorial symmetry of the buoyancy force suppresses odd-$l$ contributions at low Rayleigh number, particularly when the convective onset mode has $m$ even (here $m=8$ is preferred). The Lorentz force is large at small $l$ due to the imposed field and still large, but on par with the viscous force for all orders $l\geq8$ (the onset mode); at the smallest orders, $l\gtrsim32$, inertia becomes larger than the Lorentz force.

At slightly larger forcing, $\Ra'=1.75$, the flow strengthens and the axial field is further concentrated in the cold convective plumes (Figures \ref{fig:4.ColumnarPm1}c,d), which produces an asymmetry between the inward and outward flows. If we further increase the Rayleigh number then the $m=8$ becomes unstable and successive lower-order modes are instead preferred. At $\Ra'=4.85$, the $m=3$ mode is preferred, although the columns are unsteady, which spreads the axial field into a wider region (Figures \ref{fig:4.ColumnarPm1}e,f); these radial jets seem to correspond closely to the simulation presented by \cite{Sakuraba2007}, who instead find a $m=1$ mode to be preferred, as expected from magnetostrophic modes at smaller Ekman number \cite{Sakuraba2002}. \firstrevision{The transition to non-axisymmetric inward and outward jets, at $\Ra'\approx1.75$, is also associated with a sharp increase in the toroidal magnetic field strength (Figure~\ref{fig:4.TorPol}) which additionally serves to promote the Lorentz force into the dominant balance at all scales (Figure~\ref{fig:4.ForceBal_Pm1_d} vs.~\ref{fig:4.ForceBal_Pm1_a})}. At larger $\Ra'$, inertia and viscosity are pushed out of the dominant balance at all except the smallest length scales, leading to a clear MAC balance across scales \firstrevision{(Figure~\ref{fig:4.ForceBal_Pm1_f})}.

In contrast to magnetoconvection, dynamo simulations at $\Pm=1$ do not generate magnetic fields of comparable large-scale strength over this Rayleigh number range \firstrevision{(cf.~Fig.~\ref{fig:Ra_vs_p10_dynamos} where $p_{10}\sim0.2$), perhaps explaining why they} do not access the symmetry-breaking transition observed here only for $p_{10}=1$.

\subsection{Magnetoconvection with \texorpdfstring{$\Pm=5$}{Pm=5}}
\label{sec:4.pm5}

\begin{figure}
    \centering
    \begin{subfigure}{0.49\linewidth}
        \centering
        \includegraphics[width=\linewidth]{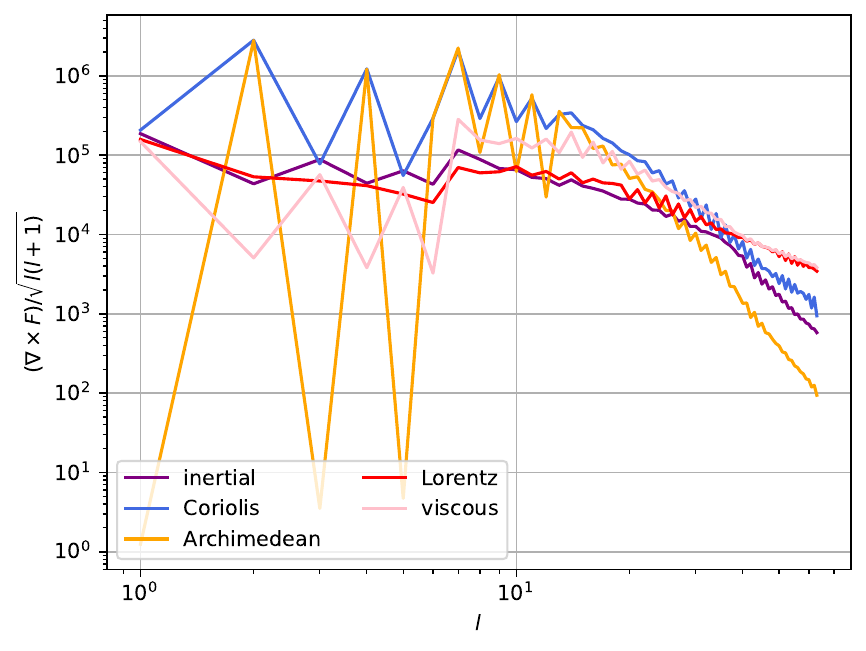}
        \caption{$\Ra'=2.5$, $p_{10}=0.1$}
        \label{fig:4.ForceBalances5c}
    \end{subfigure}
    \begin{subfigure}{0.49\linewidth}
        \centering
        \includegraphics[width=\linewidth]{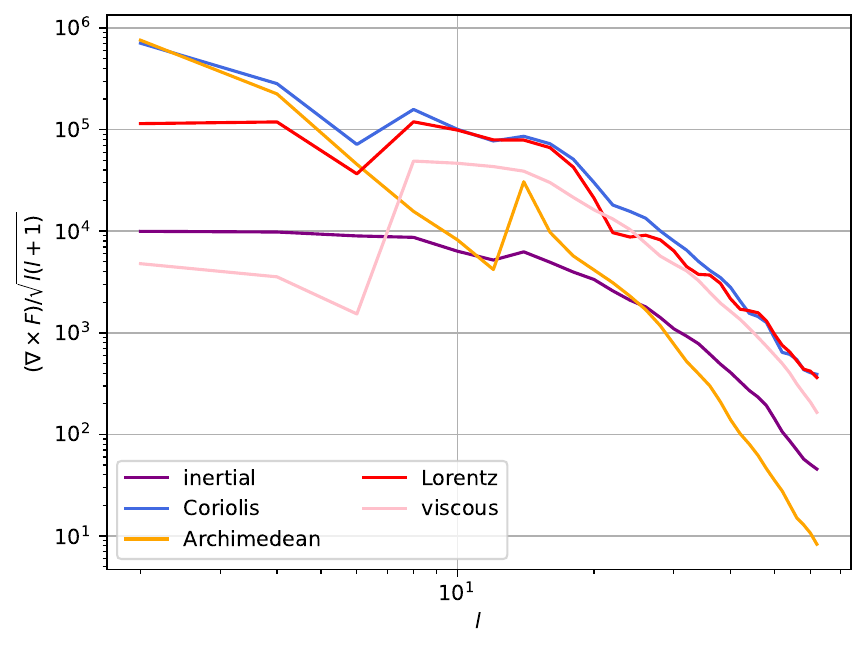}
        \caption{$\Ra'=1.6$, $p_{10}=1.0$}
        \label{fig:4.ForceBalances5a}
    \end{subfigure}
    \begin{subfigure}{0.49\linewidth}
        \centering
        \includegraphics[width=\linewidth]{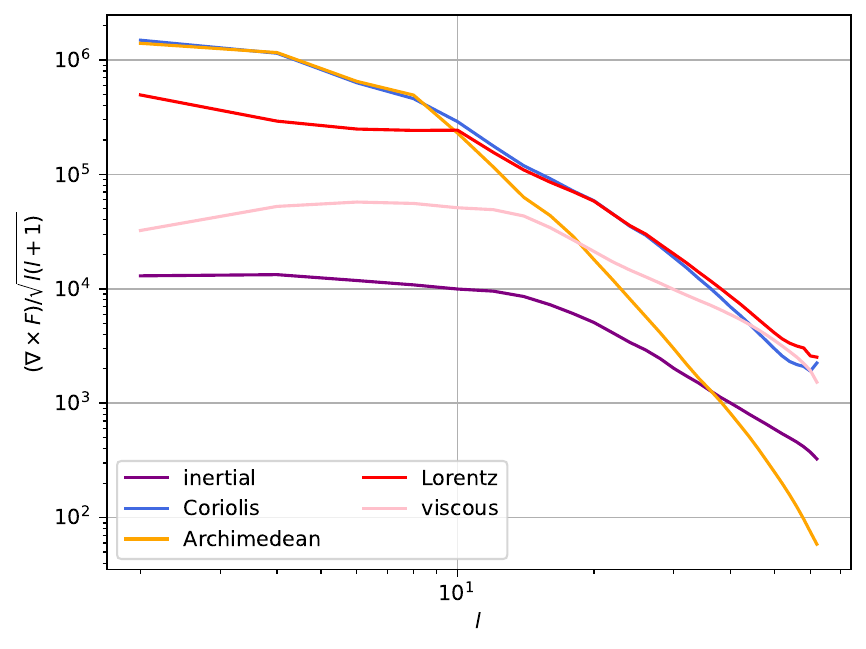}
        \caption{$\Ra'=1.8$, $p_{10}=1.0$}
        \label{fig:4.ForceBalances5b}
    \end{subfigure}
    \begin{subfigure}{0.49\linewidth}
        \centering
        \includegraphics[width=\linewidth]{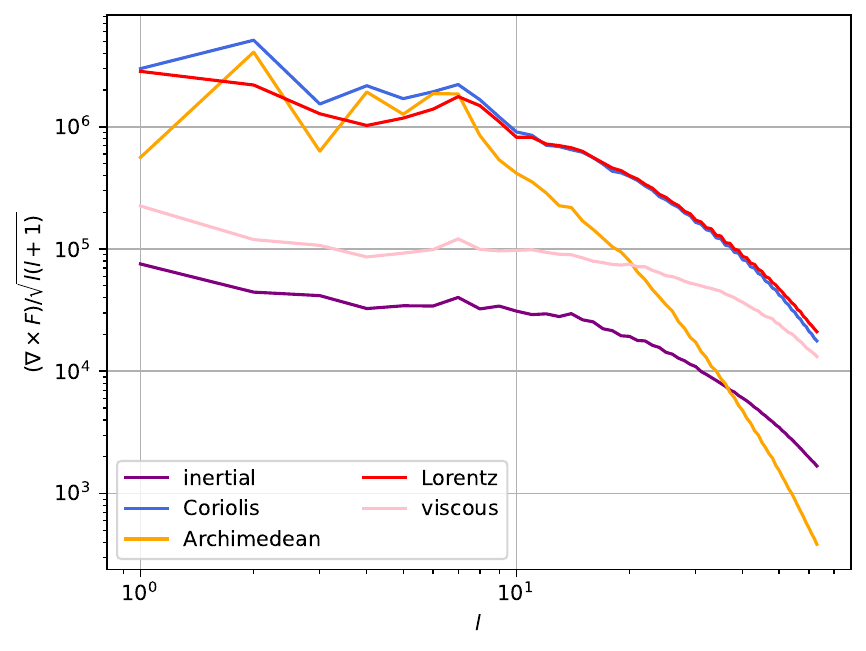}
        \caption{$\Ra'=2.5$, $p_{10}=1.0$}
        \label{fig:4.ForceBalances5d}
    \end{subfigure}
    \caption{Force spectra as in Figure~\ref{fig:4.ForcesPm1}, at $\Pm=5$, for $\Ra'=2.5$ for $p_{10}=0.1$ which is in the weak-field regime, and $\Ra'=1.6,1.8,2.5$ for $p_{10}=1.0$, which bridges the transition from the weak- to the vacillating, to the strong-field regime.}
    \label{fig:4.ForceBalances5}
\end{figure}

At $\Pm=5$, magnetoconvection exhibits a wider set of flow regimes than at $\Pm=1$ since \firstrevision{reducing magnetic dissipation leads to the development of more complex magnetic structures.} Over the range of $\Ra'$, $\Pm$, and $p_{10}$, we identify three distinct regimes as the imposed field strength is increased: (i) a weak-field columnar regime governed by a viscous–Archimedean–Coriolis (VAC) balance, (ii) an intermediate vacillating regime characterised by quasi-periodic growth and destruction of convective columns, and (iii) a strong-field regime dominated by radial jets and a magnetostrophic (MAC) balance; similar, but more pronounced than at $\Pm=1$. These regimes are most clearly distinguished by changes in columnarity, equatorial symmetry, and the partitioning of kinetic and magnetic energies shown in \firstrevision{the panels of the second column of} Figure~\ref{fig:4.Diagnostics}. We discuss each regime in turn\firstrevision{, referring to the force spectra which are combined for this $\Pm$ in Figure~\ref{fig:4.ForceBalances5}.}

\begin{figure}
    \centering
    \begin{subfigure}{0.32\linewidth}
        \centering
        \includegraphics[width=\linewidth]{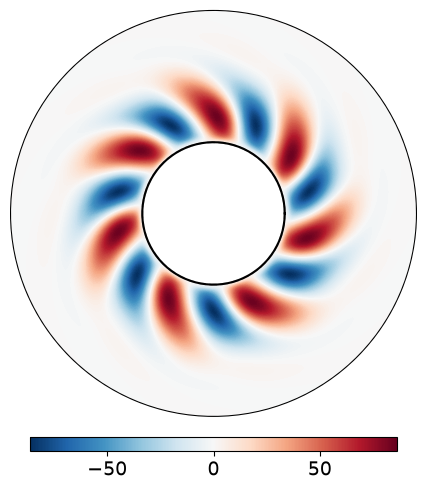}
        \caption{}
        \label{fig:4.columnar_pm5_1}
    \end{subfigure}
    \begin{subfigure}{0.32\linewidth}
        \centering
        \includegraphics[width=\linewidth]{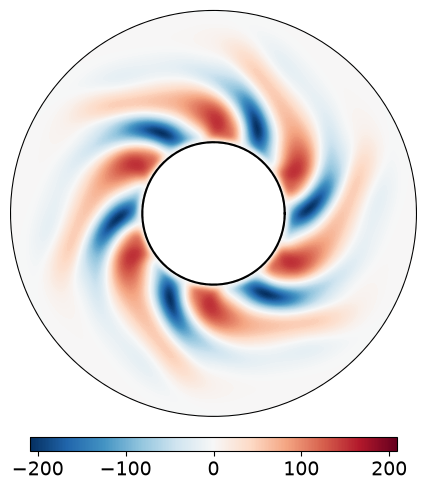}
        \caption{}
        \label{fig:4.columnar_pm5_2}
    \end{subfigure}
    \begin{subfigure}{0.32\linewidth}
        \centering
        \includegraphics[width=\linewidth]{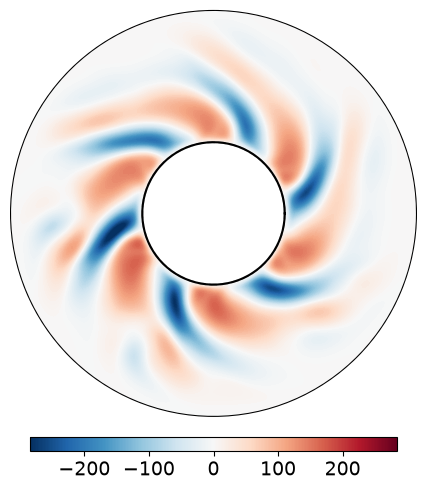}
        \caption{}
        \label{fig:4.columnar_pm5_3}
    \end{subfigure}
    \begin{subfigure}{0.32\linewidth}
        \centering
        \includegraphics[width=\linewidth]{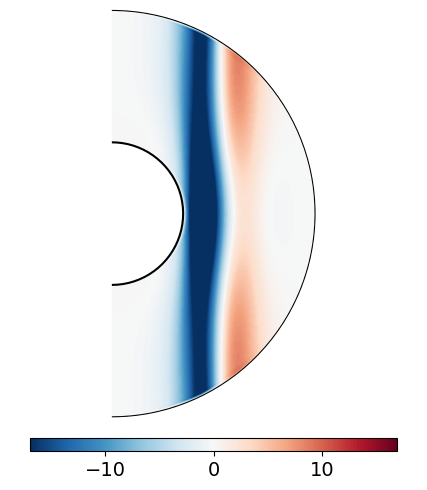}
        \caption{}
        \label{fig:4.columnar_pm5_4}
    \end{subfigure}
    \begin{subfigure}{0.32\linewidth}
        \centering
        \includegraphics[width=\linewidth]{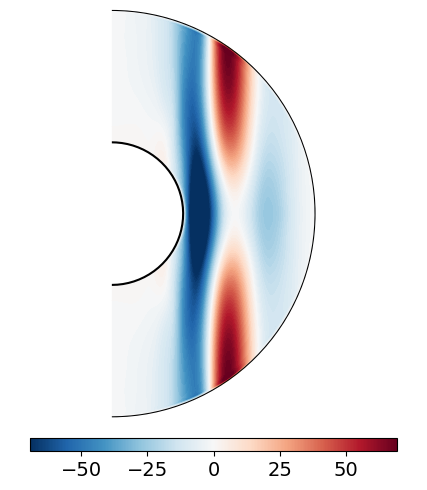}
        \caption{}
        \label{fig:4.columnar_pm5_5}
    \end{subfigure}
    \begin{subfigure}{0.32\linewidth}
        \centering
        \includegraphics[width=\linewidth]{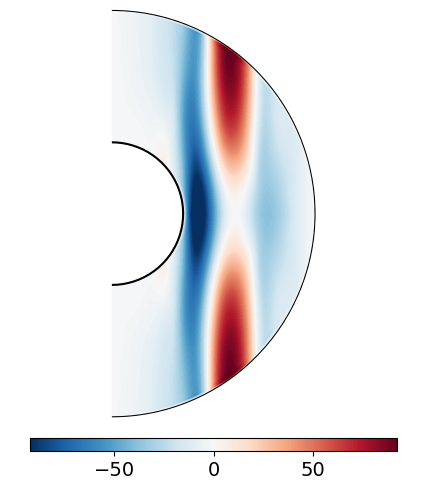}
        \caption{}
        \label{fig:4.columnar_pm5_6}
    \end{subfigure}
    \caption{Equatorial sections of the radial velocity, $\vnd{r}\cdot\vec{u}$ (a-c), and zonal averages of the zonal velocity, $\vnd{\phi}\cdot\vec{u}$ (d-f), in the columnar regime at $\Pm=5$ with $p_{10}=0.1$ (cf. $\Pm=12$, Figure \ref{fig:4.columnar}). a,d: $\Ra'=1.4$. b,e: $\Ra'=3.09$. c,f: $\Ra'=3.5$.}
    \label{fig:4.weak_pm5}
\end{figure}

\paragraph{Weak-field columnar regime.}
For small imposed field strengths ($p_{10}\lesssim0.8$, corresponding to $\Els\lesssim1$), convection takes the form of geostrophic columns closely resembling the hydrodynamic solution and a leading-order VAC balance (Figure~\ref{fig:4.ForceBalances5c} shows that these simulations have a VAC balance for $l\geq7$, with viscosity subdominant for $l$-modes below the order of the columnar mode). For $p_{10}=0.1$, the geostrophic onset mode remains stable up to $\Ra'\approx3.5$, decreasing from $m=7$ to $m=6$ as $\Ra'$ increases. Figure~\ref{fig:4.weak_pm5} illustrates the evolution of the flow structure in this regime: as thermal forcing increases, the columns extend further radially and the preferred azimuthal wavenumber decreases, as in hydrodynamic rotating convection \firstrevision{for sufficiently small driving where an equivalent columnar state can be maintained (cf.~Fig.~\ref{fig:Ecomps_hydro} for small $\Ra'$)}.

The zonal flow in this regime is weak and approximately $z$-independent at low $\Ra'$, consistent with the hydrodynamic case \citep[e.g.][]{SimitevBusse2003}. As $\Ra'$ increases, the zonal flow strengthens and acquires a modest $z$-dependence, reflecting the increasing importance of the thermal wind. This behaviour is captured by the zonal component of the curl of the momentum equation \citep[e.g.][]{Aubert2005},
\begin{equation}
    2\pdv{\left[\vnd{\phi}\cdot\vec{u}\right]_\phi}{z}=\Pm\Ra\,\pdv{\left[T\right]_\phi}{\theta}+\left[\vnd{\phi}\cdot\Del\times(\vec{j}\times\vec{B})\right]_\phi,
    \label{eqn:4.thermalwind}
\end{equation}
where \firstrevision{$\left[\right]_{\phi}$ denotes azimuthal-averaging;} the Lorentz force contribution remains subdominant in this regime. The flow remains strongly equatorially symmetric and columnarity decreases only weakly with increasing $\Ra'$ (middle panels of Figures~\ref{fig:4.Symmetryu}, \ref{fig:4.Columnarity}).

\begin{figure}
    \centering
    \includegraphics[width=0.5\linewidth]{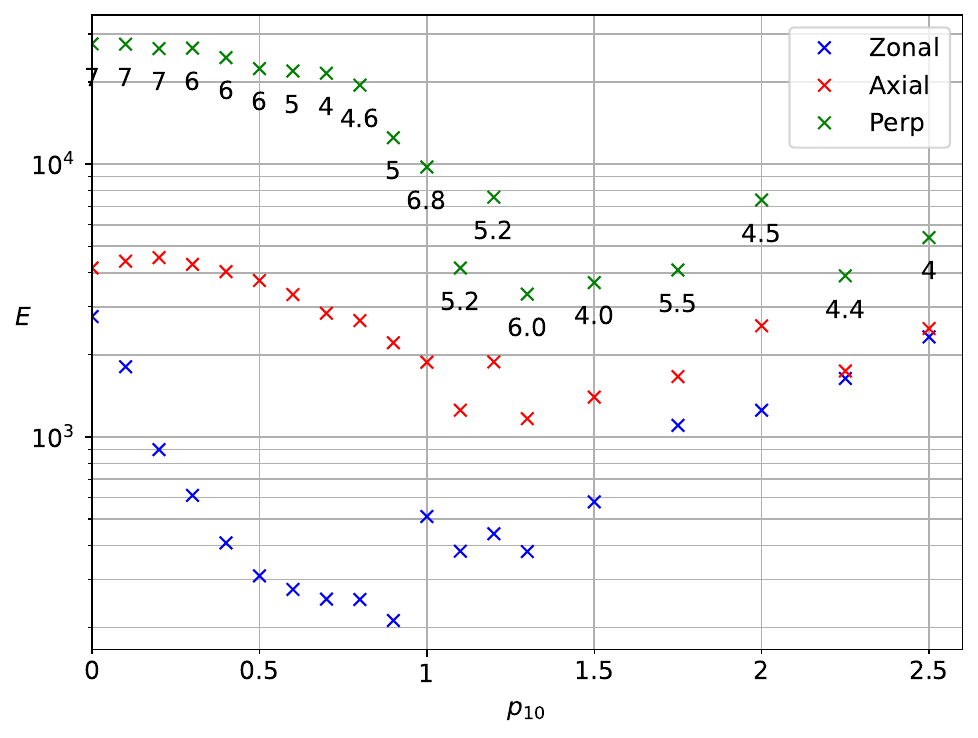}
    \caption{The time-averaged total kinetic energy as $p_{10}$ is varied when $\Ra'=1.75$ and $\Pm=5$; this has been partitioned into three components corresponding with \eqref{eqn:2.kineticenergy}. \firstrevision{Adjacent to the green curve the dominant azimuthal mode, as defined in Figure~\ref{fig:Ecomps_hydro}, is printed.}}
    \label{fig:4.ColumnarEnergySplit_a}
\end{figure}

The relative contributions of the axial, columnar, and zonal components of the flow are shown in Figure~\ref{fig:4.ColumnarEnergySplit_a} for simulations at fixed $\Ra'=1.75$ to highlight the effect of varying the magnetic field strength. As $p_{10}$ increases within the weak-field regime, the dominant azimuthal mode decreases steadily from $m=7$ to $m=4$\firstrevision{, which is broadly as expected from linear magnetoconvection studies, where a transition from geostrophic to magnetostropic modes is expected as the field strength increases \cite{Sakuraba2002}}. The energies of the axial and columnar components decrease by approximately $25\%$, while the zonal flow energy is reduced by nearly an order of magnitude\firstrevision{, consistent with Ferraro's law of isorotation. As the field strengthens, bending of field lines becomes less energetically favourable, so the flow tends toward uniform rotation along each field line; since the no-slip boundaries enforce zero flow there, the isorotation constraint increasingly suppresses zonal flow throughout the bulk as the field grows stronger \citep{Aubert2005}}.

\begin{figure}
    \centering
    \begin{subfigure}{0.244\linewidth}
        \centering
        \includegraphics[width=\linewidth]{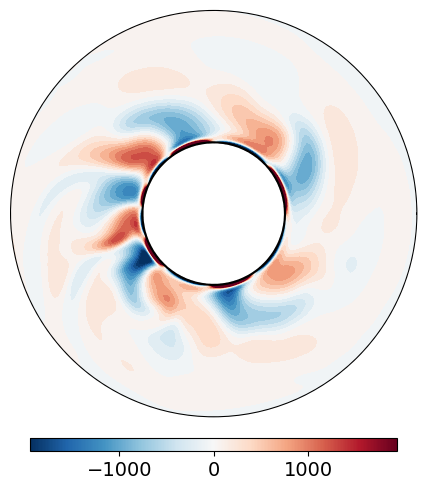}
        \caption{$\vnd{z}\cdot\vec{\omega}$, $p_{10}=1.0$}
        \label{fig:4.vacillating_a}
    \end{subfigure}
    \begin{subfigure}{0.244\linewidth}
        \centering
        \includegraphics[width=\linewidth]{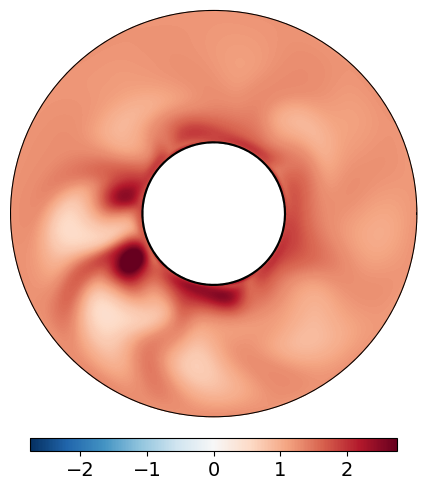}
        \caption{$B_z$, $p_{10}=1.0$}
        \label{fig:4.vacillating_b}
    \end{subfigure}
    \begin{subfigure}{0.244\linewidth}
        \centering
        \includegraphics[width=\linewidth]{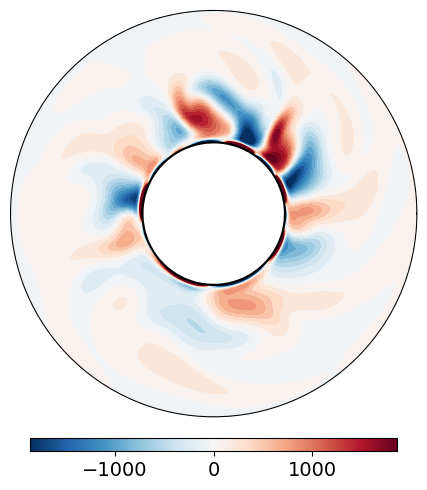}
        \caption{$\vnd{z}\cdot\vec{\omega}$, $p_{10}=1.2$}
        \label{fig:4.vacillating_c}
    \end{subfigure}
    \begin{subfigure}{0.244\linewidth}
        \centering
        \includegraphics[width=\linewidth]{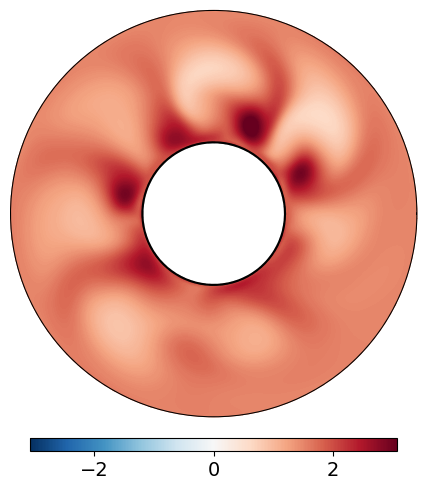}
        \caption{$B_z$, $p_{10}=1.2$}
        \label{fig:4.vacillating_d}
    \end{subfigure}
    \caption{Contours of the axial vorticity and magnetic field strength at the equator at instants in simulations with $p_{10}=1.0,1.2$ with $\Ra'=1.75$ and $\Pm=5$ after a quasi-steady state has been reached. Note that, due to the symmetry of the flow, the vorticity and magnetic field have only a $\vnd{z}$-component at the equator.}
    \label{fig:4.vacillating}
\end{figure}

\paragraph{Vacillating regime.}
For intermediate imposed field strengths ($0.8\lesssim p_{10}\lesssim1.3$), the system enters a qualitatively different regime in which convective columns grow to moderate amplitude before becoming unstable and breaking apart in a quasi-periodic manner. Figure~\ref{fig:4.vacillating} captures a moment in this process for $p_{10}=1.0$ and $1.2$ at $\Ra'=1.75$. Columns initially grow at the horizontal scale of the linear mode and propagate azimuthally around the inner core. As they propagate, axial magnetic field is swept into the negative vorticity regions, leading to magnetic field amplification. Once sufficient magnetic flux is concentrated, the Lorentz force destabilises the column, causing it to break apart into smaller-scale turbulence. This turbulence disperses the magnetic field, and then diffuses itself, allowing the columnar structure to reform and the cycle repeats.

This vacillating behaviour is absent at smaller $p_{10}$, where the magnetic field is too weak to destabilise the columns, and at larger $p_{10}$, where a strong-field state can be maintained. It therefore represents a transitional regime between weak-field VAC dynamics and strong-field MAC dynamics. In the present simulations, quasi-periodic vacillations are observed for $p_{10}=1.0$ over the range $1.7\lesssim\Ra'\lesssim2.0$.

Figures~\ref{fig:4.ForceBalances5a},\ref{fig:4.ForceBalances5b}, and \ref{fig:4.ForceBalances5d} show the force balances at $\Ra'=1.6,\ 1.8,\ 2.5$, corresponding to a level of driving below, within, and above the vacillating regime, respectively. In each case the force balance could be described as MAC, as the Lorentz force is consistently part of the dominant balance across a range of scales; however it is clear that there are subtle distinctions between the three cases. At $\Ra'=1.6$ and $\Ra'=1.8$, the Lorentz force does not enter the dominant balance at the largest scales, entering sharply at $l=8$ for $\Ra'=1.6$, when the columns are steady, and gradually at $l\approx10$, for $\Ra'=1.8$, when the columns are vacillating. %\robnote{Are the force spectra time averaged? If so, in the vacillating case, is it meaningful to look at force spectra that are time averaged over a whole cycle?}\lukenote{Discuss - the vacillations don't (generally) seem to have a regular period. Hopefully the time average period is sufficiently long that it captures all parts of the vacillation.} \robnote{Ok, that's fine. Maybe just keep as is and retain this commentary in case we need to think more about it later}.
The remaining forces are also smoother at $\Ra'=1.8$ than at $\Ra'=1.6$, consistent with the wide range of scales that emerge due to the break-up of the columns. Above the vacillating regime the Lorentz force also enters the dominant balance at smallest $l$, and viscosity and inertia are increasingly subdominant.

Despite the strongly time-dependent nature of the vacillating flow, this regime remains largely equatorially symmetric %\robnote{Time-averaged quantities may mask any significant antisymmetric components if exist during part of the cycle? Likewise for columnarity.}\lukenote{Discuss - This could in principle depend on when the time-averaging is done. I checked for the $\Pm=12$ case and found no significant chance depending on when this was done. The columnarity is not time-averaged since it comes from a single state file.} \robnote{Ok, probably fine to leave as is then. We can study this further if a reviewer asks.} 
(middle panels of Figures~\ref{fig:4.Symmetryu}, \ref{fig:4.SymmetryB}) and the columnarity decreases only gradually with increasing $\Ra'$ (middle panel of Figure~\ref{fig:4.Columnarity}), indicating that the instability is predominantly $z$-independent. The transition into this regime is, however, accompanied by a sharp increase in the toroidal magnetic energy (middle panel of Figure~\ref{fig:4.TorPol}), reflecting enhanced field generation during the growth phase of the columns.

\paragraph{\color{black} Strong-field regime.}
\firstrevision{For low driving but strong imposed fields} ($p_{10}\gtrsim1.3$, corresponding to $\Els\gtrsim2$), columnar convection is replaced by large-scale radial jets similar to those observed at $\Pm=1$ and in previous studies of imposed-field magnetoconvection \citep{Sakuraba2007}. These jets are initially irregular close to the transition but become increasingly coherent and larger in scale as $p_{10}$ is increased. Over this range, the dominant azimuthal mode decreases from $m=6$ to $m=4$, as the flow becomes strongly influenced by Lorentz forces at all but the smallest length scales \firstrevision{(Figure~\ref{fig:4.ColumnarEnergySplit_a})}.

\firstrevision{A strong-field regime, analogous to that seen in dynamo simulations, only emerges at larger convective driving; then,} the zonal flow undergoes a marked increase in amplitude, approaching the strength of the axial and columnar components \firstrevision{(Figure~\ref{fig:4.ColumnarEnergySplit_a})}. The structure of the zonal flow closely resembles that reported by \cite{MasonEtAl2022} for strongly imposed magnetic fields, in which the Lorentz force term in the zonal wind balance becomes dominant and directly drives large-scale zonal motion. Force spectra (Figure~\ref{fig:4.ForceBalances5d}) confirm the emergence of a MAC balance across a wide range of spherical harmonic degrees, with inertia and viscosity confined to the smallest scales.

Equatorial symmetry breaking occurs for all values of $p_{10}$ explored at sufficiently large $\Ra'$ (middle panel of Figure~\ref{fig:4.Symmetryu}). This contrasts with the behaviour at $\Pm=1$ and highlights the increased susceptibility of higher-$\Pm$ flows to symmetry-breaking transitions under magnetic forcing. Equatorial symmetry breaking at $\Pm=5$ also correlates with a slight bump in the poloidal kinetic and magnetic energies (middle panels of Figure~\ref{fig:4.TorPolu},\ref{fig:4.TorPol}, unfilled diamonds) which would be consistent with symmetry-breaking corresponding to the onset of polar convective modes; however it would also be consistent with the development of a large scale meridional circulation. This relation emerges better at $\Pm=12$ (see \S\ref{sec:4.pm12}), where there is greater distinction between the two regimes.

For all values of $p_{10}$ considered at $\Pm=5$, we observe a single continuous branch of magnetoconvective solutions with no evidence of bistability or hysteresis. While the vacillating regime provides a clear example of transitional dynamics between weak- and strong-field behaviour, its reliance on magnetic field concentration at scales smaller than the convection columns suggests that it is unlikely to persist at very small magnetic Prandtl numbers, such as those relevant to planetary cores.

\subsection{Magnetoconvection with \texorpdfstring{$\Pm=12$}{Pm=12}}
\label{sec:4.pm12}

\begin{figure}
    \centering
    \begin{subfigure}{0.49\linewidth}
        \includegraphics[width=\linewidth]{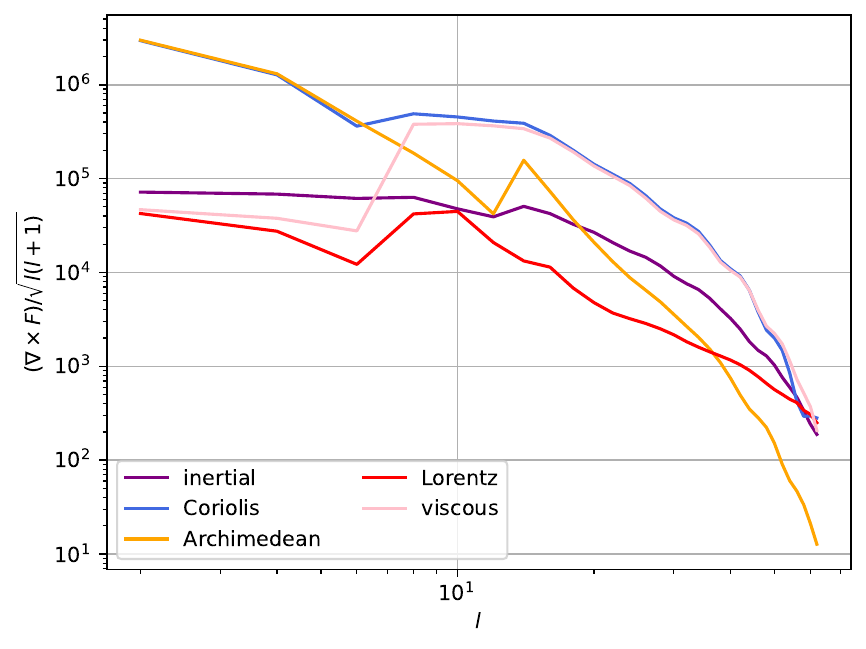}
        \caption{$\Ra'=1.3$, $p_{10}=0.1$ (even modes only).}
        \label{fig:4.ForceBal_Pm12_a}
    \end{subfigure}
    \begin{subfigure}{0.49\linewidth}
        \includegraphics[width=\linewidth]{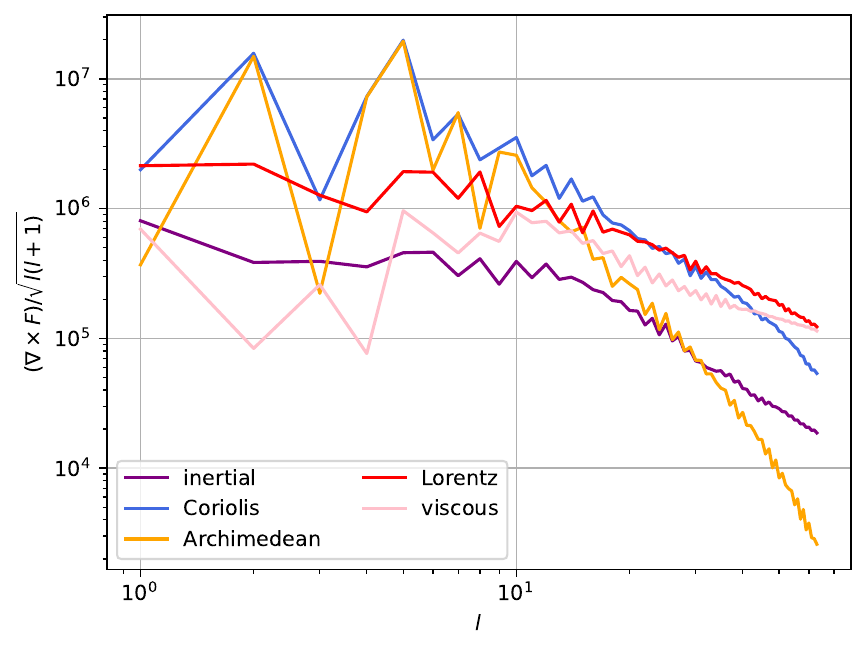}
        \caption{$\Ra'=2.4$, $p_{10}=0.1$ (weak-field, $m=5$ mode).}
        \label{fig:4.ForceBal_Pm12_b}
    \end{subfigure}

    \begin{subfigure}{0.49\linewidth}
        \includegraphics[width=\linewidth]{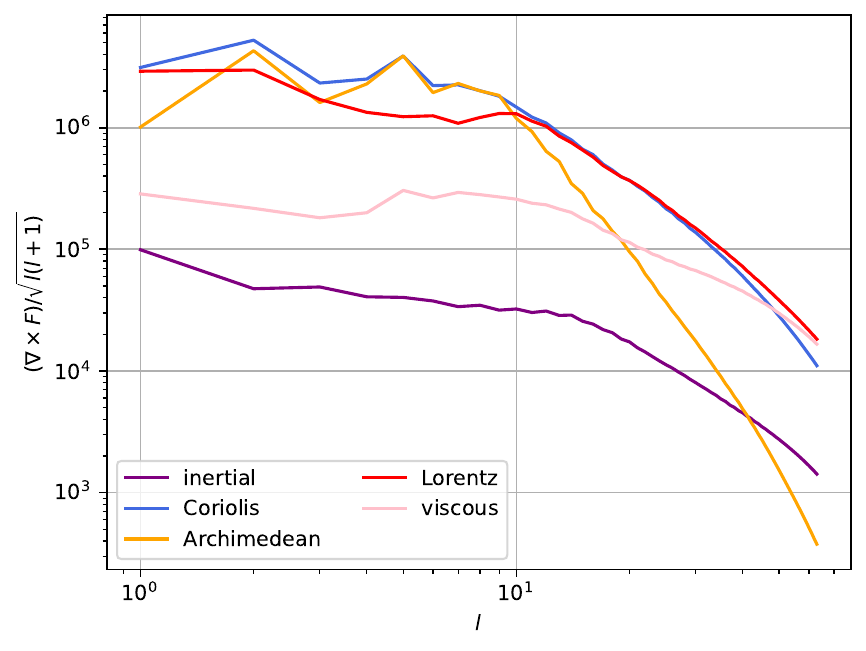}
        \caption{$\Ra'=1.6$, $p_{10}=1.0$.}
        \label{fig:4.ForceBal_Pm12_c}
    \end{subfigure}
    \begin{subfigure}{0.49\linewidth}
        \includegraphics[width=\linewidth]{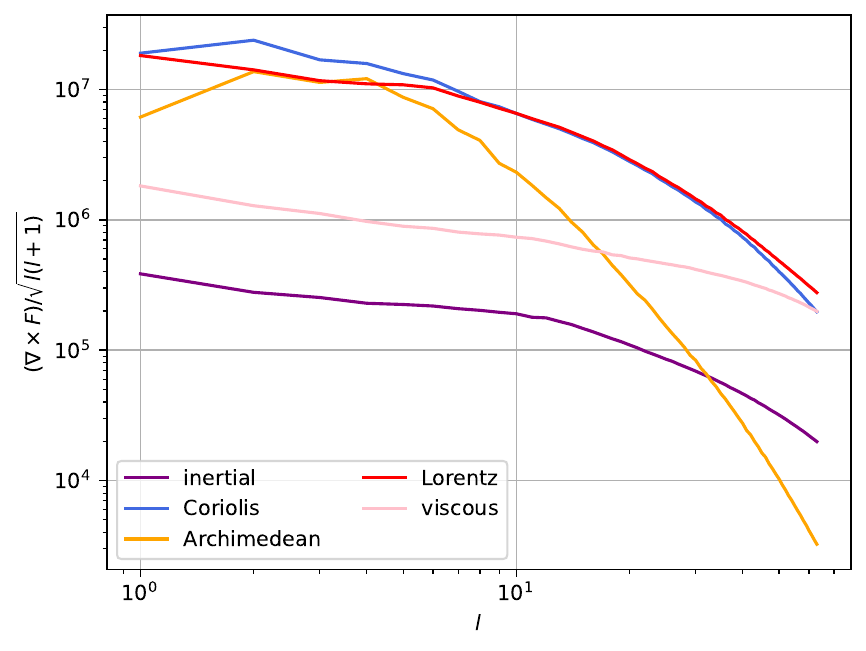}
        \caption{$\Ra'=2.0$, $p_{10}=1.0$.}
        \label{fig:4.ForceBal_Pm12_d}
    \end{subfigure}
    \begin{subfigure}{0.49\linewidth}
        \includegraphics[width=\linewidth]{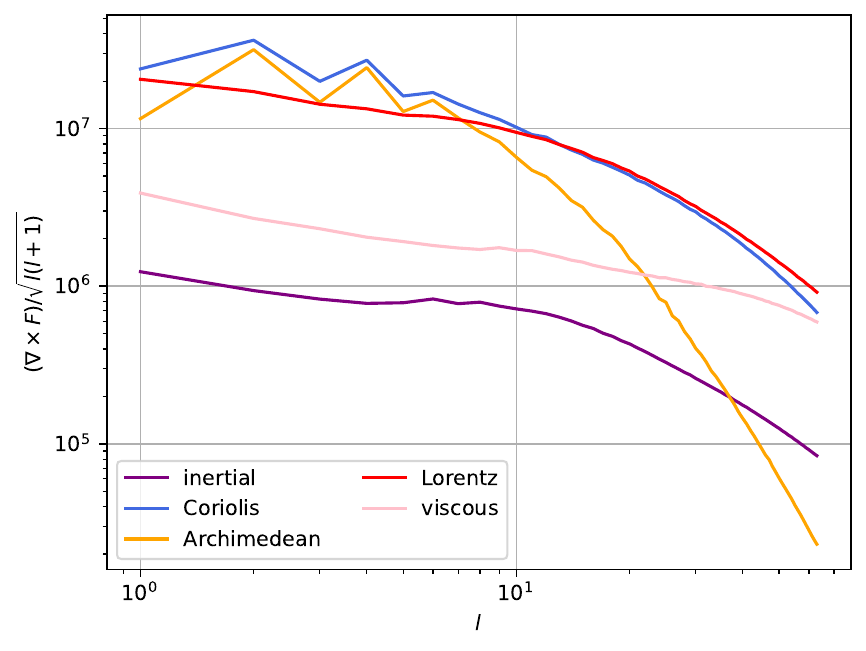}
        \caption{$\Ra'=2.4$, $p_{10}=0.1$ (strong-field).}
        \label{fig:4.ForceBal_Pm12_e}
    \end{subfigure}
    \begin{subfigure}{0.49\linewidth}
        \includegraphics[width=\linewidth]{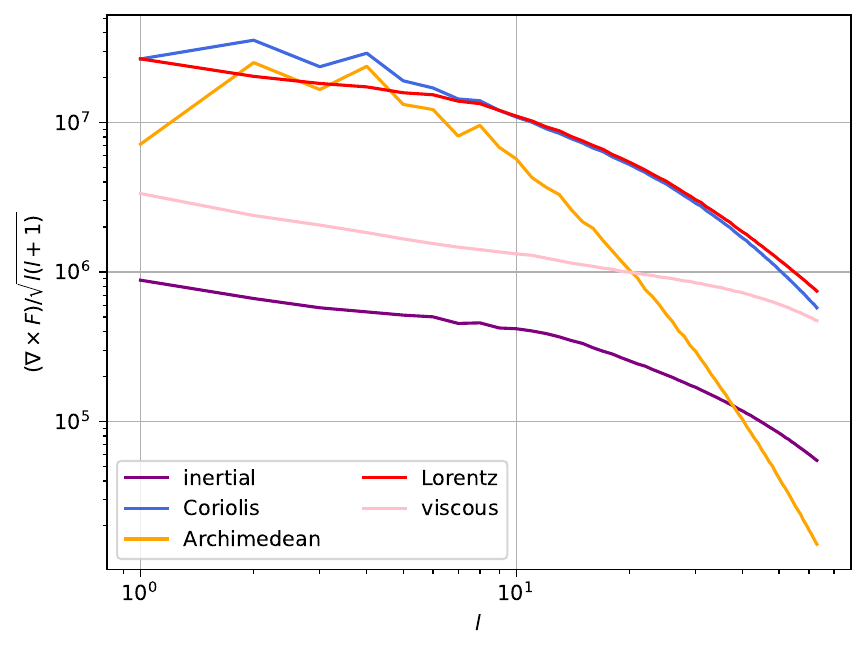}
        \caption{$\Ra'=2.4$, $p_{10}=1.0$.}
        \label{fig:4.ForceBal_Pm12_f}
    \end{subfigure}
    \caption{Force spectra as in Figure~\ref{fig:4.ForcesPm1} at $\Pm=12$. Top row: weak-field solutions. Middle: vacillating solutions. Bottom: Strong-field solutions.}
    \label{fig:4.ForcesPm12}
\end{figure}

At $\Pm=12$, magnetoconvection exhibits the same broad sequence of regimes as at $\Pm=5$, but with several key differences that stem from the reduced magnetic diffusivity. In particular, \firstrevision{we find} sharper distinctions between weak-field, transitional, and strong-field behaviour. This parameter regime is of particular interest because dynamo simulations at comparable $\Pm$ exhibit a wide region of bistability between weak- and strong-field branches\firstrevision{; we observe it in these magnetoconvection simulations too, but only when the imposed magnetic field is weak. This section is structured by considering each regime in turn, referring to the global diagnostics in panels of the right column of Figure~\ref{fig:4.Diagnostics}, and the selection of force spectra in Figure~\ref{fig:4.ForcesPm12}.}

\begin{figure}
    \centering
    \begin{subfigure}{0.32\linewidth}
        \centering
        \includegraphics[width=\linewidth]{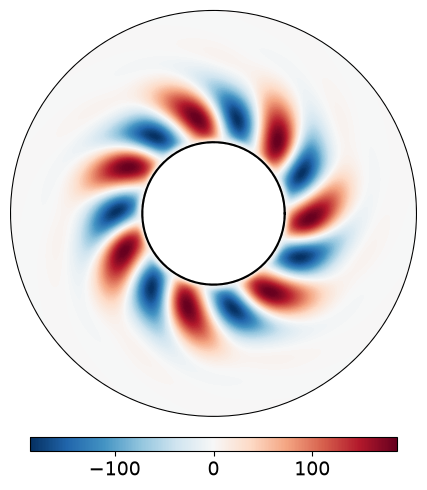}
        \caption{}
        \label{fig:4.columnarzsect1}
    \end{subfigure}
    \begin{subfigure}{0.32\linewidth}
        \centering
        \includegraphics[width=\linewidth]{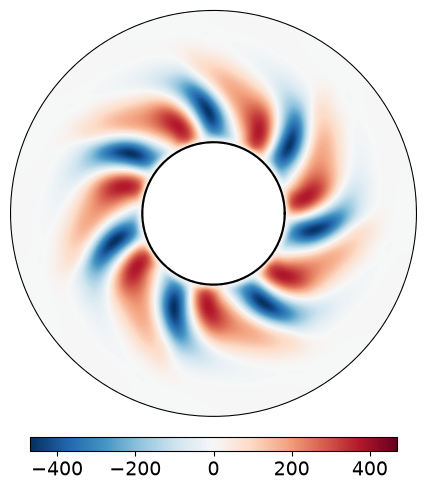}
        \caption{}
        \label{fig:4.columnarzsect2}
    \end{subfigure}
    \begin{subfigure}{0.32\linewidth}
        \centering
        \includegraphics[width=\linewidth]{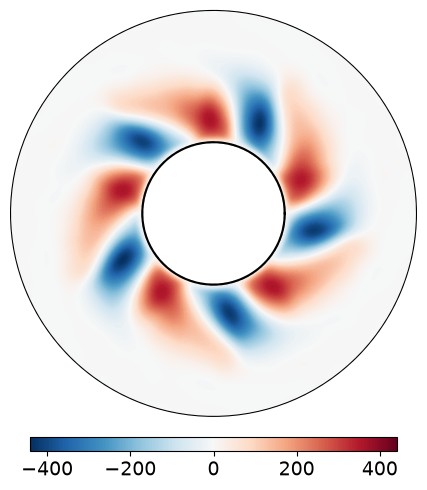}
        \caption{}
        \label{fig:4.columnarzsect3}
    \end{subfigure}
    \begin{subfigure}{0.32\linewidth}
        \centering
        \includegraphics[width=\linewidth]{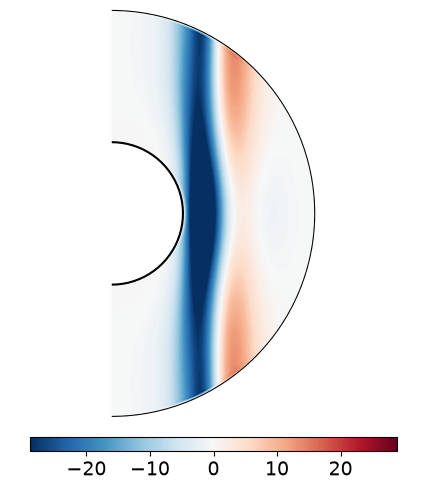}
        \caption{}
        \label{fig:4.columnarmersect1}
    \end{subfigure}
    \begin{subfigure}{0.32\linewidth}
        \centering
        \includegraphics[width=\linewidth]{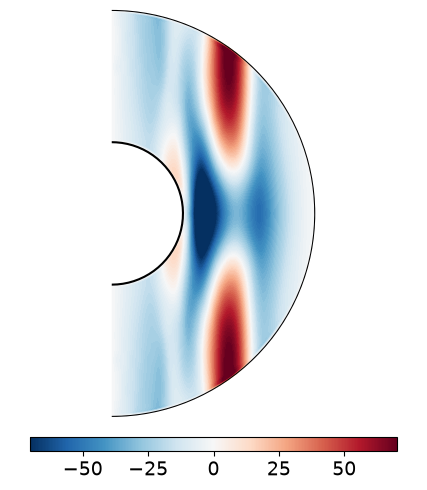}
        \caption{}
        \label{fig:4.columnarmersect2}
    \end{subfigure}
    \begin{subfigure}{0.32\linewidth}
        \centering
        \includegraphics[width=\linewidth]{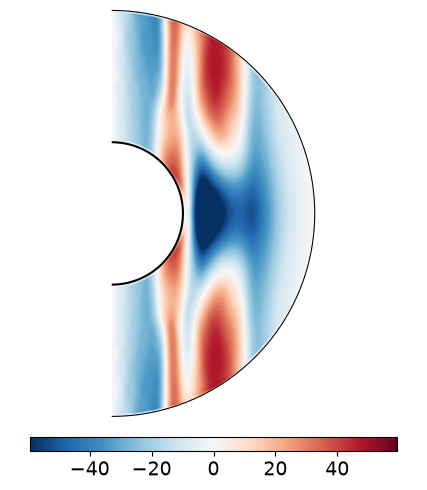}
        \caption{}
        \label{fig:4.columnarmersect3}
    \end{subfigure}
    \caption{Equatorial sections of the radial velocity, $\vnd{r}\cdot\vec{u}$ (a-c), and zonally averaged plots of the zonal velocity, $\vnd{\phi}\cdot\vec{u}$ (d-f), at a quasi-steady instant in two simulations in the columnar regime of convection, close to onset, with $p_{10}=0.1$ and $\Pm=12$. (a,d): $\Ra'=1.3$. (b,e): $\Ra'=2.4$ (pre-mode change). (c,f): $\Ra'=2.4$ (post-mode change).} 
    \label{fig:4.columnar}
\end{figure}

\begin{figure}
    \centering
    \begin{subfigure}{0.244\linewidth}
        \centering
        \includegraphics[width=\linewidth,trim=2.0cm 0cm 0.3cm 0cm, clip]{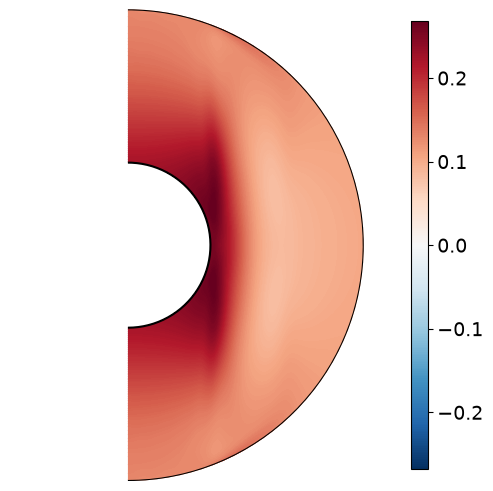}
        \caption{}
        \label{fig:4.MerField_1_Bz}
    \end{subfigure}
    \begin{subfigure}{0.244\linewidth}
        \centering
        \includegraphics[width=\linewidth,trim=2.0cm 0cm 0.3cm 0cm, clip]{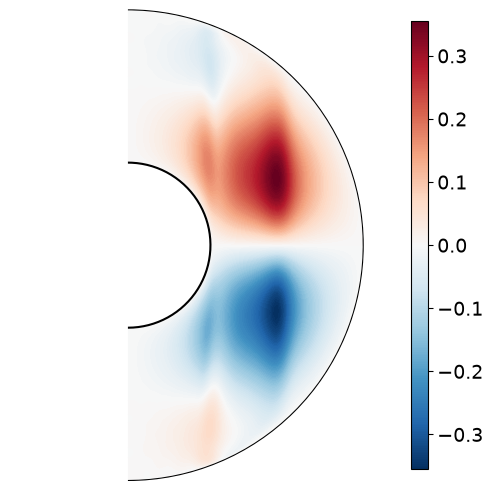}
        \caption{}
        \label{fig:4.MerField_1_Bphi}
    \end{subfigure}
    \begin{subfigure}{0.244\linewidth}
        \centering
        \includegraphics[width=\linewidth,trim=2.0cm 0cm 0.3cm 0cm, clip]{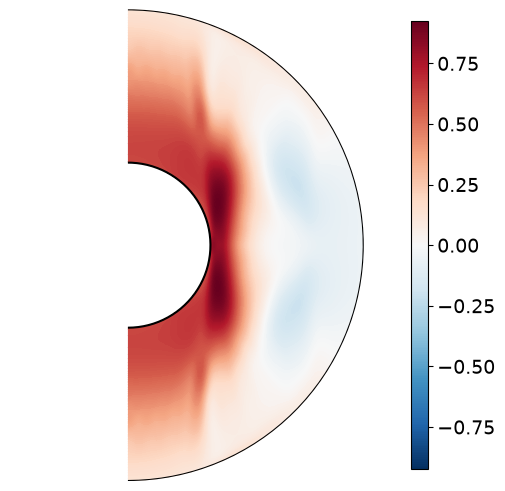}
        \caption{}
        \label{fig:4.MerField_4_Bz}
    \end{subfigure}
    \begin{subfigure}{0.244\linewidth}
        \centering
        \includegraphics[width=\linewidth,trim=2.0cm 0cm 0.3cm 0cm, clip]{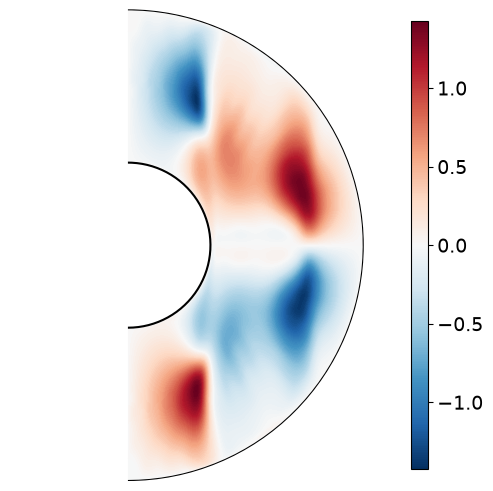}
        \caption{}
        \label{fig:4.MerField_4_Bphi}
    \end{subfigure}
    %\begin{subfigure}{0.244\linewidth}
    %    \centering
    %    \includegraphics[trim=1.8cm 0cm 0cm 0cm,clip,width=\linewidth]{Figures/mcv_OD0.1_E1e-04_2.428Rac_Pm12_WD_meridsect_B_z_phiav_m5.pdf}
    %    \caption{}
    %    \label{fig:4.MerField_3_Bz}
    %\end{subfigure}
    %\begin{subfigure}{0.244\linewidth}
    %    \centering
    %    \includegraphics[trim=1.8cm 0cm 0cm 0cm,clip,width=\linewidth]{Figures/mcv_OD0.1_E1e-04_2.428Rac_Pm12_WD_meridsect_B_phi_phiav_m5.pdf}
    %    \caption{}
    %    \label{fig:4.MerField_3_Bphi}
    %\end{subfigure}
    \caption{Meridional sections of the zonally averaged vertical, $\vnd{z}$ (a,c), and zonal, $\vnd{\phi}$ (b,d), components of the magnetic field in the weak-field regime. (a,b) $\Ra'=1.3$, $\Pm=12$, and $p_{10}=0.1$ and (c,d) $\Ra'=2.4$, $\Pm=12$, and $p_{10}=0.1$ (pre-mode change).}
    \label{fig:4.MerField}
\end{figure}

\begin{figure}
    \centering
    \includegraphics[width=0.75\linewidth]{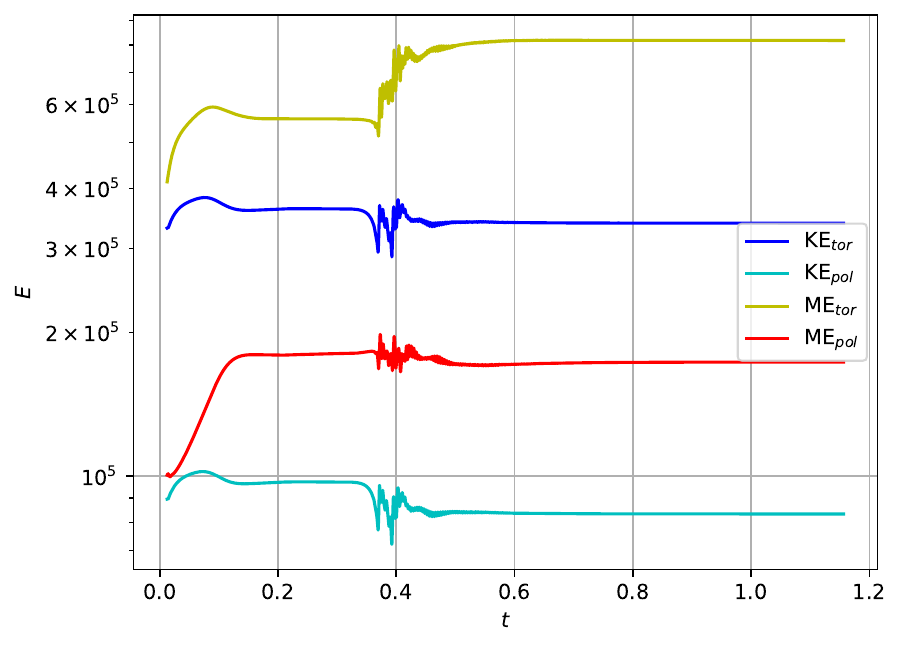}
    \caption{Kinetic and magnetic energies\firstrevision{, split into poloidal and toroidal components,} over time, for the simulation with $p_{10}=0.1$, $\Ra'=2.4$, \firstrevision{$\Pm=12$}, corresponding to the centre and right plots of Figure \ref{fig:4.columnar}.}
    \label{fig:4.ModeShiftEnergies}
\end{figure}

\begin{figure}
    \centering
    \includegraphics[width=0.7\linewidth]{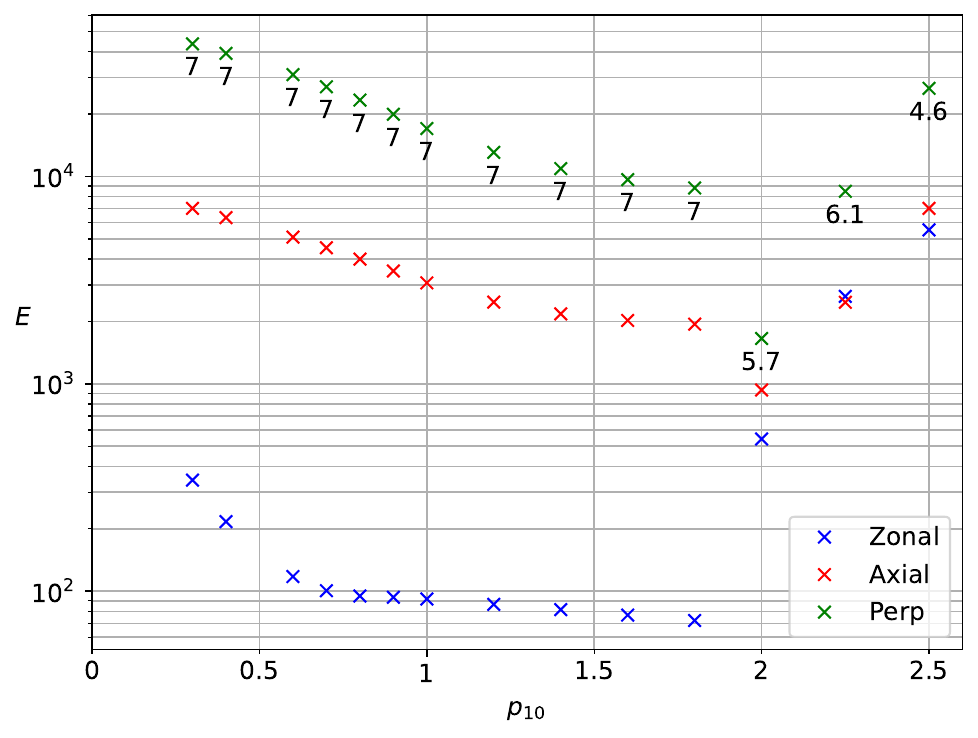}
    \caption{The time-averaged total kinetic energy as $p_{10}$ is varied when $\Ra'=1.3$ and $\Pm=12$, this has been partitioned into three components corresponding with \eqref{eqn:2.kineticenergy}. \firstrevision{Above each trio of points is the dominant azimuthal mode, as defined in Figure~\ref{fig:Ecomps_hydro}.}}
    \label{fig:4.ColumnarEnergySplit_b}
\end{figure}

\paragraph{Weak-field regime.}
We first consider weak imposed field simulations with $p_{10}=0.1$. Figure~\ref{fig:4.columnarzsect1} shows a representative case at weakly supercritical Rayleigh number, for which the flow structure is a small perturbation of the hydrodynamic solution; this is consistent with the respective force spectra (Figure~\ref{fig:4.ForceBal_Pm12_a}) which show that the Lorentz force remains subdominant to \firstrevision{the other four solenoidal forces at all except the highest degrees}. The poloidal magnetic field is largely axial and closely resembles the imposed background field, while an induced toroidal field is generated, predominantly outside the tangent cylinder (Figures~\ref{fig:4.MerField_1_Bz}, \ref{fig:4.MerField_1_Bphi}).

As the Rayleigh number is increased, the columnar convection strengthens and expands radially, while the development of an equatorial thermal anomaly modifies the zonal flow profile (Figures~\ref{fig:4.columnarzsect2}, \ref{fig:4.columnarmersect2}), in a manner analogous to lower $\Pm$ cases. At the same time, the mean poloidal magnetic field transitions from an axial to a dipolar configuration (Figure~\ref{fig:4.MerField_4_Bz}), and the induced toroidal field extends into the tangent cylinder (Figure~\ref{fig:4.MerField_4_Bphi}). This state becomes marginally unstable at $\Ra'\approx2.4$ and persists for a \firstrevision{tenth} of a magnetic diffusion time before reorganising into an $m=5$ convective mode \firstrevision{(Figures~\ref{fig:4.columnarzsect3}, \ref{fig:4.columnarmersect3}). T}his mode transition is accompanied by a substantial increase in the strength of the toroidal magnetic field \firstrevision{whilst the remaining energy components remain approximately constant (Figure~\ref{fig:4.ModeShiftEnergies}).} This increase can be interpreted as arising from an effective increase in the magnetic Reynolds number, $\Rm=U\ell/\eta$, driven primarily by the increase in the dominant horizontal length scale $\ell$, while the overall structure of the flow remains largely unchanged. It is plausible that a sequence of similar transitions, progressively increasing the convective length scale and magnetic field strength, could ultimately lead to the collapse of the weak-field branch. Notably, the instability at $\Ra'\approx2.4$ occurs just above the collapse of the weak-field branch in dynamo simulations at comparable parameters \citep[$\Ra'\approx2.38$, ][]{TeedDormy2025b}, suggesting that the large-scale field is not solely responsible for inciting the collapse of the weak-field branch, and that the alternative boundary condition restriction in our magnetoconvection models that poloidal modes with $(l,m)>(1,0)$ must vanish at the outer boundary can delay, but not suppress, \firstrevision{transition to the strong-field branch.}

\firstrevision{At lower Rayleigh numbers, e.g.~$\Ra'=1.3$ (Figure~\ref{fig:4.ColumnarEnergySplit_b}), increasing $p_{10}$ whilst not increasing the convective driving leads to the onset of magnetostrophic modes in a similar manner to that observed at $\Pm=1,5$. This clear transition, at $p_{10}=2.0$, is accompanied by a sharp increase in the zonal flow, a product of the magnetic wind, which otherwise decreases with $p_{10}$ in the thermal Rossby wave regime ($p_{10}<2.0$).}

Despite the apparently simple flow structure in this regime, the force spectra (Figure \ref{fig:4.ForceBal_Pm12_b}) reveal a complex balance: all five forces contribute at comparable orders across much of the spectrum, though inertia is consistently weak. The key distinction between this weak-field state and strong-field states is that the viscous force continues to compete with the Lorentz force at several length scales, preventing the establishment of a clear MAC balance. This hierarchy, where both Lorentz and viscous forces are important across a range of scales, is reminiscent of a `VMAC' balance observed in weak-field non-dipolar dynamo simulations \citep{TeedDormy2025a}.

\begin{figure}
    \centering
    \begin{subfigure}{0.32\linewidth}
        \centering
        \includegraphics[width=\linewidth]{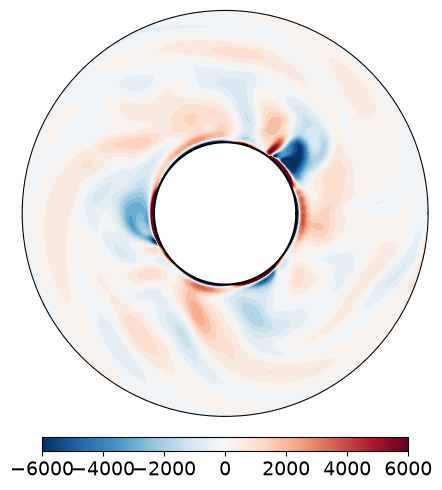}
        \caption{}
        \label{fig:4.vacill12zsect1}
    \end{subfigure}
    \begin{subfigure}{0.32\linewidth}
        \centering
        \includegraphics[width=\linewidth]{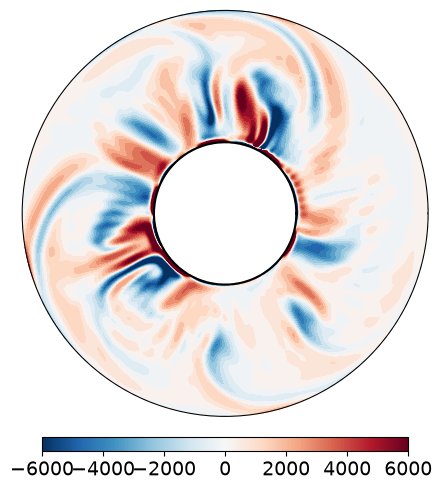}
        \caption{}
        \label{fig:4.vacill12zsect2}
    \end{subfigure}
    \begin{subfigure}{0.32\linewidth}
        \centering
        \includegraphics[width=\linewidth]{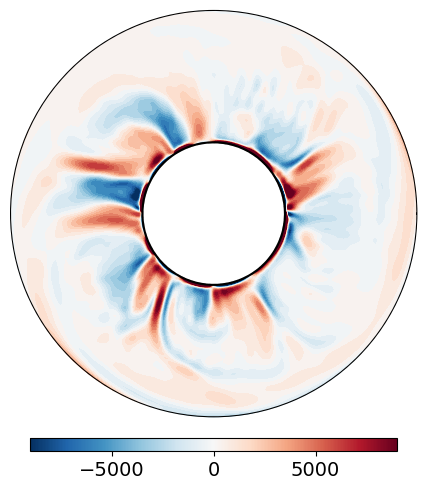}
        \caption{}
        \label{fig:4.vacill12zsect3}
    \end{subfigure}
    \begin{subfigure}{0.32\linewidth}
        \centering
        \includegraphics[width=\linewidth]{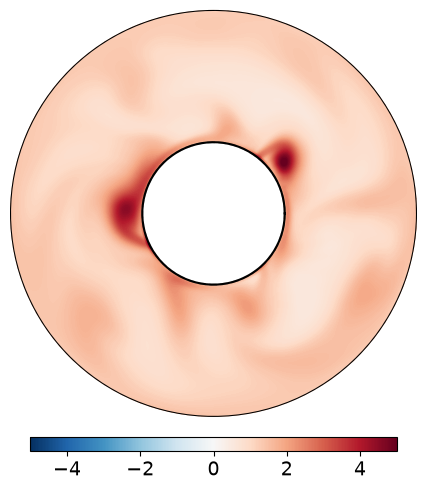}
        \caption{}
        \label{fig:4.vacill12zsect4}
    \end{subfigure}
    \begin{subfigure}{0.32\linewidth}
        \centering
        \includegraphics[width=\linewidth]{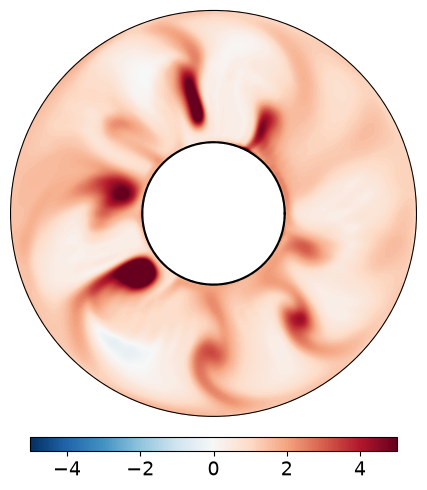}
        \caption{}
        \label{fig:4.vacill12zsect5}
    \end{subfigure}
    \begin{subfigure}{0.32\linewidth}
        \centering
        \includegraphics[width=\linewidth]{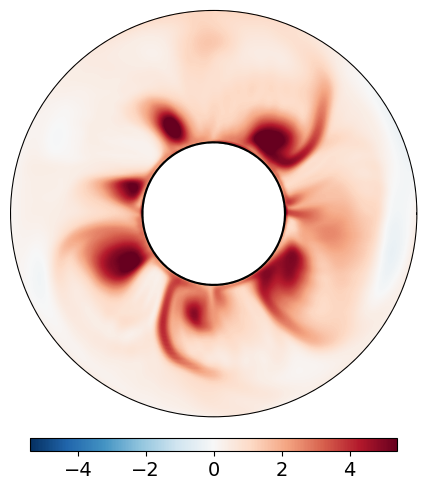}
        \caption{}
        \label{fig:4.vacill12zsect6}
    \end{subfigure}
    \caption{Equatorial sections of the axial vorticity, $\vnd{z}\cdot\vec{\omega}$ (a-c), and the axial magnetic field, $\vnd{z}\cdot\vec{B}$ (d-f), at a quasi-steady instant in two simulations in the vacillating regime, with $p_{10}=1.0$. \firstrevision{(a,d): $\Ra'=1.77$, $t=0.455$; (b,e): $\Ra'=1.77$, $t=0.469$; (c,f): $\Ra'=2.0$.}} 
    \label{fig:4.vacillating12}
\end{figure}

\begin{figure}
    \centering
    \begin{subfigure}[t]{0.48\linewidth}
        \centering
        \includegraphics[width=\linewidth]{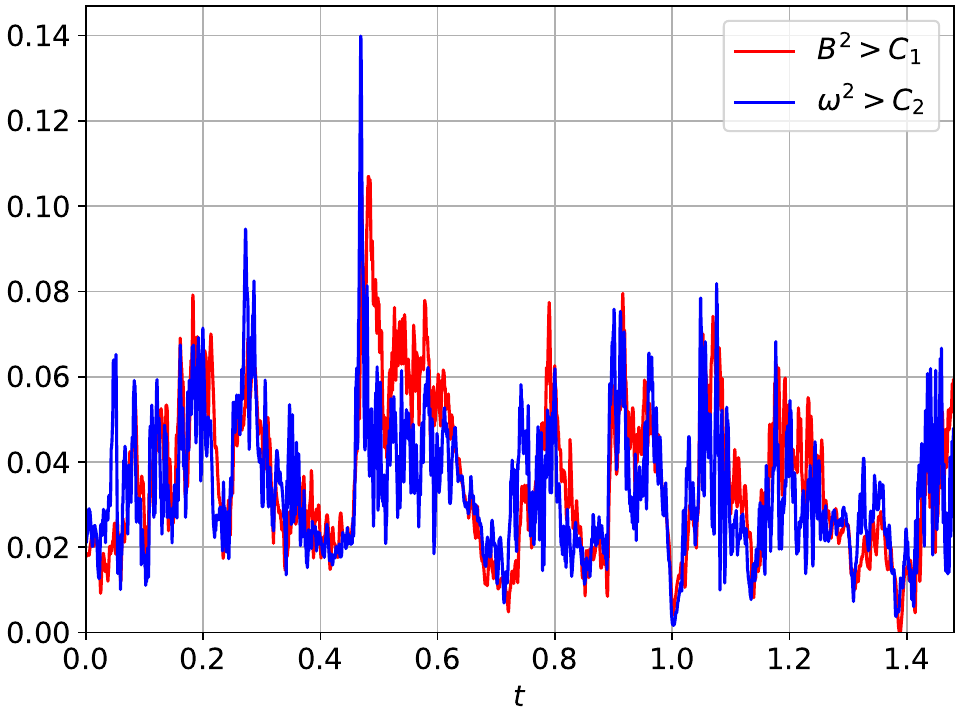}
        \caption{\firstrevision{Proportion over time of the equatorial plane (excluding the boundary layer) in which the square of the axial components of the vorticity and magnetic field are greater than a critical threshold, here $C_1=6$, $C_2=5\times10^{6}$.}}
        \label{fig:4.vacill_over_t_1}
    \end{subfigure}
    ~
    \begin{subfigure}[t]{0.48\linewidth}
        \centering
        \includegraphics[width=\linewidth]{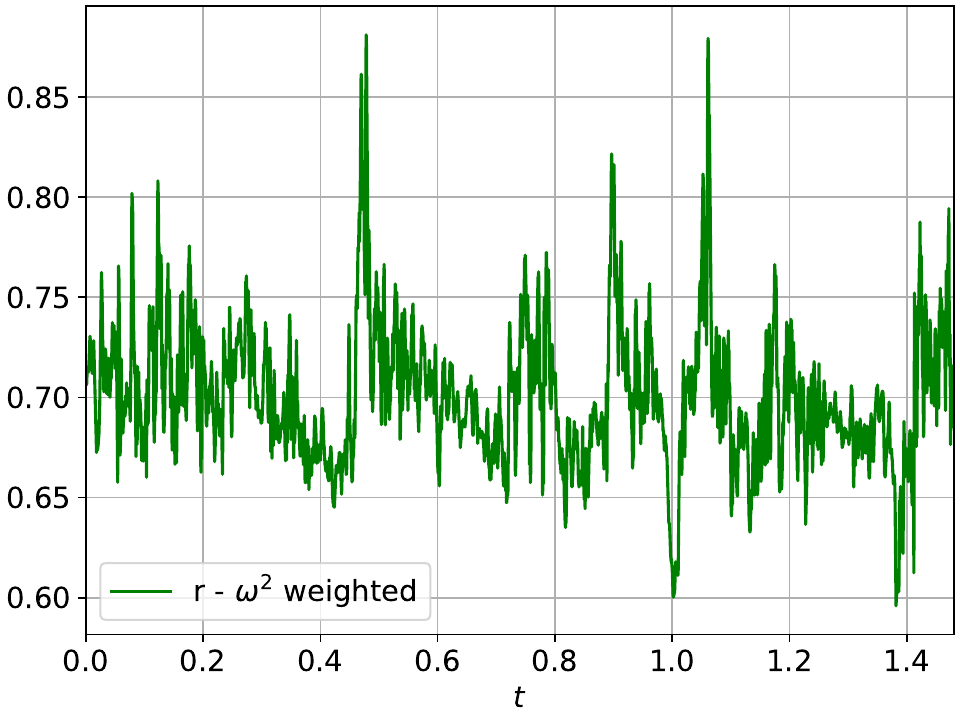}
        \caption{\firstrevision{The mean radius, weighted by the enstrophy in the equatorial plane excluding the boundary layer, defined in \eqref{eqn:4.r_enstrophy_def}.}}
        \label{fig:4.vacill_over_t_2}
    \end{subfigure}
    \begin{subfigure}{0.9\linewidth}
        \centering
        \includegraphics[width=\linewidth]{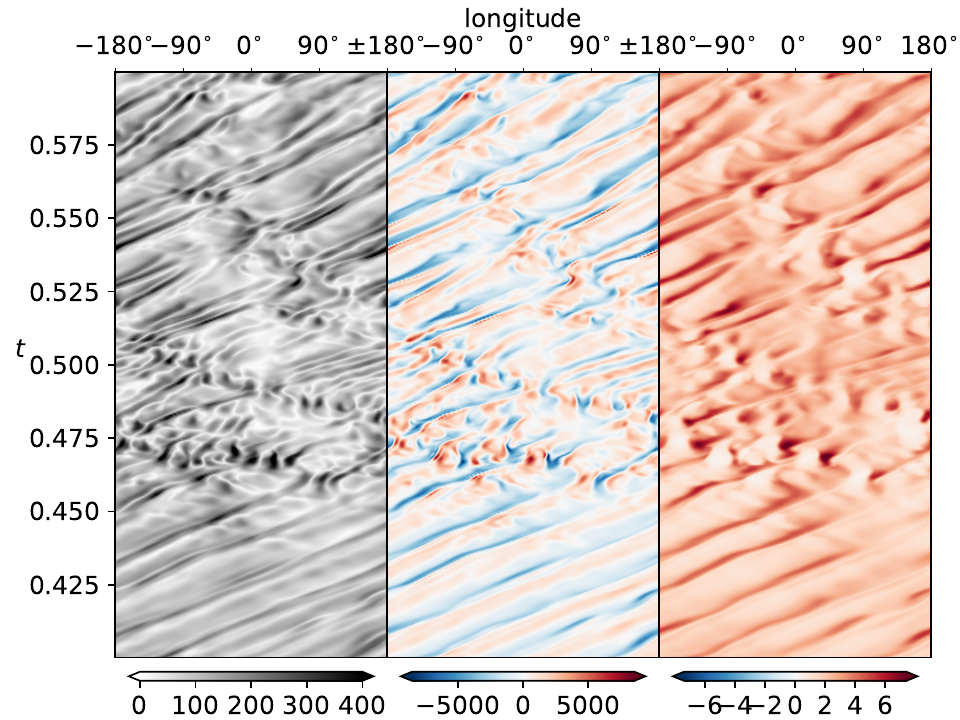}
        \caption{\firstrevision{A spatio-temporal plot in $(t,\phi)$-space for (left-to-right) the horizontal velocity, $u_h=(u_{r}^{2}+u_{\phi}^{2})^{1/2}$, the axial vorticity, $\vnd{z}\cdot\vec{\omega}$, and the axial magnetic field, $B_{z}$, at $r=0.675$ over a temporal range including the largest eruption.}}
        \label{fig:4.vacill_over_t_3}
    \end{subfigure}
    \caption{\firstrevision{Temporal plots for $p_{10}=1.0$, $\Ra'=1.77$, $\Pm=12$, a simulation in the vacillating regime. Compare with Figure~\ref{fig:4.vacill12zsect1},\ref{fig:4.vacill12zsect2},\ref{fig:4.vacill12zsect4},\ref{fig:4.vacill12zsect5} which are snapshots from the same simulation.}}
    \label{fig:4.Vacill_over_t}
\end{figure}

\paragraph{Vacillating regime.}
At intermediate imposed field strengths, most clearly illustrated at $p_{10}=1.0$, the weak-field regime transitions through a vacillating state before reaching strong-field behaviour. \firstrevision{The vacillating regime is characterised by a loss of the long-term stability of individual convection columns and an increase in the strength of toroidal magnetic field.} Figure~\ref{fig:4.vacillating12} \firstrevision{shows several snapshots of the axial vorticity and magnetic field which illustrate} this regime, showing the close correspondence between regions of intense negative axial vorticity and concentrations of strong axial magnetic field \firstrevision{in the equatorial plane as Ekman suction draws fluid into columns with negative vorticity and then away from the equator. At an arbitrary instant in time, this structure is similar to the vacillating regime at $\Pm=5$, although weaker magnetic diffusion allows for the development of a stronger induced magnetic field and finer filaments (cf.~Figure \ref{fig:4.vacillating}).}

\firstrevision{A video of the vacillating regime corresponding to the full simulation represented in Figures~\ref{fig:4.vacill12zsect1}, \ref{fig:4.vacill12zsect2}, \ref{fig:4.vacill12zsect4}, \ref{fig:4.vacill12zsect5} is included in the supplementary material and shows the quasi-periodic oscillations of columns around the inner core and their instability. As columns break apart, small-scale, highly vortical structures are formed, which can be identified in a time series by plotting the proportion of the equatorial plane (or the entire domain) in which the enstrophy density, $\left|\vec{\omega}\right|^2$, exceeds a particular threshold value.
%This is particularly effective when boundary layers are excluded from the calculation and the threshold value is large enough so as to generally exclude normal convection columns. 
This quantity corresponds to the blue curve on Figure~\ref{fig:4.vacill_over_t_1}, where it is also shown that this correlates with the magnetic energy density (red curve), calculated in a similar fashion. The vorticity and magnetic field are parallel to the axial vector at the equator due the equatorial symmetry of the state, thus the degree of correlation in Figure~\ref{fig:4.vacill_over_t_1} is unsurprising given the earlier discussion regarding the close correlation of axial vorticity and axial field. However, it is notable that there is a short delay between the peaks of the enstrophy density and magnetic field.}

\firstrevision{Figure~\ref{fig:4.vacill_over_t_3} shows a spatio-temporal plot for a time period including the largest enstrophy peak for this simulation; three quantities are plotted on a circle in the equatorial plane over time. These are the speed of the flow (left) and the axial magnetic field (right), as in \cite{Sakuraba2007} which investigates similar, but seemingly unrelated, vacillating structures, but also includes the axial vorticity (centre), motivated by the discussions above. We can see that there are long periods in which the equatorial plane is dominated by convection columns rotating prograde, interspersed with sporadic shorter periods of disruption which propagate irregularly counter to the direction of rotation of the columns themselves.

The situation is somewhat analogous to a predator-prey system between the (axial) magnetic field and the convection columns, with the caveat that the columns break apart on significantly shorter time scales than would be expected in a typical Lotka-Volterra system. These instances of collapse of the convection columns are further examined in Figure~\ref{fig:4.vacill_over_t_2} using the enstrophy-weighted radius,
\begin{equation}
    r_{\omega^2}=\frac{\int_{A}r|\vec{\omega}|^2~\dx{A}}{\int_{A}|\vec{\omega}|^2~\dx{A}},\label{eqn:4.r_enstrophy_def}
\end{equation}
where $A$ is the equatorial plane, excluding a layer close to the boundary. This parameter is used to estimate the radius of the most vigorous parts of the flow. The quasi-periodic instability of columns coincides with sharp increases in the weighted radius, indicating that these destruction events are ``eruptions'' away from the inner core. Figures~\ref{fig:4.vacill12zsect2} and \ref{fig:4.vacill12zsect5} correspond to an instant ($t=0.469$) during one of these eruptions, which can be compared with Figure~\ref{fig:4.Vacill_over_t}. The spherical geometry and in particular the presence of the inner core enforce this feature of the disruption as the critical radius of convection is, in this case, close to the inner core. This could differ in simulations with internal heating rather than differential, or with different core sizes, where the critical radius can vary \cite{SeeleyEtAl2025}.

Following each eruption, activity slowly decays back towards the inner core. Convection is only weakly supercritical, particularly farther from the inner core; while convection can be supported by the magnetic field, a true strong-field state is inaccessible and there are regions where the field is weak. This results in the overall decay of small-scale vortices, more rapidly at larger radii; these can also combine, forming larger convection columns, and eventually structures on the typical convective scale can reform, restarting the process.}

\firstrevision{At smaller Rayleigh numbers (such as $\Ra'=1.77$ for the case discussed above), regions of condensed magnetic field form and disperse regularly as the amplified field destabilises the convection columns. However,} as $\Ra'$ increases, the magnetic field amplification becomes sufficiently strong to modify the convective structures themselves, producing \firstrevision{vacillating regimes with} longer-lived \firstrevision{and less columnar structures as shown in Figures~\ref{fig:4.vacill12zsect3} and \ref{fig:4.vacill12zsect6} for $\Ra'=2$.} As $\Ra'$ increases further, the vacillating regime transitions smoothly to the strong-field regime \firstrevision{as the flow becomes more vigorous, allowing for longer excursions into the strong-field-like regime}. Distinguishing these two regimes on the basis of force spectra alone is difficult between $\Ra'\approx2.0$ and $2.4$\firstrevision{, in part due to the time-averaging of this data, which includes times where the simulation approaches, but cannot sustain, a strong-field-like state; notably, however}, equatorial symmetry is broken \firstrevision{during this transition (right panel of Figure \ref{fig:4.Symmetryu})}.

\begin{figure}
    \centering
    \begin{subfigure}{0.275\linewidth}
        \centering
        \includegraphics[width=\linewidth]{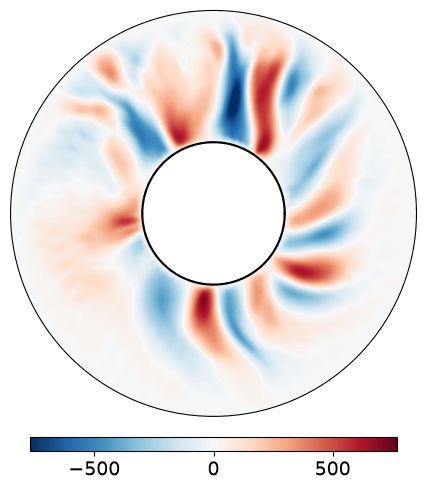}
        \caption{}
        \label{fig:4.strong_2.649_a}
    \end{subfigure}
    \begin{subfigure}{0.275\linewidth}
        \centering
        \includegraphics[width=\linewidth]{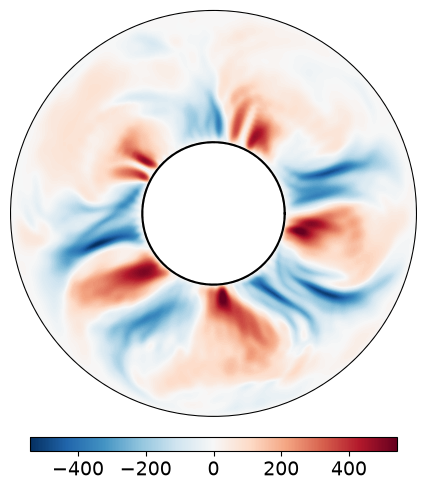}
        \caption{}
        \label{fig:4.strong_2.649_b}
    \end{subfigure}
    \begin{subfigure}{0.21\linewidth}
        \centering
        \includegraphics[width=\linewidth,trim=2.0cm -2.4cm 0.3cm 0cm, clip]{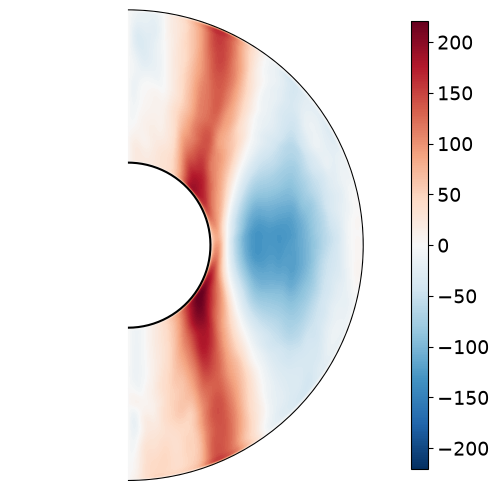}
        \caption{}
        \label{fig:4.strong_2.649_c}
    \end{subfigure}
    \begin{subfigure}{0.21\linewidth}
        \centering
        \includegraphics[width=\linewidth,trim=2.0cm -2.4cm 0.3cm 0cm, clip]{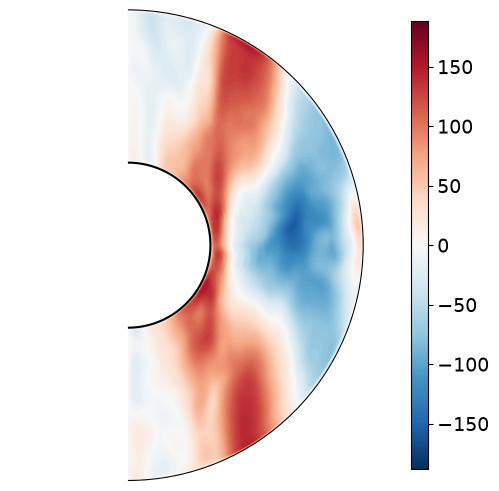}
        \caption{}
        \label{fig:4.strong_2.649_d}
    \end{subfigure}
    \caption{Equatorial sections of the radial velocity, $\vnd{r}\cdot\vec{u}$ (a,b), and zonally averaged plots of the zonal velocity, $\vnd{\phi}\cdot\vec{u}$ (c,d), at a quasi-steady instant in two simulations in the strong (MAC) regime, with \firstrevision{$\Pm=12$}, $\Ra'=2.65$. (a,c): $p_{10}=0.1$. (b,d): $p_{10}=1.0$.}
    \label{fig:4.strong_2.649}
\end{figure}

% \begin{figure}
%     \centering
%         \begin{subfigure}{0.275\linewidth}
%         \centering
%         \includegraphics[width=\linewidth]{Figures/mcv_OD0.1_E1e-04_4.0Rac_Pm12_01_0001_eqsect_u_r.png}
%         \caption{}
%         \label{fig:4.strong_4.0_a}
%     \end{subfigure}
%     \centering
%         \begin{subfigure}{0.275\linewidth}
%         \centering
%         \includegraphics[width=\linewidth]{Figures/mcv_OD1.0_E1e-04_4.0Rac_Pm12_02_0004_eqsect_u_r.png}
%         \caption{}
%         \label{fig:4.strong_4.0_b}
%     \end{subfigure}
%     \centering
%         \begin{subfigure}{0.21\linewidth}
%         \centering
%         \includegraphics[width=\linewidth,trim=2.0cm -2.4cm 0.3cm 0cm, clip]{Figures/mcv_OD0.1_E1e-04_4.0Rac_Pm12_01_0001_meridsect_u_phi_phiav.png}
%         \caption{}
%         \label{fig:4.strong_4.0_c}
%     \end{subfigure}
%     \centering
%         \begin{subfigure}{0.21\linewidth}
%         \centering
%         \includegraphics[width=\linewidth,trim=2.0cm -2.4cm 0.3cm 0cm, clip]{Figures/mcv_OD1.0_E1e-04_4.0Rac_Pm12_02_0004_meridsect_u_phi_phiav.png}
%         \caption{}
%         \label{fig:4.strong_4.0_d}
%     \end{subfigure}
%     \caption{As in Figure \ref{fig:4.strong_2.649} with stronger forcing, $\Ra'=4.0$.}
%     \label{fig:4.strong_4.0}
% \end{figure}

\paragraph{Strong-field regime.}
\firstrevision{With large enough driving, a strong-field regime emerges for all values of $p_{10}$ tested. When the imposed field is weak, $p_{10}=0.1$, ``strong-field'' behaviour emerges abruptly as $\Ra'$ is increased and coincides with the breaking of equatorial symmetry. This sharp transition is associated with the sudden excitation of a new set of modes \citep{GusevaEtAl2025}, which significantly enhances the strength of the poloidal flow (right panel of Figure~\ref{fig:4.TorPolu}) and hence the poloidal magnetic field (right panel of Figure~\ref{fig:4.TorPol}). This increase in magnetic field strength then allows the flow to reorganise into a structure closely resembling that found in strong-field dynamo simulations.
With a stronger imposed field, $p_{10}=1.0$, ``strong-field'' behaviour instead emerges gradually from the vacillating regime (found for $1.6\lesssim\Ra'\lesssim2.0$) with equatorial symmetry broken at $\Ra'\approx2.2$; however, this symmetry breaking has little immediate impact on the overall structure of the flow.

At Rayleigh numbers $\Ra'\gtrsim2.4$, both the compensated-curl force spectra (Figures~\ref{fig:4.ForceBal_Pm12_e},\ref{fig:4.ForceBal_Pm12_f}) and the flow morphology (Figure~\ref{fig:4.strong_2.649}) are similar to the strong-field dynamo regime. Comparing the $p_{10}=0.1$ and $p_{10}=1.0$ cases, we find that a stronger imposed magnetic field weakens the flow, reducing inertia, however the Coriolis force remains at a similar order so that the first-order balance between the Coriolis and Lorentz forces is separated from the second-order terms. We also find that a stronger poloidal magnetic field significantly weakens the zonal flow (Figures~\ref{fig:4.strong_2.649_c}, \ref{fig:4.strong_2.649_d}), allowing the pattern of the equatorial thermal anomaly to emerge, as at $\Pm=5$. With $p_{10}=1.0$, we find little qualitative difference in the force balances of the vacilliating regime at $\Ra'=2.0$ and strong-field regime at $\Ra'=2.4$ (Figures~\ref{fig:4.ForceBal_Pm12_d},\ref{fig:4.ForceBal_Pm12_f})%, although equatorial symmetry is broken in this range.
.}

\begin{figure}
    \centering
    \begin{subfigure}{0.24\linewidth}
        \centering
        \includegraphics[width=\linewidth,trim=2cm 0cm 2cm 0cm, clip]{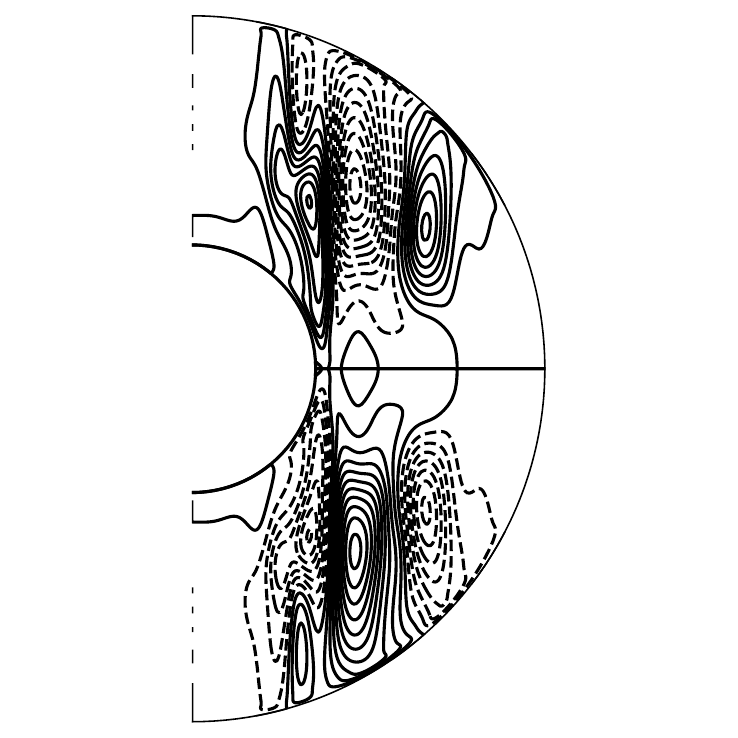}
        \caption{}
        \label{fig:4.Streamlines12a}
    \end{subfigure}
    \begin{subfigure}{0.24\linewidth}
        \centering
        \includegraphics[width=\linewidth,trim=2cm 0cm 2cm 0cm, clip]{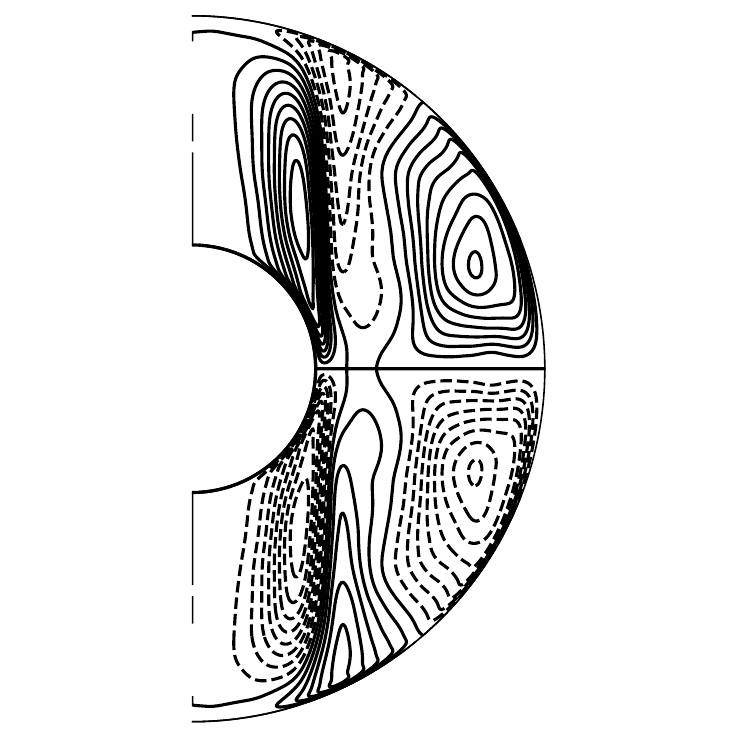}
        \caption{}
        \label{fig:4.Streamlines12b}
    \end{subfigure}
    \begin{subfigure}{0.24\linewidth}
        \centering
        \includegraphics[width=\linewidth,trim=2cm 0cm 2cm 0cm, clip]{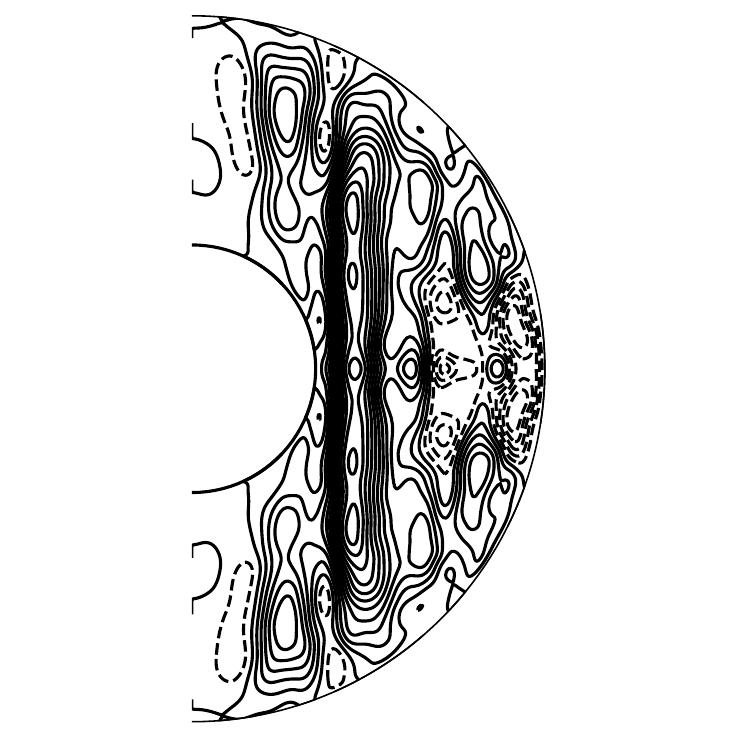}
        \caption{}
        \label{fig:4.Streamlines12c}
    \end{subfigure}
    \begin{subfigure}{0.24\linewidth}
        \centering
        \includegraphics[width=\linewidth,trim=2cm 0cm 2cm 0cm, clip]{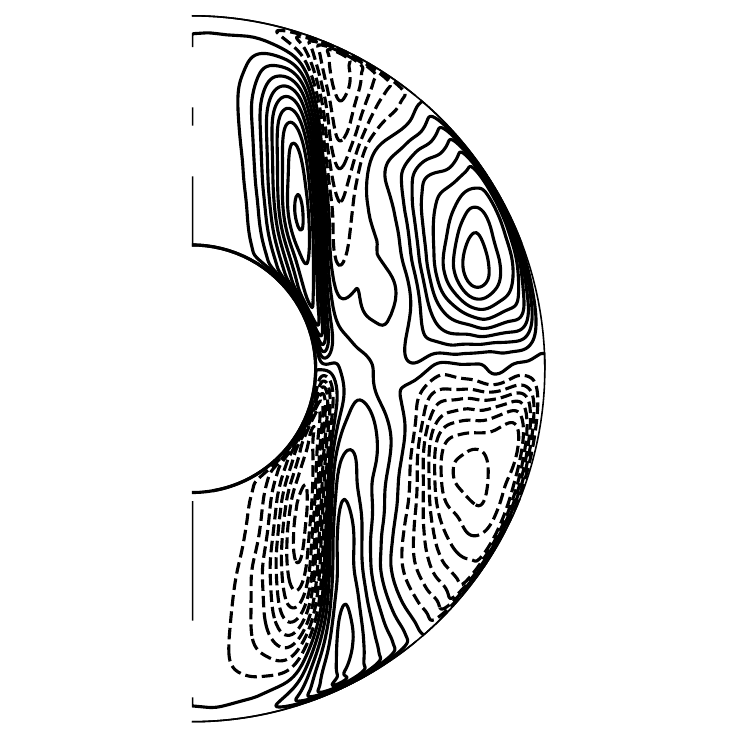}
        \caption{}
        \label{fig:4.Streamlines12d}
    \end{subfigure}
    \caption{Streamlines of the temporally- and zonally-averaged poloidal flow for bistable solutions at $\Ra'=2.4$, $\Pm=12$, $p_{10}=0.1$. (a) The weak-field (equatorially symmetric) branch. (b)-(d) The strong-field branch, displayed as the streamlines of the equatorially symmetric and antisymmetric, and total flow, respectively. Solid streamlines represent clockwise-rotating flow, and dashed streamlines, anti-clockwise.}
    \label{fig:4.Streamlines12}
\end{figure}

\firstrevision{To investigate the role of the breaking of equatorial symmetry, Figure \ref{fig:4.Streamlines12} compares the zonally averaged streamlines for a location in parameter space that is bistable between weak- and strong-field solutions.}
%Figure~\ref{fig:4.Streamlines12a} shows the zonally averaged poloidal streamlines for a weak-field solution at a bistable point in parameter space. 
In Figure~\ref{fig:4.Streamlines12a}, for the weak-field solution, the streamlines are equatorially symmetric and primarily reflect the axial circulation within convection columns, with neighbouring counter-rotating cells persisting after zonal averaging due to slight asymmetries between cyclonic and anti-cyclonic columns.

On the strong-field branch, \firstrevision{Figure~\ref{fig:4.Streamlines12d} shows that} the zonally averaged streamlines remain largely equatorially symmetric, consistent with $s_u>0$ (Figure~\ref{fig:4.Symmetryu}), but the circulation cells extend much further into the equatorial region and inside the tangent cylinder. This reflects the increased magnetic field strength, which allows the Lorentz force to overcome the rotational constraint in regions where the boundaries are closest and (hence) the Coriolis force is most strongly felt.

By decomposing the flow into its symmetric and antisymmetric components (Figures~\ref{fig:4.Streamlines12b}, \ref{fig:4.Streamlines12c}), we also identify the development of a meridional circulation aligned with a cylindrical surface at approximately half the outer radius. \firstrevision{The increase in axial velocity at the equator, where the axial vorticity is typically largest, increases the helicity there, which is otherwise identically zero in perfectly symmetric states. Increased helicity can strengthen or modify the existing magnetic field and, as such, may provide support for the maintenance of a large-scale magnetic field in the strong-field regime.}
\section{Discussion}
\label{sec:conclusion}

% WHY MAGNETOCONVECTION
\firstrevision{Our motivation for this work is a better understanding of how patterns in convective flows, particularly those in rapidly-rotating systems, respond to variation of the magnetic field. Consequently, our study is performed using magnetoconvection simulations where, in contrast to dynamo simulations, the outer boundary field is imposed rather than dynamically-generated. This allows for fine control of the magnetic field strength, alleviates the need for dynamo generated fields (although our model still allows for this), and, consequently, provides access to magnetically-influenced flow regimes in regions of parameter space where dynamo action would otherwise fail. Magnetoconvection is typically a computationally cheaper method since the time required for stabilisation of the numerical state is reduced, and also broadens the parameter space, allowing us to navigate around abrupt transitions where the direction of cause and effect can become difficult to distinguish.

% FORCE BALANCES
A key motivation for this work is the weak-to-strong field transition which has been observed in dynamo simulations \cite{SimitevBusse2009,Dormy2025}. In dynamos, this transition is characterised by a significant strengthening of the dipolar component of the magnetic field. Using magnetoconvection simulations, we are able to decouple how more subtle changes in the magnetic field strength affect properties of the flow. One commonly used measure of the dynamical state of the system is the force balance, obtained, for example, by calculation of the compensated-curls of individual terms of the Navier-Stokes equation \cite{TeedDormy2023}.
%; there are several ways to measure this, each of which with their own merits \cite{TeedDormy2023}, but we have opted to use the compensated-curl of forces, which is straightforward to numerically calculate and excludes the geostrophic flow, which mathematically acts only as a Lagrange multiplier on the Navier-Stokes operator. 
In such force balances, we broadly show that sharp changes observed in simulations of the weak-to-strong dynamo transition (\S\ref{sec:3.dynamo}) can be separated into those due to the emergence of a wider set of onset modes, once the forcing becomes sufficiently large that the primary onset mode can no longer suppress them, and those due to the growth of the Lorentz force, which is a more subtle transition that depends on $\Pm$ and $p_{10}$ and varies across scales.

% CONNECTION WITH DYNAMOS
\firstrevision{We have broadly mapped the parameter space of magnetoconvection near to the known weak-to-strong field transitions in equivalent dynamo simulations. At $\Pm=1$, as in dynamo simulations, we have not found a strong-field regime, even when the imposed field is large. Instead, with strong imposed fields, we obtain magnetostrophic modes at low forcing, similar to Mason et al. \cite{MasonEtAl2022} and jet-like structures at higher forcing, similar to those of Sakuraba \cite{Sakuraba2007}. With a larger magnetic Prandtl number ($\Pm\geq5$) we are able to access the strong-field regime at all of the imposed field strengths studied although, at $\Pm=5$, the magnetic field diffuses more rapidly than at $\Pm=12$, so we generally find that the Lorentz force, and its associated features, become significant at lower forcing, $\Ra'$, and $p_{10}$.
We have not explored parameter space for $\Ra'>5$, but in dynamo models larger forcing typically results in either (depending on $\Pm$) a collapse of the weak-field regime and the emergence of multipolar dynamos with a CIA force balance \cite[e.g.][]{ChristensenAubert2006} or a consolidation of the strong-field regime \cite{TeedDormy2025b}.}

% COLUMNS
An important measure of the flow is its columnarity measuring the prevalence of vertically-aligned structures, which are preferred in rapidly-rotating systems in situations where no other force is able to break the Taylor-Proudman constraint \cite{SoderlundEtAl2012}. A reduction in columnarity correlates well with the rise of the Lorentz force since this is the primary term responsible for breaking this constraint at our applied level of driving, where inertia remains small. In principle a new constraint can be formulated in terms of the magnetic field \cite[e.g.][]{Sakuraba2002}, however the field is, in practice, too complex for this constraint to be meaningful, particularly at larger magnetic Prandtl numbers. In dynamo models we showed that the weak-to-strong transition coincides with an abrupt reduction in columnarity (compared to non-magnetic flows) and a loss of equatorial symmetry in the previously columnar, weak-field state. However, despite this, we find that columnarity alone can be insufficient as an indicator for the weak-to-strong field transition through which it can vary in an apparently continuous manner. The columnarity calculations from our magnetoconvection models
%in Figure~\ref{fig:Columnarity_dynamos},\ref{fig:4.Columnarity} 
indicate that the loss of equatorial symmetry need not emerge as an instability of the columns themselves, since the columnar character of the flow in the bulk can be lost at Rayleigh numbers significantly below those at which symmetry breaking occurs. Our magnetoconvection results indicate that at large magnetic Prandtl number and with strong imposed fields columns can become unstable, here clearly exemplified in the form of the vacillating regime, but without an associated breaking of equatorial symmetry. Instead, equatorial symmetry is broken by the separate onset of convective modes inside the tangent cylinder, whose eigenfunctions have no preference for equatorial symmetry or antisymmetry (Appendix~\ref{app:symmetry}).}

% POLES APART
Gilman \cite{Gilman1975} considered the onset of convection inside the tangent cylinder, which \firstrevision{is typically a quiescent region until long after convection has set in elsewhere in the domain} as a result \firstrevision{of the correlation between the radial and axial directions and} the enhanced constraint of rapid rotation due to the closeness, in $z$, of the \firstrevision{no-slip} inner and outer boundaries. The Lorentz force, in rapidly-rotating magnetoconvection, can overcome this Taylor-Proudman constraint at significantly lower Rayleigh numbers than the viscous force alone \citep{Fearn1979b}. Sreenivasan \& Jones \cite{SreenivasanJones2006} find that an axial magnetic field reduces the critical Rayleigh number by around a factor of three; the further addition of a strong toroidal magnetic field, which, as we demonstrated, is also present in the tangent cylinder in models of weak-field magnetoconvection, would only compound this effect. For instance, at $\Pm=5$, symmetry-breaking can occur at around $\Ra'=3.3$, (three times less than in purely hydrodynamic simulations where symmetry-breaking must wait for inertial effects), whilst at $\Pm=12$, this occurs at $\Ra'=2$.

An analysis of the linear problem, including a tracking of individual onset modes, was performed by \cite{Sakuraba2002}. However, further consideration of the linear problem may be useful to isolate the onset of the polar convective modes and the effect of variation of the magnetic Prandtl number. Accurately determining the critical Rayleigh number for the onset of polar (magneto)convection in a spherical shell may enable  prediction of the onset of the strong-field branch and suggest regions of bistability in dynamo simulations, without the need to perform suites of costly simulations.

\firstrevision{
\begin{figure}[ht]
    \centering
    \includegraphics[width=0.8\linewidth]{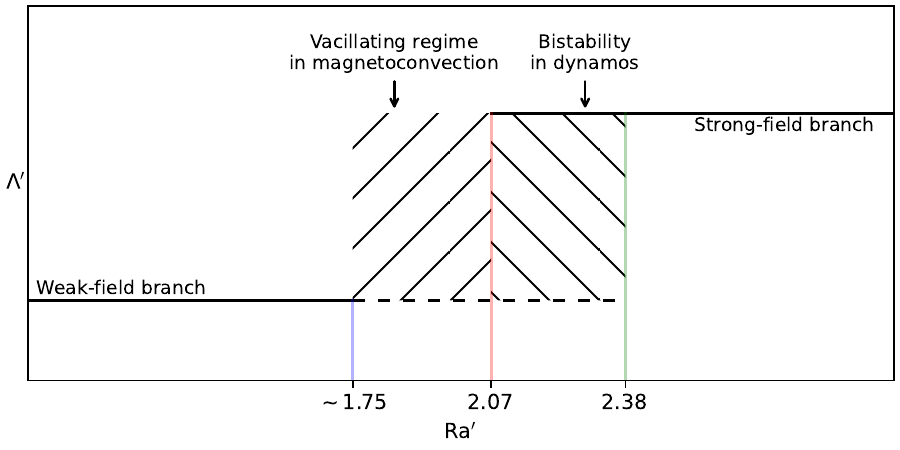}
    \caption{\firstrevision{Sketch of the vacillating magnetoconvection regime in the $\Pm=12$ case, and its relation to the weak- and strong-field dynamo branches. Solid lines represent stable branches in both magnetoconvection and dynamo simulations and the dashed part of the weak-field branch is stable for dynamos, but unstable in magnetoconvection with a sufficiently strong imposed field. The blue line represents the location in $\Ra'$-space of the lower terminus of the vacillating regime in magnetoconvection simulations. The red (green) line represents the location in $\Ra'$-space of the lower (upper) terminus of the strong-field (weak-field) dynamo branch. Dynamo regime thresholds taken from \cite{TeedDormy2025b}.}}
    \label{fig:vacill_sketch}
\end{figure}

% VERY VIGOUROUS VACILLATION
Another interesting feature of the parameter space of magnetoconvection is the emergence of a vacillating regime, in which convection columns intermittently form and break apart. The term vacillating has been used in the past to describe periodic or quasi-periodic oscillations of thermal Rossby waves, corresponding to a weakly supercritical (hydrodynamic) instability \cite{BusseOr1986,OrBusse1987,Zhang1992}. Our state clearly arises through a very different instability, magnetic rather than hydrodynamic, but is similar in essence, nonetheless. We find this state when the magnetic diffusion is small (the magnetic Prandtl number is large) and when the imposed magnetic field is of the order of that produced by a strong-field dynamo. A possible sketch of the emergence of this regime is given in Figure~\ref{fig:vacill_sketch} using values for the paradigmatic $\Pm=12$ case.

The strong-field branch can first emerge in dynamos at $\Ra'\simeq2.07$ when it is possible for a self-sustaining feedback mechanism to persist between the Lorentz force and a large-scale, non-columnar flow, overcoming the Taylor-Proudman constraint; this is the red line in Figure~\ref{fig:vacill_sketch}. The green line denotes the Rayleigh number at which the weak-field, columnar branch becomes unstable in dynamo simulations, and forms the uppermost bound for the bistable region. In dynamo simulations, the large-scale field of the weak-field branch is considerably weaker than on the strong-field branch; thus, when these columns become unstable to their own induced (small-scale) magnetic field, the weak-field branch becomes untenable and the flow reorganises towards the other stable branch, the strong-field branch.

In magnetoconvection, when the imposed field is equivalent in strength to that of a strong-field dynamo (cf. Figure~\ref{fig:Ra_vs_p10_dynamos}), the strong-field-like state stabilises at a similar value of Rayleigh number (the red line). The weak-field branch, however, is characterised by columnar convection, and these columns become unstable at much smaller Rayleigh numbers when the imposed field is large (here in the range $1.6<\Ra'<1.75$ when $p_{10}=1.0$), which is denoted by the blue line in Figure~\ref{fig:vacill_sketch}. At these Rayleigh numbers there is no stable strong-field branch to which the magnetic field can reorganise. This leads to the emergence of the vacillating regime, where columns periodically break up and reform once the induced magnetic field, which is produced by the convection columns, decays due to their absence.

A key question is the relevance of the vacillating regime, in nature and in models. The necessary ingredients for these magnetically-induced vacillations appear to be weakly supercritical convection (to maintain columnar structures) and large magnetic Prandtl number (so the magnetic field can grow sufficiently large inside columns to trigger an instability). For our set-up we find the requirement of $\Pm\gtrsim5$, with no such state found at $\Pm=1$. The cross-vortex length scale of the columns is set by a balance of viscous and Coriolis forces; hence the existence of this type of vacillation will depend directly (though not necessarily linearly) on $\Pm$.
%We are unaware of a regime satisfying these criteria existing in nature, and it seems likely to be rare due to 
The natural properties of liquid metals, which typically diffuse magnetic fields much more rapidly than momentum, heat or composition may not satisfy the constraints we have found on $\Pm$ for the existence of a vacillating regime at the values of the other parameters used in this study. Nevertheless, it is possible that the vacillating regime could exist in magnetoconvection (or dynamo) models for $\Pm<1$ as parameters are lowered systematically as part of distinguished limit \cite{Dormy2016}.}
%except in particularly hot plasmas, which are likely to be in a state well past the onset of convection. Therefore, this new regime is likely confined to models, but does still shine a light on the processes underlying the weak-to-strong field transition, particularly at large magnetic Prandtl numbers. 

\section{Conclusion}
\label{sec:conclusion2}

% WE REST OUR CASE FOR MAGNETOCONVECTION
\firstrevision{
In this work we investigated the effect of the magnetic field strength on convective fluid flows in spherical geometry of Earth's outer core with a particular focus on the transition between solutions where the induced magnetic field is either ``weak'' or ``strong''; we find a correlation in parameter space between these regimes in magnetoconvection simulations and their equivalents in dynamo simulations. Key dynamical and structural indicators of the flow such as the force balance, equatorial symmetry, and level of vertical dependence can be subtly or drastically influenced by the strength of imposed magnetic field and magnetic diffusion. Our work aids understanding of the role the magnetic field plays in the transition from columnar weak-field to chaotic strong-field states. We further highlighted the ability of magnetoconvection studies to tease out why such transitions do not occur under some parameter regimes in dynamo models. The transition from weak-field dipolar to multipolar dynamo solutions at large driving and the apparent lack of such a transition for the strong-field dipolar regime \cite{Petitdemange2018,TeedDormy2025b} may also be worthy of investigation.}
%Two further transitions are \firstrevision{also} important in understanding the solution space of (Boussinesq, spherical-shell) dynamo simulations: these are The transition between regimes with multipolar and  dipolar magnetic fields, 
% and the transition \firstrevision{of strong-field states at} $\Pm>1$ to those $\Pm\ll1$, . These two problems can both be better understood making use of magnetoconvection simulations, and could be the subject of future work.

\firstrevision{With dynamo models now making use of large magnetic Prandtl numbers to promote strong-field dynamos \cite{TeedDormy2025b} and reversals \cite{JonesTsang2025}, there is an ongoing need to reduce $\Pm$ below unity, which is important for correctly capturing the effect of small-scale turbulence in models \citep{Tobias2021}.
%so as to properly capture the dynamics and regime transitions in convection. 
Transitioning from $\Pm>1$ to $\Pm\ll1$, however, whilst maintaining strong-field dynamos is a highly challenging task requiring significant computing power, partly as a result of the need to also reduce the Ekman number in tandem. Alternative models, such as magnetoconvection may overcome this limitation since states with strong fields could be explored at numerically achievable Ekman numbers yet at values of $\Pm$ below those needed to sustain a dynamo. Our work highlighted the utility of magnetoconvection studies in replicating, and enhancing understanding of, transitions in dynamo models; it could therefore enable the study of properties of strong-field states with  with $\Pm\ll1$, helping us to better understand the extreme regime in which the geodynamo operates.}

\appendix
\section{Magnetic boundary conditions}
\label{app:MagBCs}

\firstrevision{In this paper, we compare two classes of MHD simulation: dynamo and magnetoconvection simulations. Since in our work magnetoconvection is implemented via altering the magnetic boundary condition at the outer boundary, in this Appendix we discuss three types of boundary conditions that can be implemented. First, dynamo simulations (\S\ref{sec:3.dynamo}) correspond with taking boundaries to be electrically-insulating so that any magnetic field emerges only from induction of the conducting fluid in the domain. Second, for the magnetoconvection simulations of our work (S\ref{sec:4.columnar}), we enforce that the toroidal component of the magnetic field vanishes (as at an insulating boundary) and the radial component of the field is fixed to be sinusoidal with latitude with variable magnitude (the remaining component is determined dynamically). Similar conditions have been used in previous studies; \cite{TeedEtAl2015}, for example. Third, magnetoconvection has also been used to describe simulations with a fixed external source, as in \cite{MasonEtAl2022}. Whilst this is easiest to motivate physically, it allows for less fine control of the large-scale magnetic field strength, which we hope to determine the effect of. We do not implement this third condition in our work but it is discussed here for completeness. The mathematical definition of each of these three boundary conditions is given below.}

\paragraph{Electrically-insulating boundaries.}

To a reasonable approximation, the Earth's mantle is electrically insulating. In an insulating medium, the current vanishes,
\begin{equation}
    \vec{j}=\vec{0}.
\end{equation}
Since the magnetic field and the current (its curl) are both solenoidal fields, they can be decomposed into poloidal and toroidal components as
\begin{subequations}
\begin{align}
    \vec{B}&=\curl\curl\left(B_p\vec{r}\right)+\curl\left(B_t\vec{r}\right),\\
    \vec{j}&=\curl\curl\left(j_p\vec{r}\right)+\curl\left(j_t\vec{r}\right),
\end{align}
\end{subequations}
respectively. By definition, $\vec{B}$ and $\vec{j}$ are related by
\begin{equation}
    j_p=B_t,\qquad j_t=-\laplacian B_p,
\end{equation}
and hence, the magnetic field in an insulating medium must satisfy
\begin{equation}
    B_t=\laplacian B_p=0.
\end{equation}
Given that $B_p$, $B_t$, can be decomposed into spherical harmonics as
\begin{equation}
    B_p=\sum_l\sum_m P^{m}_{l}\left(r\right)Y^{m}_{l}\left(\theta,\phi\right),\quad B_t=\sum_l\sum_m T^{m}_{l}\left(r\right)Y^{m}_{l}\left(\theta,\phi\right),
\end{equation}
then
\begin{subequations}
\begin{align}
    T_{l}^{m}&=0,\\\left[r^2\odv{^2}{r^2}+2r\odv{}{r}-l\left(l+1\right)\right]P_{l}^{m}&=0,\label{eqn:ext_P_cdn}
\end{align}
\end{subequations}
for all $l,m$. Equation \eqref{eqn:ext_P_cdn} admits solutions $P^{m}_{l}\sim r^{\alpha}$ with $\alpha$ satisfying $\alpha\left(\alpha+1\right)=l\left(l+1\right)$ which has particular solutions $\alpha=l$ (corresponding to external sources, since $l>0$) and $\alpha=-l-1$ (corresponding to internal sources).

If there are no external sources then the exterior magnetic field is produced entirely by fluid in the spherical shell and $P^{m}_{l}=Ar^{-l-1}$, for some constant $A$; the boundary condition on the internal field, matching with this solution is
\begin{equation}
    \left(r\odv{}{r}+l\left(l+1\right)\right)P^{m}_{l}\biggr|_{r=\router}=0,\quad\forall~l,m.\label{eqn:PolBC1}
\end{equation}
At the inner boundary the role of internal and external sources is reversed by the geometry in relation to increasing $r$, i.e., sources correspond to $\alpha=l$, therefore, the no-source condition at the inner boundary is
\begin{equation}
    \left(r\odv{}{r}-l\right)P^{m}_{l}\biggr|_{r=\rinner}=0,\quad\forall~l,m.
\end{equation}
\firstrevision{These conditions are used for our dynamo simulations of \S\ref{sec:3.dynamo}.}

\paragraph{Fixed radial magnetic field.}

A significant difference between weak- and strong-field dynamos is the strength of the dipolar component of the magnetic field. To better understand the direct effect this has on the flow, we impose a fixed morphology and strength at the outer boundary. To match with the electrically-insulating conditions above, we maintain the zero condition on the toroidal scalar of the magnetic field as zero but fix the poloidal scalar, i.e., the radial component of the field.

The radial component of the magnetic field is given by
\begin{equation}
    B_r\left(r,\theta,\phi\right)=\sum_{l,m}l\left(l+1\right)\frac{P_{l}^{m}\left(r\right)}{r}\cdot\hat{Y}^{m}_{l}\left(\theta,\phi\right)\label{eqn:radialpol}.
\end{equation}
\firstrevision{Say that we set the dipolar component of the poloidal function, $P_{1}^{0}(\router)=p_{10}\in\mathbb{R}$, with $P^{m}_{l}=0$, for all $\left(l,m\right)\neq\left(1,0\right)$. Then}
\begin{equation}
    B_{r}\left(\router,\theta,\phi\right)\iffalse{=2\frac{C}{\router}\left[\frac{1}{2}\odv{}{\mu}\left(\mu^2-1\right)\right]_{\mu=\cos{\theta}}}\fi=\frac{2\firstrevision{p_{10}}}{\router}\cos{\theta}=\frac{13\firstrevision{p_{10}}}{10}\cos{\theta},
\end{equation}
%\robnote{Here we are using the generating function but we haven't really explained that and we didn't use it in the external sources section above, even though we could have done.}
where $\router=20/13$, when using radius ratio, $\chi=0.35$, and normalised shell thickness. Fixing the poloidal scalar at the boundary is equivalent to perfectly electrically-conducting conditions on the poloidal component of the magnetic field, but not on the toroidal component, where $T_l^m=0$ from the insulating conditions is retained. \firstrevision{These conditions are used for our magnetoconvection simulations of \S\ref{sec:4.columnar}.}

\paragraph{External sources.}

Say an external source at the outer boundary induces a uniform axial magnetic field of the form $\vec{B}=B_0\vnd{z}$ where $\vnd{z}=\cos{\theta}\vnd{r}-\sin{\theta}\vnd{\theta}$. Then, given $B_r=r^{-1}L^2B_p$ (where $L^2=l(l+1)$, is the angular component of the spherical Laplacian), the axial field is
\begin{equation}
    B_p(r,\theta,\phi)=P^{0}_{1}(r)\cos{\theta},\quad P^{0}_{1}(r)=\frac{B_0}{2}r.
\end{equation}
The exterior magnetic field is then a compound of the internal and external sources, i.e.,
\begin{equation}
    P^{0}_{1}=\frac{B_0}{2}r+\frac{A^{0}_{1}}{r^2},\qquad P^{m}_{l}=A^{m}_{l}r^{-l-1},~\left(l,m\right)\neq\left(1,0\right),\label{eqn:PolBC2}
\end{equation}
with $A^{m}_{l}$ arbitrary constants. This implies that the boundary conditions on the $(1,0)$ mode are modified to
\begin{equation}
    \left(\odv{}{r}+\frac{2}{r}\right)P_{1}^{0}(r)\biggr|_{r=\router}=\frac{3B_0}{2},
\end{equation}
whilst the boundary conditions on all other modes remain the same.

\firstrevision{These conditions are not used in our work but} this form of magnetoconvection is adopted by \cite{MasonEtAl2022}, for example.

\section{Equatorial symmetry}
\label{app:symmetry}

In this appendix, we explain the origin of the terms \emph{equatorial} and \emph{polar} symmetry and their connection with the odd and even modes of (magneto)convection (sometimes called modes with Roberts/Busse symmetry) that are reported by many authors \citep[e.g.][]{Fearn1979b,Sakuraba2002,SeeleyEtAl2025}.

\paragraph{Full sphere}
Roberts \cite{Roberts1968} derived the equations governing the onset of hydrodynamic convection in a sphere via internal heat sources; he noted that, after considering solutions in the form of spherical harmonics, the structure of the resultant equations leads to two distinct families of solutions emerging, which are characterised by the parity of $l+m$, where $l,m$ are the spherical harmonic degree and order, respectively, of $U$ and $T$, where
\begin{equation}
    \vec{u}=\curl\curl{U\vec{r}}+\curl{V\vec{r}}.
\end{equation}
The coupled harmonics of $V$ have opposite parity. Hence, it can be seen that for even modes (modes with $l+m$ of $U$, even) $u_r$, $u_\phi$, and $T$ will be symmetric about the equator, whilst $u_\theta$, $u_z$ will be antisymmetric, corresponding to a mirror symmetry of $\vec{u}$ about the equator. Similarly, odd modes correspond with mirror-antisymmetry about the equator.

In the axisymmetric, $m=0$, case, \cite{Roberts1968} show that the odd modes are the most unstable - this solution corresponds to a global meridional circulation, with streamlines corresponding to hemispherical rings. Roberts similarly calculates the critical Rayleigh numbers for odd modes with $m>0$. For this reason, odd modes can be referred to as Roberts symmetry.

Busse \cite{Busse1970} then tackles a perturbation analysis for the spherical convection problem and calculates the critical Rayleigh numbers of the even modes, showing that these are preferred at onset for $m>0$ and generally preferred when considering all $m$, for small Ekman numbers. These modes include  tall thin columnar structures appearing at onset as thermal Rossby waves. This symmetry is sometimes called Busse symmetry and is frequently taken as an assumption in linear onset convection studies to simplify the analysis.

Meridional convection is efficient at transporting heat since a significant portion of the velocity is directed in the radial direction; however, the gradients of the velocity parallel to the rotation axis will also be large, so at onset the flow must contest with the stabilising effect of the Taylor-Proudman constraint. Thermal Rossby waves are less efficient at transporting heat since the flow is only partly radial, except at the equator; however, they can almost take the form of perfect Taylor columns, with only small perturbations required at the outer boundaries. Hence, for sufficiently small Ekman numbers, it is safe to expect that the Busse modes will have a smaller critical Rayleigh number.

\begin{figure}
    \centering
    \begin{tikzpicture}
        % =========================
% LEFT FIGURE
% =========================
% Outer sphere boundary (half)
\draw (0,0) arc (-90:90:3);
% Inner core
\draw (0,2) arc (-90:90:1);
% Tangent cylinder
\draw[dashed] (1,6) -- (1,-0);

% Circulation cell (outer)
\draw[blue,postaction={decorate},decoration={markings, mark=at position 0.1 with {\arrow{>}}, , mark=at position 0.6 with {\arrow{>}}}] (2,3) ellipse (0.8cm and 1.5cm);

% =========================
% RIGHT FIGURE
% =========================
% Shift
\begin{scope}[xshift=5cm]

% Outer sphere boundary (half)
\draw (0,0) arc (-90:90:3);
% Inner core
\draw (0,2) arc (-90:90:1);
% Tangent cylinder
\draw[dashed] (1,6) -- (1,-0);

% Circulation cells (top, bottom)
\draw[blue,postaction={decorate},decoration={markings, mark=at position 0.1 with {\arrow{<}}, , mark=at position 0.6 with {\arrow{<}}}] (0.35,4.95) ellipse (0.2cm and 0.75cm);
\draw[blue,postaction={decorate},decoration={markings, mark=at position 0.1 with {\arrow{<}}, , mark=at position 0.6 with {\arrow{<}}}] (0.35,1.05) ellipse (0.2cm and 0.75cm);

\end{scope}
    \end{tikzpicture}
    \caption{Streamlines of (a) Meridional convection outside the tangent cylinder. (b) Meridional convection inside the tangent cylinder.}
    \label{fig:MeridConv}
\end{figure}

\paragraph{Effect of an inner core}
If an inner core is present, global meridional convection becomes impossible, and large scale meridional convection must be restricted to the outside of the tangent cylinder (Figure \ref{fig:MeridConv}a). Smaller convection cells could also develop inside the tangent cylinder (Figure \ref{fig:MeridConv}b); these latter modes can be efficient in transporting heat, since streamlines abut both the inner and outer boundaries, however, they are impeded by viscosity due to their small scale. Gilman \cite{Gilman1975} refers to these modes as \emph{polar convective modes}. In the absence of the region outside the tangent cylinder the two polar regions would be disconnected and there would be no preference for either equatorially symmetric or antisymmetric modes; if we consider the outer continuation of this mode, then the Roberts mode, which aligns with the meridional circulation in Figure \ref{fig:MeridConv}a, will be slightly preferable at onset \citep{Sakuraba2002}.

The critical radius for thermal Rossby waves will be outside the tangent cylinder, where heat from the inner core can be transported radially at the equator; these modes also peak in amplitude at the equator hence \cite{Gilman1975} refers to these as \emph{equatorial modes}. These will onset before the polar mode for the reasons already discussed.

An equatorial mode will significantly suppress other equatorial modes (of either symmetry), leading to strongly columnar states just above the onset of convection (Figure \ref{fig:Columnarity_hydro}). However, the polar modes might be expected to onset almost independently \citep{Gilman1975} - particularly if the inner core is large. The nonlinear terms of the momentum equation will preserve equatorial symmetry (but not antisymmetry); hence, the onset of these polar modes is important in breaking the equatorial symmetry of the onset state (Figure \ref{fig:Symmetry_hydro}.

\paragraph{Effect of a magnetic field}
Fearn \cite{Fearn1979b} shows that a toroidal field can destabilise the system so that convection occurs at a lower critical Rayleigh number, provided the magnetic field strength is large enough. The most unstable modes in this case are modes with Busse symmetry which are either modified Rossby modes or magnetostrophic modes when $\Els$ is sufficiently large. Sakuraba \cite{Sakuraba2002} investigates the effect of an axial magnetic field and shows that the polar convective mode can be destabilised at much smaller $\Els$ (in fact, Sakuraba finds that an antisymmetric mode can sometimes be the preferred mode of convection).

In conclusion, we use the terms equatorial and polar modes to describe the modes which peak in amplitude at the respective locations; the most unstable modes in each class are equatorially symmetric and antisymmetric respectively. One cause of confusion is that the magnetic field is a pseudovector, and hence magnetic field lines have the opposite symmetry to the flow.

\bibliography{References}{}
\bibliographystyle{unsrt}

\end{document}